%% file: PHYS_REP_Main.tex
\documentclass[preprint,12pt]{elsarticle}

\usepackage{amssymb}
\usepackage{amsmath}

\journal{Physics Report}
\newcommand{\rmvect}[1]{\boldsymbol{\mathrm{#1}}} 
\newcommand{\vect}[1]{\boldsymbol{#1}} 
\newcommand{\oper}[1]{\boldsymbol{\mathsf{#1}}} 
\newcommand{\ket}[1]{\ensuremath{\left|{#1}\right\rangle}} 
\newcommand{\bra}[1]{\ensuremath{\left\langle{#1}\right |}} 
\newcommand{\sinc}{\ensuremath{\mathrm{sinc}}} 

\newcommand{\mmod}{\ensuremath{{\mathrm{mod}}}}
\begin{document}

\begin{frontmatter}


\title{Spatial correlations in parametric down-conversion}
\author[ufrj]{S. P. Walborn}
\author[ufmg]{C. H. Monken}
\author[ufmg]{S. P\'adua}
\author[ufrj]{P. H. Souto Ribeiro}
\address[ufrj]{Instituto de F\'{\i}sica, Universidade Federal do Rio de
Janeiro, Caixa Postal 68528, Rio de Janeiro, RJ 21941-972, Brazil}

\address[ufmg]{Universidade Federal de Minas Gerais, Caixa Postal 702, Belo Horizonte, MG 30123-970, Brazil}

\begin{abstract}
The transverse spatial effects observed in photon pairs produced by
parametric down-conversion provide a robust and fertile testing
ground for studies of quantum mechanics, non-classical states of
light, correlated imaging and quantum information.  Over the last 20
years there has been much progress in this area, ranging from technical
advances and applications such as quantum imaging to investigations of fundamental
aspects of quantum physics such as complementarity relations, Bell's inequality violation and entanglement.  The field has grown
immensely:  a quick search shows that there are hundreds of papers
published in this field, some with hundreds of citations.  The
objective of this article is to review the building blocks
and major theoretical and experimental advances in the field, along with some possible technical
applications and connections to other research areas.   
\end{abstract}
\end{frontmatter}

\tableofcontents
\input{section1}

\input{section2}
\input{section3}
\input{section4}
\input{section5}

\input{section6}
\input{section7}
\input{section8}

\input{section9}
\input{section10}



The authors acknowledge financial support from the Brazilian funding
agencies CNPq, CAPES, PRONEX, FAPEMIG and FAPERJ.
This work was performed as part of the Brazilian National Institute of Science and Technology for Quantum Information.

\bibliographystyle{apsrev}


\end{document}

%% file: section1.tex
\section{Introduction}
\label{sec:intro}

Correlations between optical fields produced in spontaneous parametric down-conversion (SPDC) were first observed in 1970 by Burnham and Weinberg \cite{BurnhamWeinberg}.
In this early experiment, they investigated the temporal and spatial correlation between pairs of photons. 
From the theoretical point of view, it was Zeldovich, Krindach and Klyshko who first studied the
statistical properties of the light produced in this process, as early as 1969 \cite{zeldovich69,klyshko69b}. The temporal correlations were studied and utilized in the pioneering series of work of Prof. Leonard Mandel and his group \cite{friberg85,hong85,hong86,ghosh87,ou88b}, giving rise to the use of the term {\em twin photons}, since the temporal correlation indicates that pairs of photons are born simultaneously.  These correlations are of fundamental importance in all experiments with twin photons. They provide the time correlation which allows for the post-selection of photon pairs born from the same pump photon, which may present correlations in other degrees of freedom, such as the transverse position and momentum.  Both the time and spatial correlations have been proven to be non-classical. The simple observation of a coincidence rate above the rate of accidental coincidence counts already leads to the violation of a classical inequality \cite{zou91}.  
In the early Burnham and Weinberg experiment \cite{BurnhamWeinberg},
spatial correlations were already observed. They demonstrated that the transverse profile of the coincidence distribution in the far field is narrower than the single photon intensity distribution. However, more comprehensive investigations were made only in the 1990s. For instance, Grayson and Barbosa \cite{GraysonBarbosa} investigated spatial correlations for use in induced coherence without induced emission \cite{zou91b}. Joobeur et al. \cite{joobeur94} studied general spatial and temporal coherence properties, and Souto Ribeiro et al. \cite{Ribeiro94a}
started a long series of experiments using double-slit apertures and coincidence detection,
initially dedicated to the study of spatial coherence properties of twin photons. 
Strekalov et al.\cite{strekalov94}
performed a novel double-slit experiment to observe {\em ghost} interference fringes and Pittman et al. \cite{pittman95}
initiated a series of experiments on what was called {\em Quantum Imaging}, with the two-photon field from SPDC.  
Monken et al.\cite{monken98a}
demonstrated that the spatial correlations could be controlled through the shaping of the angular spectrum of the pump beam. This control allowed the use of the spatial correlations in a series of applications, ranging from fundamental aspects of Quantum Mechanics such as the measurement of the photonic de Broglie wavelength of a two-photon wave packet \cite{fonseca99b} and the observation of the spatial anti-bunching of photons \cite{nogueira01,nogueira02},
to applications to quantum information such as, for instance, the production of spatial qudits \cite{neves05}.  Howell et al. \cite{howell04}
and D'Angelo et al. \cite{dangelo04}
formally demonstrated entanglement between the spatial properties of the twin photons.
\par
In this review, we will discuss the spatial correlations between twin photons produced in parametric down-conversion, 
starting with the basic concepts and basic theory for the description of the down-conversion process and temporal simultaneity. We will 
introduce some basic concepts related to the classical spatial coherence of light fields and propagation of paraxial light beams. 
The theoretical framework for the description of quantum spatial correlations will be introduced and the remainder of the 
text will be dedicated to the description and discussion of several key experiments and applications: double-slit experiments, quantum images, 
demonstration of non-classicality, spatial entanglement, transverse modes and applications to quantum information.

%% file: section2.tex
\section{Fundamentals of parametric down-conversion}
\label{sec:fundamentals}
\begin{figure}
\begin{center}
\includegraphics[width=4cm]{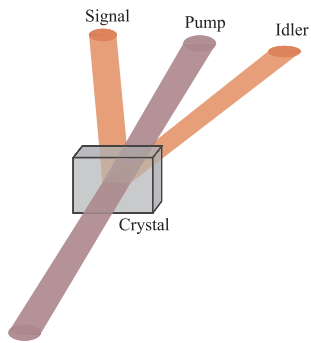}
\caption{Pumping a non-linear crystal with an intense laser beam produces low intensity signal and idler beams by the non-linear process of spontaneous parametric down-conversion.}
\label{fig:SPDC}
\end{center}
\end{figure}
In SPDC, a nonlinear birefringent crystal is typically pumped by an intense pump laser beam, producing low intensity signal and idler fields, as illustrated in Fig. \ref{fig:SPDC}.  The pump laser has frequency $\omega_p$ and is typically in the UV or violet region, while the down-converted signal and idler fields have frequencies $\omega_s$ and $\omega_i$ that are usually in the visible or near infra-red region of the spectrum.   
There have been many theoretical treatments of SPDC, beginning with
the early work of Klyshko \cite{klyshko69a} and later by Hong and
Mandel \cite{hong85}.  A detailed account of these calculations is out
of the scope of this review, but can be found in
the Ph.D. thesis of L. Wang, \cite{wang92} and the books by D. N. Klyshko \cite{klyshko88}, Mandel and Wolf\cite{mandel95} and Ou\cite{ou06}.  
In this section, we follow the treatment presented in Refs.  \cite{hong85,wang92} to introduce the
main features related to the temporal correlations between twin photons produced in SPDC.
Even though our main concern is the spatial correlation, their time correlation is essential in all experiments.

\subsection{Classical and quantum parametric interaction}
\label{subsec:parametricinteraction}

The quantum theory of parametric down-conversion can be
derived from the classical description of the nonlinear interaction, followed by 
the quantization of the electromagnetic field.  Up to second order, the components of the electric polarization of a
nonlinear and non centrosymmetric optical medium when an electric
optical field $\rmvect{E}$ propagates through it is
\cite{bloembergen77,shen84}
\begin{eqnarray}
\label{nonlinear}
{P}_{i}(\rmvect{r},t) &= &
\,\epsilon_{0}\int\limits_{0}^{\infty}dt^{\prime} \chi^{(1)}_{ij}(t^{\prime})
{E}_{j}(\rmvect{r},t-t^{\prime}) + \\
&& \int\limits_{0}^{\infty}dt^{\prime}\int\limits_{0}^{\infty}dt^{\prime\prime}
\chi^{(2)}_{ijk}(t^{\prime},t^{\prime\prime}){E}_{j}(\rmvect{r},t-t^{\prime})
{E}_{k}(\rmvect{r},t-t^{\prime\prime}),\nonumber
\end{eqnarray}
where ${E}_j(\rmvect{r},t)$ is the
$j$ component of the electric field vector and $\chi^{(1)}$ and $\chi^{(2)}$ are the first and second order
susceptibility tensors, respectively. Summation over all combinations of the
tensor components and electric field components is assumed.
For light sources with low electric field strengths
compared with the characteristic atomic electric field strength
$E_{at} \approx 6 \times 10^{11}$ V/m \cite{boyd92}, only the linear
 term in Eq. \eqref{nonlinear} is significant.  Increasing
the field intensity by using laser sources, for example, increases the
nonlinear terms in the polarization. We will focus on
the second order non-linear interaction.
\par
In order to quantize the field, we begin with the
electromagnetic field Hamiltonian in the
dielectric medium of volume $V$ /cite{mandel95}
\begin{equation}
\mathcal{H}(t) =
\frac{1}{2}\int\limits_{V}d\rmvect{r}\left [ \rmvect{D}(\rmvect{r},t)\cdot\rmvect{E}(\rmvect{r},t) +
\rmvect{B}(\rmvect{r},t)\cdot\rmvect{H}(\rmvect{r},t) \right ], \label{eq:fun10}
\end{equation}
where $\rmvect{D}$ is the displacement vector, $\rmvect{B}$ is the
magnetic induction and $\rmvect{H}$ is the magnetic field. If we use the
definition
$\rmvect{D}(\rmvect{r},t) = \epsilon_{0}\rmvect{E}(\rmvect{r},t) + \rmvect{P}(\rmvect{r},t)$
in Eq. \eqref{eq:fun10}, and use Eq. \eqref{nonlinear}, we can rewrite the Hamiltonian as
\begin{equation}
\mathcal{H}(t) = \mathcal{H}_{0}(t) + \mathcal{H}_{I}(t),
\end{equation}
where $ \mathcal{H}_{0}(t)$  contains the interaction of
the electric field and the first order linear component of the
electric polarization. The ``perturbation" $\mathcal{H}_{I}(t)$ is the
nonlinear interaction Hamiltonian, given by
\begin{align}
\mathcal{H}_{I}(t) = & 
\frac{1}{2}\int\limits_{V}d\rmvect{r}\,\rmvect{E}(\rmvect{r},t)\cdot\rmvect{P}_{\mathrm{nl}}(\rmvect{r},t)
 \nonumber \\
= &
\frac{1}{2}\int\limits_{V}d\rmvect{r}\int\limits_{0}^{\infty}dt^{\prime}\int\limits_{0}^{\infty}
dt^{\prime\prime}
\chi_{ijk}^{(2)}(t^{\prime},t^{\prime\prime})E_{i}(\rmvect{r},t)E_{j}(\rmvect{r},t-t^{\prime})
E_{k}(\rmvect{r},t-t^{\prime\prime}),
\label{eq:fun11}
\end{align}
where $\rmvect{P}_{\mathrm{nl}}(\rmvect{r},t)$ is the nonlinear component of the
electric polarization, and summation on repeated indices is understood.  In
order to avoid unnecessary difficulties in the quantization of the
field, it is considered that there are no electromagnetic boundaries
between the medium and the air. In all situations discussed here, the effects related to refraction and birefringence
can be included classically, after the field is quantized. A more rigorous approach is not yet available and is still subject of research.
\par
From this point on we concern ourselves only with the nonlinear
portion of the Hamiltonian.  It is also convenient to
suppose that just after the crystal there are two interference filters limiting the frequency spectrum of the  
down-converted fields, and that the pump beam has a narrow frequency spectrum. The two-photon quantum state 
will then include the functions describing the signal and idler interference filters.  As a part of the quantization procedure,
we expand the classical optical electric field in terms of plane
waves:
\begin{equation}
\label{eq:fun12}
\rmvect{E}(\rmvect{r},t) = \rmvect{E}^{+}(\rmvect{r},t) + \rmvect{E}^{-}(\rmvect{r},t),
\end{equation}
with
\begin{equation}
\rmvect{E}^{+}(\rmvect{r},t) =
\frac{1}{\sqrt{\mathcal{V}}}\sum\limits_{\rmvect{k},{\sigma}}
\rmvect{e}_{\rmvect{k},{\sigma}}\varepsilon_{\rmvect{k},{\sigma}}\alpha_{\rmvect{k},{\sigma}}G(\omega)
\exp\left[i(\rmvect{k}\cdot\rmvect{r}-\omega t)\right] =
\left[\rmvect{E}^{-}(\rmvect{r},t)\right]^{*}, \label{eq:fun13}
\end{equation}
where 
\begin{equation}
\varepsilon_{\rmvect{k},{\sigma}} = \sqrt{\hbar
\omega(\rmvect{k},{\sigma})/2 \epsilon_o n^{2}(\rmvect{k}, 
{\sigma})},
\end{equation}
$\epsilon_{0}$ is the free space permittivity,  $G(\omega)$ is the transmission function of the interference filter, 
$\mathcal{V}$ is the quantization volume,  $\rmvect{k}$
is the wave vector,  $\rmvect{e}_{\rmvect{k},{\sigma}}$
is the two-dimensional polarization vector, $\omega
$ is the frequency, and $\alpha_{\rmvect{k},\sigma} $ is the mode
amplitude.  The index ${\sigma}$ is summed over orthogonal components of a  two-dimensional 
polarization vector and  $\rmvect{k}$ is summed over all possible wave vectors.   
\par
We adopt the usual method of  quantization of the electric field,
letting $\alpha_{\rmvect{k},{\sigma}} \longrightarrow
\oper{a}_{\rmvect{k},{\sigma}}$, where $\oper{a}_{\rmvect{k},{\sigma}}$ is
the photon annihilation operator.  Then, the electric field
amplitude becomes a field operator, given by
\begin{eqnarray}
\rmvect{E}^{+}(\rmvect{r},t) \longrightarrow
{\oper{E}}^{+}(\rmvect{r},t)&=&\frac{1}{\sqrt{\mathcal{V}}}
\sum\limits_{\rmvect{k},{\sigma}}
\vec{e}_{\rmvect{k},{\sigma}}\varepsilon_{\rmvect{k},\sigma}\oper{a}_{\rmvect{k},{\sigma}} G(\omega)
\exp\left[i(\rmvect{k}\cdot\rmvect{r}-\omega t)\right]\nonumber\\
&=&\left[{\oper{E}}^{-}(\rmvect{r},t)\right]^{\dagger}.
\label{eq:fun14}
\end{eqnarray}
Substituting expression \eqref{eq:fun14} into the classical
Hamiltonian \eqref{eq:fun11} we have a quantum Hamiltonian operator
\begin{eqnarray}
\oper{H}_{I} &=& \frac{1}{2\mathcal{V}^{3/2}}
\sum\limits_{\rmvect{k}_{\sigma},{\sigma}_{s}}
\sum\limits_{\rmvect{k}_{i},{\sigma}_{i}}
\sum\limits_{\rmvect{k}_{p},{\sigma}_{p}}
g^{*}_{\rmvect{k}_{s},\sigma_{s}}\ g^{*}_{\rmvect{k}_{i},\sigma_{i}}\
g_{\rmvect{k}_{p},\sigma_{p}}
\oper{a}^{\dagger}_{\rmvect{k}_{s},{\sigma}_{s}}
\oper{a}^{\dagger}_{\rmvect{k}_{i},{\sigma}_{i}}
\oper{a}_{\rmvect{k}_{p},{\sigma}_{p}} \nonumber \\
&&\times \exp\left[i(\omega_{s}+\omega_{i}-\omega_{p})t\right]
\chi_{ijk}(\rmvect{e}_{\rmvect{k}_{s},{\sigma}_{s}})^{*}_{i}
(\rmvect{e}_{\rmvect{k}_{i},{\sigma}_{i}})^{*}_{j}
(\rmvect{e}_{\rmvect{k}_{p},{\sigma}_{p}})_{k}\nonumber\\
&&\times \int\limits_{V}
\exp\left[-i(\rmvect{k}_{s}+\rmvect{k}_{i}-\rmvect{k}_{p})\cdot 
\rmvect{r}\right]d\rmvect{r} + H.C.,
\label{eq:fun15}
\end{eqnarray}
where we have distinguished the three fields as \textit{pump} ($p$), 
\textit{signal} ($s$) and \textit{idler} ($i$),
\begin{equation}
g_{\rmvect{k}_{j},\sigma_{j}}=i\sqrt{\frac{\hbar\omega(\rmvect{k}_j, \sigma_j)}
{2 \epsilon_o n^{2}(\rmvect{k}_j, {\sigma}_j)}} G[\omega(\rmvect{k}_j, \sigma_j)],
\end{equation}
$V$ is the interaction (crystal) volume, and $H.C.$ stands for
Hermitian Conjugate.  Here, $n(\rmvect{k}_j, {\sigma}_j)$ is the linear
refractive index of the (anisotropic) crystal.  We have also eliminated all terms
which do not conserve energy and have defined
\begin{equation}
\chi_{ijk} \equiv \chi_{ijk}^{(2)}(\omega_{p} = \omega_{s} +
\omega_{i}) + \chi^{(2)}_{ijk}(\omega_{i} = \omega_{s} + \omega_{p}) +
\chi^{(2)}_{ijk}(\omega_{s} = \omega_{i} + \omega_{p}),
\label{eq:fun16}
\end{equation}
with
\begin{equation}
\chi_{ijk}^{(2)}(\omega = \omega^{\prime} + \omega^{\prime\prime}) =
\int\limits_{0}^{\infty} dt^{\prime}
\int\limits_{0}^{\infty}dt^{\prime\prime}
\chi_{ijk}^{(2)}(t^{\prime},t^{\prime\prime})
\exp\left[-i(\omega^{\prime}t^{\prime} +
\omega^{\prime\prime}t^{\prime\prime})]\right].
\label{eq:fun17}
\end{equation}
To find the quantum state at a given time $t$, we assume that the
nonlinear interaction is turned on at time $t_{0}=0$ when the system
is in the initial state $\ket{\psi(0)}$.   The state at time $t$ is given by
the time evolution of some initial state at $t_{0}=0$:
\begin{equation}
\ket{\psi(t)} = \oper{U}(t)\ket{\psi(0)}
\label{eq:fun18}
\end{equation}
where
\begin{equation}
\oper{U}(t) = \exp \left(\frac{1}{i\hbar}\int\limits_{0}^{t}d\tau\oper{H}_{I}(\tau) \right),
\label{eq:fun19}
\end{equation}
is the time evolution operator.
\par
If the pump field is sufficiently weak, such that the interaction time
is small compared to the average time between down-conversions, then
we can expand Eq.  \eqref{eq:fun19} in power series and keep only the
two-photon term:
\begin{equation}
\oper{U}(t) = 1 +  \left(\frac{1}{i\hbar}\int\limits_{0}^{t}d\tau\oper{H}_{I}(\tau) \right) + \cdots.
\label{eq:fun20}
\end{equation}
The integral can be expressed in terms of Eq. \eqref{eq:fun15} as:
\begin{eqnarray}
\int_{0}^{t}d\tau\,\oper{H}_{I}(\tau) &=&
\frac{1}{2\mathcal{V}^{3/2}} \sum\limits_{\rmvect{k}_{s},\sigma_{s}}
\sum\limits_{\rmvect{k}_{i},\sigma_{i}}
\sum\limits_{\rmvect{k}_{p},\sigma_{p}}
g^{*}_{\rmvect{k}_{s},\sigma_{s}}\ g^{*}_{\rmvect{k}_{i},\sigma_{i}}\
g_{\rmvect{k}_{p},\sigma_{p}}
\oper{a}^{\dagger}_{\rmvect{k}_{s},\sigma_{s}}
\oper{a}^{\dagger}_{\rmvect{k}_{i},\sigma_{i}}
\oper{a}_{\rmvect{k}_{p},\sigma_{p}} \nonumber \\
&& \times\, 
\exp\left[i(\omega_{s}+\omega_{i}-\omega_{p})t/2\right]
\chi_{ijk}(\rmvect{e}_{\rmvect{k}_{s},\sigma_{s}})^{*}_{i}
(\rmvect{e}_{\rmvect{k}_{i},\sigma_{i}})^{*}_{j}
(\rmvect{e}_{\rmvect{k}_{p},\sigma_{p}})_{k} \nonumber \\
&&\times\,  t\,\sinc{\left[(\omega_{s}+\omega_{i}-\omega_{p})t/2\right]}
\int\limits_{V}d\rmvect{r}\,
\exp\left[-i(\rmvect{k}_{s}+\rmvect{k}_{i}-\rmvect{k}_{p})\cdot 
\rmvect{r}\right]
 \nonumber \\
&& + H.C.
\label{eq:fun21}
\end{eqnarray}
Integration in $\rmvect{r}$ leads to a $\sinc$ function
involving the wave vectors, which provides the conservation of
momentum condition:
\begin{eqnarray}
\int_{0}^{t}d\tau\,\oper{H}_{I}(\tau) &=&
\frac{Vt}{2\mathcal{V}^{3/2}}
\sum\limits_{\rmvect{k}_{s},\sigma_{s}}
\sum\limits_{\rmvect{k}_{i},\sigma_{i}}
\sum\limits_{\rmvect{k}_{p},\sigma_{p}}
g^{*}_{\rmvect{k}_{s},\sigma_{s}}\ g^{*}_{\rmvect{k}_{i},\sigma_{i}}\
g_{\rmvect{k}_{p},\sigma_{p}}
\oper{a}^{\dagger}_{\rmvect{k}_{s},\sigma_{s}}
\oper{a}^{\dagger}_{\rmvect{k}_{i},\sigma_{i}}
\oper{a}_{\rmvect{k}_{p},\sigma_{p}} \nonumber \\
&&\times\, 
\chi_{ijk}(\rmvect{e}_{\rmvect{k}_{s},\sigma_{s}})^{*}_{i}
(\rmvect{e}_{\rmvect{k}_{i},\sigma_{i}})^{*}_{j}
(\rmvect{e}_{\rmvect{k}_{p},\sigma_{p}})_{k}
 \, \sinc{\left[(\omega_{s}+\omega_{i}-\omega_{p})t/2\right]}\nonumber \\
&&\times\exp\left[i(\omega_{s}+\omega_{i}-\omega_{p})t/2\right]
\prod_{m}\sinc\left[ (\rmvect{k}_s +\rmvect{k}_i - \rmvect{k}_p)_m 
l_{m}/2\right] \nonumber \\
&& \times\exp \left[ -i(\rmvect{k}_s +\rmvect{k}_i - \rmvect{k}_p)_z 
l_z/2\right] + H.C. ,
\label{eq:fun22}
\end{eqnarray}
where $V=l_x \times l_y \times l_z$ and $l_m$ is the dimension of the nonlinear medium in direction $m$ ($m=x,y,z$).
\par
The quantum state at time $t$ can  be finally obtained using
Eq. \eqref{eq:fun18}, considering the initial state as
the vacuum state, and using the interaction Hamiltonian in the form of
Eq. \eqref{eq:fun22}:
\begin{align}
\ket{\psi(t)} = \ket{vac} &+ 
\frac{Vt}{2i\hbar\mathcal{V}^{3/2}} \sum\limits_{\rmvect{k}_{s},\sigma_{s}}
\sum\limits_{\rmvect{k}_{i}, \sigma_{i}}
\sum\limits_{\rmvect{k}_{p},\sigma_{p}}
g^{*}_{\rmvect{k}_{s},\sigma_{s}}\ g^{*}_{\rmvect{k}_{i},\sigma_{i}}\
g_{\rmvect{k}_{p},\sigma_{p}}  v_p({\rmvect{k}_{p},\sigma_{p}})
\nonumber \\
& \times 
\chi_{ijk}(\rmvect{e}_{\rmvect{k}_{s},\sigma_{s}})^{*}_{i}
(\rmvect{e}_{\rmvect{k}_{i},\sigma_{i}})^{*}_{j}
(\rmvect{e}_{\rmvect{k}_{p},\sigma_{p}})_{k}\,\sinc\,{\left[(\omega_{s}+\omega_{i}-\omega_{p})
t/2\right]} \nonumber \\
& \times 
\exp\left[i(\omega_{s}+\omega_{i}-\omega_{p})t/2\right]\prod_{m}
\sinc{\,[(\rmvect{k}_s +\rmvect{k}_i - \rmvect{k}_p)_m 
l_m/2}]\nonumber \\
& \times \exp \left[ -i(\rmvect{k}_s +\rmvect{k}_i - \rmvect{k}_p)_z 
l_z/2\right]\ket{\rmvect{k}_{s},\sigma_{s}}
\ket{\rmvect{k}_{i},\sigma_{i}},
\label{eq:fun23}
\end{align}
where $\ket{\rmvect{k}_{s},\sigma_{s}}$ and
$\ket{\rmvect{k}_{i},\sigma_{i}}$ are single photon Fock states in
modes ($\rmvect{k}_{s},\sigma_{s}$) (signal) and
($\rmvect{k}_{i},\sigma_{i}$) (idler) respectively, and
$v_p({\rmvect{k}_{p},\sigma_{p}})$ is a classical amplitude 
corresponding to the plane wave component ($\rmvect{k}_{p},\sigma_{p}$)
of the pump beam.  Quantum effects in the pump field are neglected in this treatment, since the pump depletion is much smaller than its average intensity and its quantum state is effectively constant.  The annihilation operator for
the pump modes was replaced in Eq. (\ref{eq:fun23}) by a classical amplitude.

In order to simplify expression \eqref{eq:fun23}, it is convenient to
make the following approximations:

(\textit{a}) The interaction time is long enough, so that the term $\sinc\,
(\omega_{s}+\omega_{i}=\omega_{p}) t/2$ is significant only when
$\omega_{s}+\omega_{i}=\omega_{p}$.  
This assumption can be justified by the use of a moderate power
pump laser so that the time interval between two
down-conversions is large compared to the detection resolving time.

(\textit{b}) The frequency spread of the detectable down-converted
fields is small compared to the central frequencies, so
that the dispersion of the refractive indices around the central
frequencies $\bar{\omega}_{j}$ is small and a linear approximation can be
used.  This assumption is justified by the use of narrow-band
interference filters in front of the detectors.  

(\textit{c}) The terms $g_{\rmvect{{k}}_j,\sigma_j}$ and
$\tilde{\chi}_{ijk}^{(2)}$ are slowly-varying functions of
$\rmvect{k}_{j}$, so that they may be taken as constants in the intervals 
considered for $\rmvect{k}_{j}$.

(\textit{d}) The pump beam propagates along the $z$ axis and the
crystal is large enough in the $x$ and $y$ directions to contain the
whole pump beam transverse profile.  In this case, $l_{x}$ and $l_{y}$
can be extended to infinity and the last term in the third line of
expression \eqref{eq:fun23} is proportional to
\begin{displaymath}
\delta(\rmvect{q}_{s}+\rmvect{q}_{i}-\rmvect{q}_{p})\,  
\sinc\,[(k_{sz}+k_{iz}-k_{pz})L/2],
\end{displaymath}
where $\rmvect{q}_{j}=(k_{jx},k_{jy})$ is the transverse ($xy$)
component of $\rmvect{k}_{j}$ and $L=l_{z}$ is the crystal thickness.

(\textit{e}) The quantization volume is large enough to justify the
replacement of summations over $\rmvect{k}$ by integrals.  

(\textit{f}) The pump beam contains only extraordinary polarization.
It is implicit in this assumption that we are dealing with negative
birefringent crystals.  

Under the above assumptions, Eq. \eqref{eq:fun23} is written as
\begin{eqnarray}
\label{psi2}
\ket{\psi}=\ket{vac}&+&\sum_{\sigma_{s},\sigma_{i}}\int\!d\omega_{s}
\int\!d\omega_{i} \int\!d\rmvect{q}_{s} \int\!d\rmvect{q}_{i}\
\Phi_{\sigma_{s}\sigma_{i}}(\rmvect{q}_{s},\rmvect{q}_{i},\omega_{s},\omega_{i})
\nonumber\\
&&\times\,\ket{\rmvect{q}_{s},\omega_{s},\sigma_{s}}\ket{\rmvect{q}_{i},\omega_{i},
\sigma_{i}},
\end{eqnarray}
where $\ket{\rmvect{q}_{j},\omega_{j},\sigma_{j}}$ represents a one-photon state
in the mode defined by the transverse  component $\rmvect{q}_{j}$
of the wave vector, by the frequency $\omega_{j}$ and by the
polarization $\sigma_{j}$.  The amplitude $\Phi$ is now reduced to
\begin{equation}
\label{ampli2}
\Phi_{\sigma_{s}\sigma_{i}}\approx C_{\sigma_{s}\sigma_{i}}\ 
G_{s}(\omega_{s})G_{i}(\omega_{i})\ 
v(\rmvect{q}_{s}+\rmvect{q}_{i},\omega_{s}+\omega_{i})\
\sinc\,[(k_{sz}+k_{iz}-k_{pz})L/2],
\end{equation}
where $C_{\sigma_{s}\sigma_{i}}$ is a coupling constant which depends
on the nonlinear susceptibility tensor, and $G(\omega_{j})$ is the
spectral function defined by the narrow bandwidth filters placed in
front of the detectors.  If the anisotropy of the medium is neglected
(which is not always convenient), the longitudinal wave vector
mismatch $k_{sz}+k_{iz}-k_{pz}$ is written as
$\sqrt{|\rmvect{k}_{s}|^{2}-|\rmvect{q}_{s}|^{2}}+
\sqrt{|\rmvect{k}_{i}|^{2}-|\rmvect{q}_{i}|^{2}}-
\sqrt{|\rmvect{k}_{p}|^{2}-|\rmvect{q}_{s}+\rmvect{q}_{i}|^{2}}$.

\subsection{The coincidence count rate}
\label{sec:cc}
The state given by Eq. \eqref{eq:fun23} can be simplified, considering
approximations that are appropriate to given specific experimental
situations.  In this section, let us consider that signal and idler
fields are detected through small apertures at fixed positions far
enough from the source, so that only one spatial mode is selected by
each detector.  In this case, when narrow bandwidth filters are used,
the quantum state takes a very simple form:
\begin{equation}
\label{eq:cc1}
\ket{\psi(t)} = C_1 \ket{vac} + C_2 \int\!d\omega_s\int\! d\omega_i\, 
G_{s}(\omega_{s})G_{i}(\omega_{i}) v_p(\omega_i+\omega_s)
\ket{\omega_s}\ket{\omega_i},
\end{equation}
where $C_1$ and $C_2$ are normalization constants. 

The electric field operator can also be simplified to
\begin{equation}
\label{eq:cc2}
\oper{E}^{+}(t+\tau)= C \int d\omega\, \oper{a}(\omega)\exp{[-i\omega(t+\tau)]},
\end{equation}
where $C$ is a constant.

The state in Eq. \eqref{eq:cc1} and field operator in Eq. \eqref{eq:cc2} were obtained
under several approximations. However, they adequately
describe the main features of time correlations between twin photons.
The time correlation is always assumed and used
in all twin-photon experiments with coincidence detection, even when the spatial correlations are
the focus of the investigation. Therefore, it is important and interesting to
calculate the coincidence count rate in this simplified point of view.
\par
The coincidence count rate is given by the fourth order (in the fields)
correlation function:
\begin{eqnarray}
\label{eq:cc3}
R_{c}(t+\tau_i,t+\tau_s) &=& \langle\oper{E}^-_s(t+\tau_s)
\oper{E}^-_i(t+\tau_i) \oper{E}^+_i(t+\tau_i)\oper{E}^+_s(t+\tau_s)
\rangle\nonumber\\
&=& |\oper{E}^+_i(t+\tau_i)\oper{E}^+_s(t+\tau_s)\ket{\psi(t)}|^2.
\end{eqnarray}
Using Eqs. \eqref{eq:cc1} and \eqref{eq:cc2}, we obtain:
\begin{eqnarray}
\label{eq:cc4}
R_{c}(t+\tau_i,t+\tau_s) &=& \eta^2 
\bigg|\int\!d\omega_1\int\!d\omega_2\, \oper{a}(\omega_1)\oper{a}(\omega_2)
\exp{\{-i[\omega_1(t+\tau_i)+\omega_2(t+\tau_s)]\}}
 \nonumber\\
&&\times \int\!d\omega_i\int\!d\omega_s\, G_{i}(\omega_{i})G_{s}(\omega_{s})v_p(\omega_i+\omega_s)
\ket{\omega_i}\ket{\omega_s}\bigg|^2 ,
\end{eqnarray}
where $\eta$ is a constant that depends on the square root of the pump beam power, the efficiency of
the detectors and of the parametric down conversion process.
After the action of the operators on the quantum state
and integration over $\omega_1$ and $\omega_2$, we obtain:
\begin{eqnarray}
\label{eq:cc5} R_{c}(t+\tau_i,t+\tau_s) &=& \eta^2 \bigg|\int\!
d\omega_i \int\!d\omega_s G_{i}(\omega_{i})G_{s}(\omega_{s})v_p(\omega_i+\omega_s)\nonumber\\ 
&&\times\exp{[-i(\omega_i+\omega_s)t]}\exp{[-i(\omega_i\tau_i+\omega_s\tau_s)]}\bigg|^2 .
\end{eqnarray}
\par
In the limit where the pump spectrum can be approximated by a delta function
$v_p(\omega)\rightarrow \delta(\omega_{p}-\omega)$, the
time correlation is:
\begin{eqnarray}
\label{eq:cc6}
R_{c}(\tau_i,\tau_s) &=& \eta^2 \bigg|\int d\omega\, 
G_{i}(\omega)G_{s}(\omega_{p}-\omega)\exp{[i\omega(\tau_s-\tau_i)]}\bigg|^2\nonumber\\
&=& \eta^2 \left|F(\tau_{s}-\tau_{i})\right|^{2},
\end{eqnarray}
where $F(t)$ is the convolution of the Fourier transforms of the filter functions  
$G_{i}(\omega)$ and $G_{s}(\omega_{p}-\omega)$. Usually, the width of 
$F(t)$ is on the order of femtoseconds. This means that detections are simultaneous 
within the window of a few femtoseconds.

%% file: section3.tex
\section{Fundamentals of spatial correlations}
\label{sec:fundamentals}

The investigation of the transverse spatial properties of
light has accompanied the study of the nature of light itself. The
development of the theory of diffraction and all resulting
applications represent an impressive example of how finely a
physical theory can describe nature.
Despite the incredible success of  classical optics, it has
been found in the last few decades that amendments should be made in
order to accurately describe the quantum effects which arise when dealing with
some special light sources. In this respect, the transverse
correlations between the twin photons produced in the parametric
down-conversion process have largely contributed to this
end. The quantum optics theory of spatial effects which has been developed so far is strongly based on 
classical diffraction theory, and has been shown to describe a
large array of quantum phenomena in optics.
In the following, we will briefly discuss some important topics in
classical diffraction theory which will help in the development of
the quantum theory of spatial correlations.

\subsection{Transverse coherence and partial coherence: classical optics}

In this section we will review some fundamental aspects of
transverse coherence and partial coherence in classical optics.
These concepts were essential in the development of quantum
coherence theory and are still essential for the design
and understanding of many quantum optics experiments.
We introduce the discussion by analyzing
the Young's double-slit experiment and its relationship with the van
Cittert-Zernike \cite{wolf07} theorem, and the concept of coherence
area.

\subsubsection{Double slit experiment: perfect coherence}

\begin{figure}
\begin{center}
\includegraphics[width=0.3\linewidth]{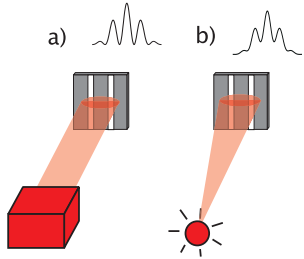}
\caption{Double-slit experiment. The double-slit is
illuminated by a perfect coherent light source in a) and by a
partially coherent light source in b).} \label{dslit}
\end{center}
\end{figure}

Fig. \ref{dslit} shows a  light source illuminating a double-slit aperture, and the resulting intensity pattern
which is observed at the detection screen. When
the light emitted by the source is perfectly coherent,
interference fringes are observed with high visibility or contrast, as in Fig.
\ref{dslit}a). Many of the laser light sources available today are almost perfectly
coherent, but many other important light sources are incoherent or partially coherent. Incoherent
light sources can also be used to perform interference
experiments under certain conditions. All real light
 beams present partial coherence, depending on the
geometrical properties and propagation of the beam. Thus, interference fringes
can still be observed, but with reduced visibility, as illustrated in Fig.
\ref{dslit}b).
The light intensity distribution after a double-slit is
given by \cite{BWolf,wolf07}:
\begin{equation}
\label{Int:dslit}
I(x,y) = I_0(x,y) [1+ |\gamma_{12}(d)| \cos({\bf k} \cdot 
\boldsymbol{\rho} + \delta)],
\end{equation}
where $I_0(x,y)$ gives the single slit diffraction pattern, ${\bf k}$
is the wave vector, $\boldsymbol{\rho}$ gives the position in the transverse plane, $\gamma_{12}$ is the normalized mutual coherence
of the light field in the plane of the slits, $d$ is the distance
between slits and $\delta$ is a fixed phase factor.  When the source
is perfectly coherent, $|\gamma_{12}| = 1$ and the contrast of the interference fringes is maximum.  The contrast can be quantified by the visibility $V$, defined as $V=(I_{\textrm{max}}-I_{\textrm{min}})/(I_{\textrm{max}}+I_{\textrm{min}})$, where $I_{\textrm{max}}$ and $I_{\textrm{min}}$ are the intensities at the interference maximum and minimum.   
When the coherence is partial, $0\leq |\gamma_{12}| \leq 1$, 
and can be calculated from geometrical and statistical properties of the source \cite{wolf07}.

One important result in the calculation of $\gamma_{12}$ is known as
the van Cittert-Zernike theorem. We will discuss this theorem in more detail.

\subsubsection{Partial coherence and the van Cittert-Zernike theorem}
\label{sec:vCZ}
\begin{figure}
\begin{center}
\includegraphics[width=6cm]{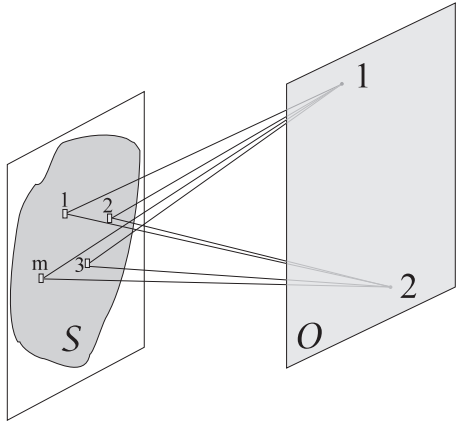}
\caption{Finite source $S$ illuminating an observation screen $O$.}
\label{citzern1}
\end{center}
\end{figure}

Let us consider the situation sketched in Fig.  \ref{citzern1}, where
a light source $\cal{S}$ is illuminating the observation screen
$\cal{O}$.  We assume that the source is constituted by infinitesimal
independent light emitters, such as a thermal source or parametric
down-conversion, for example.  The non-normalized mutual coherence function
between points 1 and 2 on the screen is defined as:

\begin{equation}
\label{cohra:12}
\Gamma_{12}(d,\tau) = \sum_m \langle E_{m1}(t+\tau) \, E_{m2}^\ast(t) \rangle
+ \sum\sum_{m\neq n} \langle E_{m1}(t+\tau) \, E_{n2}^\ast(t) \rangle,
\end{equation}
where the summations are performed over the emitting points of the source,
$d$ is the distance between points 1 and 2 at the screen, $\tau$
is the time difference due to different propagation distances from
point $m$ at the source to point 1(or point 2) at the screen, $E^\ast$
is the complex conjugate of $E$, and the
brackets indicate time average.

If the field is quasi-monochromatic, the mutual coherence function can be written as
\begin{equation}
\label{cohrb:12}
\Gamma_{12}(d,\tau) = \Gamma_{12}(d,0) \, \mbox{e}^{ikc\tau},
\end{equation}
where $k=|\bf{k}|$ and $c$ is the speed of light. This approximation is valid when the 
time difference $\tau$ is much smaller than the coherence time of the source.
In other words, the phase relationship between $E_{m1}(t+\tau)$ and
$E_{m2}(t)$ is preserved. However, for spatially incoherent sources such as
thermal sources and SPDC, there is no phase relationship between
$E_{m1}(t+\tau)$ and $E_{n2}(t)$ if $m \neq n$  and the second term
in the left side of Eq. \eqref{cohra:12} averages to zero. Therefore, we end up
with:
\begin{equation}
\label{cohrc:12}
\Gamma_{12}(d,0) = \sum_m \langle E_{m1} \, E_{m2}^\ast \rangle.
\end{equation}

It is convenient to take the continuum limit and suppose that each
point of the source emits a spherical wave. Thus,
Eq. \eqref{cohrc:12} takes the form:

\begin{equation}
\label{cohrd:12}
\Gamma_{12}(d,0) = \frac{1}{R^2}\int\limits_{\mathcal{S}} d\rmvect{r}_{0}\ 
I(\rmvect{r}_{0}) \mbox{e}^{ik(R_1 - R_2)},
\end{equation}
where $I(\rmvect{r}_{0})$ is the intensity distribution of the source,
$R_1=|\rmvect{r}_{0}-\rmvect{r}_{1}|$ and
$R_2=|\rmvect{r}_{0}-\rmvect{r}_{2}|$ are the distances between point
$\rmvect{r}_{0}$ in the source and points $\rmvect{r}_{1}$ and
$\rmvect{r}_{2}$, respectively, and it was assumed that $R_1\sim R_2 = R
\gg (R_1-R_2)$.

\begin{figure}
\begin{center}
\includegraphics[width=7cm]{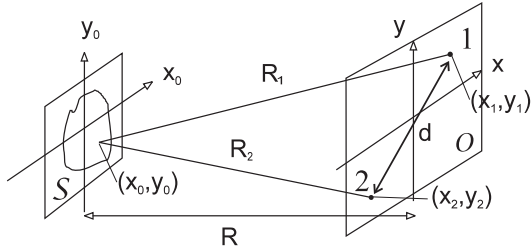}
\caption{Finite source illuminating an observation screen.}
\label{citzern2}
\end{center}
\end{figure}

It is interesting to write the difference $R_1 - R_2$
explicitly in terms of the coordinates of the generic
source point $(x_0,y_0)$ and the coordinates of the points 1
$(x_1,y_1)$ and 2 $(x_2,y_2)$ (see Fig. \ref{citzern2}) and use
again the relation $R_1\sim R_2 = R \gg (R_1-R_2)$.   We then obtain the
result known as the van Cittert-Zernike theorem:
\begin{equation}
\label{cohre:12}
\Gamma_{12}(d,0) = 
\frac{\mbox{e}^{i\alpha}}{R^2}\iint\limits_{\mathcal{S}} dx_0 
dy_0\ I(x_0, y_0)
\mbox{e}^{ik(px_0 + qy_0)},
\end{equation}
where $\alpha$ is a constant phase and
\begin{equation}
p = \frac{(x_1 - x_2)}{R}\,\, \mbox{and}\,\, q = \frac{(y_1 - y_2 )}{R}.
\end{equation}
The dependence on the distance $d$ appears through $p$ and $q$.
The van Cittert-Zernike theorem shows that the mutual coherence is
given by the Fourier transform of the intensity distribution of the
light source.  This result is valid if the emitting points in the
source are independent and if the mutual coherence is considered in a
plane far enough from the source, so that Fraunhofer diffraction
regime can be assumed.
The normalized degree of coherence $\gamma_{12}$ is:
\begin{equation}
\label{cohrf:12}
\gamma_{12}(d,0) = \frac{\mbox{e}^{i\alpha}\iint dx_0 dy_0 I(x_0, y_0)
\mbox{e}^{ik(px_0 + qy_0)}}{\iint dx_0 dy_0 I(x_0, y_0)}.
\end{equation}
The quantity $|\gamma_{12}(d)|$ gives the visibility of the interference
fringes in a double slit experiment, Eq. \eqref{Int:dslit}, where we remember that $d$ is
the distance between slits.
\subsubsection{The coherence area}

\begin{figure}
\begin{center}
\includegraphics[width=7cm]{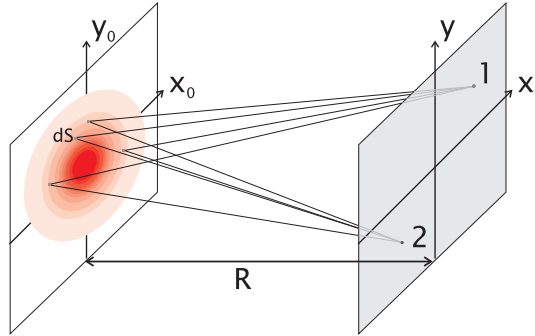}
\caption{Gaussian source illuminating an observation screen.}
\label{ca:gauss}
\end{center}
\end{figure}

Let us illustrate the application of the van Cittert-Zernike theorem
in the calculation of the coherence area of a light source.  Referring to  Fig.  \ref{ca:gauss}, we start
by calculating the degree of coherence for a light source with a Gaussian
intensity distribution.   The
integrals in Eq. \eqref{cohrf:12} are  computed using the initial intensity distribution
\begin{equation}
\label{aca:12a} I(x_0,y_0) = A \exp{\left(-\frac{x^2_0 +
y^2_0}{\sigma^2}\right)}, 
\end{equation}
and we obtain
\begin{equation}
\label{aca:12b}
\gamma_{12}(d,0) = \exp{\left(-\frac{k^2 \sigma^2 d^2}{4R^2}\right)}.
\end{equation}
The coherence area is the area within
which the value of $|\gamma_{12}|$ is appreciable and the field
can be considered spatially coherent. From Eq. \eqref{aca:12b}, we can
see that, on the one hand, the degree of coherence decreases with
the wave number $k$, source width $\sigma$ and
distance $d$ between points 1 and 2. On the other hand, the degree
of coherence increases with the propagation distance $R$. As the
coherence function is Gaussian, it never reaches zero and we
can take the value $\gamma_{12}(d,0) = 1/{e}$ as a lower
bound for the coherence. Then the circular coherence area ${\cal
A}_c = \pi (\frac{d}{2})^2$ can be deduced from Eq. \eqref{aca:12b} by
imposing $\gamma_{12}(d,0) = 1/{e}$ :
\begin{equation}
\label{acb:12}
{\cal A}_c = \frac{\pi R^2}{k^2\sigma^2}.
\end{equation}
In terms of a double slit experiment, this result indicates that there will be interference fringes with considerable visibility as long as the slits
are placed within the area given by Eq. \eqref{acb:12}.
\par
The double-slit experiment provides a method for determining the coherence area of a light field at some distance from the source.   
A double-slit aperture with variable slit separation $d$ is placed at the plane of interest, the visibility of the resulting
interference patterns is measured in the far-field.  By plotting the visibilities of the interference patterns as a function of 
$d$, a measurement of the transverse coherence length of the field in the one direction is obtained. Repeating the procedure
in the perpendicular direction in the transverse plane, we obtain a measurement of the coherence area.  This method was used to measure the 
transverse coherence length of one of the down-converted fields produced by SPDC 
\cite{soutoribeiro94b}, and it was observed that the coherence length is the same as that predicted for a thermal source.  The same method was also applied to the two-photon beam produced by SPDC, and it was shown that the transverse coherence length of the two-photon field is much larger than the one-photon field \cite{fonseca99a}. A quantum multimode theory was necessary to explain the experimental results in this case. 
\subsection{The Fresnel and paraxial approximations}
Another important and basic issue in spatial correlations concerns the
Fresnel and paraxial approximations. Most of the quantum
optics theory concerning SPDC was developed within these approximations and
describes very well the vast majority of the experimental situations
reported so far. In the following sections, we will derive the quantum state
for the transverse degrees of freedom of the twin photons produced
by parametric down-conversion, according to the Fresnel
and paraxial approximations.
\par
\label{sec:parax} 
Let us assume that a monochromatic light beam is
propagating along the $z$ direction in an isotropic medium.  We can write the
$k_{z}$ component as
\begin{equation}
k_{z}=\sqrt{k^{2}-q^{2}}\approx k\left(1-\frac{q^{2}}{2k^{2}}\right).
\label{eq:fresnel}
\end{equation}
The above approximation, known as the Fresnel approximation, is obtained by simple Taylor expansion and is valid when $q^{2} \ll \, k^{2}$.
\par The Fresnel approximation is essentially a particular application
of the more general paraxial approximation.  In geometric optics,
where light is represented by rays, paraxial rays are those that lie
at small angles to the optical axis of the optical system under
consideration.  If we were to draw rays from the origin to points
$\rmvect{k}$ in $k$-space satisfying the approximation in expression
\eqref{eq:fresnel}, they would be paraxial rays.  In this respect the
paraxial approximation and the Fresnel approximation are essentially
the same.  Throughout this work we will refer to both as the paraxial
approximation.
A thorough account of the paraxial approximation can be found in
most classical optics textbooks \cite{saleh91,goodman96}. Here we
will present only the bare essentials. The paraxial approximation
can be extended to wave optics if one considers only waves whose
wavefront normals are paraxial rays. In wave optics, an optical
wave $E(\rmvect{r},t)$ satisfies the wave equation:
\begin{equation}
\nabla^2E(\rmvect{r},t) -\frac{1}{c^2}\frac{\partial^{2}}{\partial t^2}E(\rmvect{r},t)=0. 
\end{equation}
If one considers that the optical wave is monochromatic with
harmonic time dependence, so that
$E(\rmvect{r},t)=\mathcal{E}(\rmvect{r})\exp(-i\omega t)$, where
$\omega$ is the angular frequency, one arrives at the Helmholtz
equation:
  \begin{equation}
\nabla^2\mathcal{E}(\rmvect{r}) +k^2\mathcal{E}(\rmvect{r})=0. 
\label{eq:helmholtz}
\end{equation}
If we now consider only paraxial waves propagating near the $z$
axis, we can write $\mathcal{E}(\rmvect{r}) =
\mathcal{U}(\rmvect{r})\exp(ikz)$, where $\mathcal{U}(\rmvect{r})$
is a slowly varying function of $\rmvect{r}$ such that
$\mathcal{E}(\rmvect{r})$ maintains a plane wave structure for
distances within that of a wavelength.  Using this form of
$\mathcal{E}(\rmvect{r})$ in the Helmholtz equation, one arrives
at the paraxial Helmholtz equation
  \begin{equation}
\left(\frac{\partial^2}{\partial x^2}+\frac{\partial^2}{\partial y^2}+2 i k \frac{\partial}{\partial z}\right)\mathcal{U}(\rmvect{r}) =0.
\label{eq:paraxhelm}
\end{equation}
In obtaining \eqref{eq:paraxhelm}, we have used the fact that
the term $\partial^2\mathcal{U}(\rmvect{r})/\partial z^2$ is very
small within distances of a wavelength: $\partial^2\mathcal{U}(\rmvect{r})/\partial z^2 \ll k \partial\mathcal{U}(\rmvect{r})/\partial z$.  The paraxial Helmholtz
equation admits several well known solutions, including the
Hermite-Gaussian and Laguerre-Gaussian beams. Quantum fields
prepared in these modes have been proposed for several
applications and will be discussed in more detail later. It has
been shown by several authors that the paraxial Helmholtz equation
is analogous to the Schr\"odinger equation in quantum mechanics
\cite{bacry81,stoler,marcuse}. In reference \cite{bacry81} an alternative
derivation of equation \eqref{eq:paraxhelm} is provided which
requires that the optical wave is only nearly monochromatic.

\subsection{The angular spectrum and its propagation}
\label{sec:propag}

The quantum theory which accounts for the transverse correlations of
photons makes use of several techniques of Fourier Optics, in
particular the propagation of the angular spectrum.  Fourier Optics
\cite{goodman96} provides a useful method of calculating the
propagation of an optical field through a given optical system.  Here
we consider a monochromatic scalar field far from its source,
satisfying the Helmholtz equation \eqref{eq:helmholtz}, for which the field amplitude can be represented \cite{goodman96} as
 \begin{equation}
\mathcal{U}(\vect{\rho},z) 
=\frac{1}{(2\pi)^{2}}\int\limits_{\mathbb{R}^{2}} d\rmvect{q}\ 
v(\rmvect{q},z) \exp\left[i\rmvect{q}\cdot\vect{\rho}\right],
\label{eq:fourierfield}
 \end{equation}
where we assume that the field is propagating along the $z$ direction
and have defined $\vect{\rho}$ and $\rmvect{q}$ as the transverse
components of $\rmvect{r}$ and $\rmvect{k}$, respectively.  The
angular spectrum $v(\rmvect{q},z)$ is the inverse Fourier transform of
the electric field:
  \begin{equation}
v(\rmvect{q},z) = \int\limits_{\mathbb{R}^{2}}d\vect{\rho}\ 
\mathcal{U}(\vect{\rho},z)\exp\left[-i\rmvect{q}\cdot
\vect{\rho}\right].
\label{eq:angspect}
 \end{equation}
 One can also understand the angular spectrum by recognizing equation
 \eqref{eq:fourierfield} as an expansion of
 $\mathcal{U}(\vect{\rho},z)$ in terms of plane waves
 $\exp(i\rmvect{q}\cdot\vect{\rho})$, in which the angular spectrum
 $v(\rmvect{q},z) $ acts as a weight function.  \par

As the field propagates, its angular spectrum
changes accordingly.  Combining Eqs.  \eqref{eq:paraxhelm} and
\eqref{eq:fourierfield}, it is easy to show that if the angular 
spectrum of a field is known at $z=0$, it propagates as
\begin{equation}
v(\rmvect{q},z)=v(\rmvect{q},0)\exp(ik_{z}z).
\end{equation}
In order to be useful, $k_{z}$ must be expressed in terms of the transverse component
$\rmvect{q}$.  For isotropic media, this can be done using Eq. (\ref{eq:fresnel}).  
\par
Dealing with anisotropic media, which is the case of the nonlinear
crystals used for parametric down-conversion, the expressions for
$k_{z}$ in terms of $\rmvect{q}$ can be quite complicated
\cite{BWolf}.  We will restrict our analysis to uniaxial media for two
reasons: first, in the majority of the work on spatial correlation
properties of two-photon states, the photon pairs are generated by
spontaneous parametric down-conversion in uniaxial crystals.  Second,
the physics of down-conversion in biaxial crystals, though more
involved, to our knowledge does not contain any essentially new effect.  
\begin{figure}
\begin{center}
\includegraphics[width=4cm]{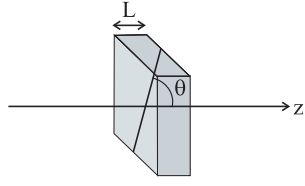}
\caption{Non-linear crystal and its optical axis. $z$ axis represents the direction of propagation of light. }
\label{fig:slab}
\end{center}
\end{figure}
We will
consider the geometry depicted in Fig.  \ref{fig:slab}, where a slab
of a uniaxial crystal of thickness $L$ is cut to have its optic axis
making an angle $\theta$ with its normal, which is oriented parallel
to the $z$ axis.  In this medium, the \textit{wave vector surface} has
two sheets, defined by
\begin{equation}\label{eq:k-ordinary}
\frac{q_{x}^{2}+q_{y}^{2}+k_{z}^{2}}{n_{o}^{2}}=\frac{\omega^2}{c^2},
\end{equation}
which allows us to immediately write
\begin{equation}
k_{z}=\sqrt{\left(n_{o}\frac{\omega}{c}\right)^2-|\rmvect{q}|^2}\approx 
n_{o}\frac{\omega}{c}- 
\frac{c}{2n_{o}\omega}|\rmvect{q}|^2,
\end{equation}
and
\begin{align}\label{eq:k-extraordinaty}
&\left(\frac{\cos^{2}\theta}{n_{e}^{2}}+\frac{\sin^{2}\theta}{n_{o}^{2}}\right)q_{x}^{2} + \frac{q_{y}^{2}}{n_{e}^{2}}+
\left(\frac{\cos^{2}\theta}{n_{o}^{2}}+\frac{\sin^{2}\theta}{n_{e}^{2}}\right)k_{z}^{2} \nonumber
\\
&+\left(\frac{1}{n_{e}^{2}}-\frac{1}{n_{o}^{2}}\right)\sin (2\theta) q_{x}\, k_{z}
= \frac{\omega^2}{c^2},
\end{align}
where $n_{o}$ and $n_{e}$ are the ordinary and extraordinary 
refractive indices, respectively. Eq. \eqref{eq:k-ordinary} applies 
to plane waves with ordinary polarization (orthogonal to the optical axis), whereas Eq. 
\eqref{eq:k-extraordinaty} applies to plane waves with extraordinary 
polarization. In the latter case, Eq. \eqref{eq:k-extraordinaty} can 
be solved for $k_{z}$, leading to
\begin{align}
k_{z}&=\alpha q_{x}+\sqrt{\left(\eta\frac{\omega}{c}\right)^2-\beta^{2} 
q_{x}^{2}-\gamma^{2}q_{y}^{2}}\nonumber\\
&\approx \alpha q_{x}+ \eta \frac{\omega}{c}-\frac{c}{2\eta\omega}(\beta^{2} 
q_{x}^{2}-\gamma^{2}q_{y}^{2}),
\end{align}
where 
\begin{eqnarray}
\alpha&=&\frac{(n_{o}^{2}-n_{e}^{2})\sin\theta 
\cos\theta}{n_{o}^{2}\sin^{2}\theta+n_{e}^{2}
\cos^{2}\theta},\\
\beta&=&\frac{n_{o}n_{e}}{n_{o}^{2}\sin^{2}\theta+
n_{e}^{2}\cos^{2}\theta},\\
\gamma&=&\frac{n_{o}}{\sqrt{n_{o}^{2}\sin^{2}\theta+n_{e}^{2}\cos^{2}\theta}},\\
\eta&=&\frac{n_{o}n_{e}}{\sqrt{n_{o}^{2}\sin^{2}\theta+
n_{e}^{2}\cos^{2}\theta}}.
\end{eqnarray}
The term $\alpha$ is responsible for the so-called \textit{walk-off}.
The terms $\beta$ and $\gamma$ are close to 1 and cause a slight
astigmatism in the beams propagating through the uniaxial medium.
Their effect is marginal, and both $\beta^{2}$ and $\gamma^{2}$ can be
approximated by $1$.  Therefore, for extraordinary polarization,
\begin{equation}
k_{z}\approx \eta \frac{\omega}{c}+\alpha 
q_{x}-\frac{c}{2\eta\omega}|\rmvect{q}|^{2}.
\end{equation}

\subsection{Quantum state of  photon pairs: spatial degrees of freedom}
\label{sec:state} 
The spatial properties of the two-photon state depend strongly on the birefringence of the non-linear crystal and type of phase matching \cite{rubin94,rubin96,atature02b,torres05,fedorov07}.   Let us consider first the case of type I phase matching, where the
pump field has extraordinary polarization and the down-converted
fields have ordinary polarization ($e\rightarrow oo$). In this case,
\begin{align}
k_{pz}&\approx \eta_{p}\frac{\omega_{p}}{c}-\alpha_{p}
q_{px}-\frac{c}{2\eta_{p}\omega_{p}}|\rmvect{q}_{s}+\rmvect{q}_{i}|^2,
\label{kpz}\\
k_{sz}&\approx 
n_{s}\frac{\omega_{s}}{c}-\frac{c}{2n_{s}\omega_{s}}|\rmvect{q}_{s}|^2,
\label{ksz}\\
k_{iz}&\approx 
n_{i}\frac{\omega_{i}}{c}-\frac{c}{2n_{i}\omega_{i}}|\rmvect{q}_{i}|^2,
\label{kiz}
\end{align}
where $n_{j}$ ($j=s,i$) stands for the ordinary refractive index.

Using Eqs.  \eqref{kpz}-\eqref{kiz} in Eq.  \eqref{ampli2} and
considering a monochromatic pump beam, we arrive at the two-photon
detection amplitude for type I phase matching:
\begin{eqnarray}
\label{amplioo}
\Phi_{oo}(\rmvect{q}_{s},\rmvect{q}_{i})&\approx& C_{oo}\ G_{s}(\omega_{s})G_{i}(\omega_{i})\
v(\rmvect{q}_{s}+\rmvect{q}_{i})\delta(\omega_{s}+\omega_{i}-\omega_{p})\nonumber\\
&&\times\,\sinc\,\left[\mu_{oo}+l_{t}(q_{sx}+q_{ix})+
\frac{L}{4K}\left|\frac{\rmvect{q}_{s}}{r_{s}} -
\frac{\rmvect{q}_{i}}{r_{i}}\right|^2\right]\nonumber\\
&&\times \exp[-il_{t}(q_{sx}+q_{i}x)],
\end{eqnarray}
where $K=\eta_{p}\omega_{p}/c$ is the pump beam wave number inside the
crystal, $r_{s}=\sqrt{\omega_{s}/\omega_{i}}$,
$r_{i}=\sqrt{\omega_{i}/\omega_{s}}$,
$\mu_{oo}=(\bar{n}-\eta_{p})KL/2\eta_{p}$, and $l_{t}=\alpha_{p}L/2$ is the 
\textit{transverse walk-off length}\cite{molina03,torres05}.  The
ordinary refractive index $\bar{n}$ is calculated at the frequency
$\omega_{p}/2$. In collinear phase matching, $\mu_{oo}=0$.  Eq.
\eqref{amplioo} is accurate up to first order in
$(\omega_{s}-\omega_{i})/2\omega_{p}$.

For each particular set of frequencies $\omega_{s}$, $\omega_{i}$, the
function $\Phi_{oo}(\rmvect{q}_{s},\rmvect{q}_{i})$ can be regarded as
the normalized angular spectrum of the two-photon field.  Note that
the angular spectrum of the pump beam is present in $\Phi_{oo}$,
meaning that it is transferred to the fourth-order spatial correlation
properties of the two-photon state \cite{monken98a}.  The amplitude
$\Phi_{oo}$ is not a separable function of $\rmvect{q}_{s}$, and
$\rmvect{q}_{i}$, that is to say, $\Phi_{oo} \neq
F_{s}(\rmvect{q}_{s}) F_{i}(\rmvect{q}_{i})$.  This non-separability
is responsible for many of the nonlocal and non-classical effects
observed with the state \eqref{eq:spdc2}.

For simplicity, let us assume that the crystal is thin enough to allow us to ignore the effects of  birefringence.  If narrow-band interference filters selecting $\omega_s=\omega_i=\omega_p/2$
are used, the two-photon state generated by SPDC in quasi-collinear phase
matching ($\mu_{oo}\approx 0$) can be written in a very simple form as
\cite{monken98a,monken02}
\begin{equation}
\ket{\psi}_\textsc{SPDC} = C_{\text{v}}\ket{vac} + 
C_{2}\ket{\psi}_{\textsc{I}},
\label{eq:spdc1}
\end{equation}
where
\begin{equation}
\ket{\psi}_\textsc{I}=\iint
d\rmvect{q}_{s}
d\rmvect{q}_{i}\ 
v(\rmvect{q}_{s}+\rmvect{q}_{i})\gamma(\rmvect{q}_{s}-\rmvect{q}_{i})\ket{\rmvect{q}_{s},\rmvect{e}_{o}}
\ket{\rmvect{q}_{i},\rmvect{e}_{o}},
\label{eq:quantumstate}
\end{equation}
$\gamma(\rmvect{q})=\sqrt{2L/\pi^2 K}\sinc(Lq^2/4K)$,
and $\ket{\rmvect{q}_{j},\rmvect{e}_{o}}$ represent Fock states in
plane wave modes labeled by the transverse component $\rmvect{q}_{j}$
of the wave vector $\rmvect{k}_{j}$, and ordinary polarization vector $\rmvect{e}_{o}$. 
If the nonlinear crystal is sufficiently thin, so that the width of the sinc
function in Eq.  \eqref{amplioo} can be much greater than the width of
the pump beam angular spectrum, the sinc function can then be approximated by unity.  This is known as the \emph{thin crystal approximation}.  
The quantum state (\ref{eq:quantumstate}) simplifies to 
\begin{equation}
\ket{\psi}_{\mathrm{tc}}=\iint\limits_{D}
d\rmvect{q}_{s}
d\rmvect{q}_{i}\ 
v(\rmvect{q}_{s}+\rmvect{q}_{i})\ket{\rmvect{q}_{s},\rmvect{e}_{o}}
\ket{\rmvect{q}_{i},\rmvect{e}_{o}},
\label{eq:spdc2}
\end{equation}
$D$
is a domain in the $\rmvect{q}$ space within which the thin-crystal approximation is valid, meaning that the
two-photon angular spectrum $\Phi_{oo}(\rmvect{q}_{s},\rmvect{q}_{i})$
mimics the pump beam angular spectrum $v(\rmvect{q})$ in the sum of
transverse wave vectors $\rmvect{q}_{s}+\rmvect{q}_{i}$.  Since all
the information about a monochromatic beam is contained in its angular
spectrum, the transverse and longitudinal spatial properties of the
pump beam are transferred to the two-photon field, in this context,
defining a \emph{correlation beam}.

In the general case, however, when the crystal length is not
negligible, the sinc function in Eq.  \eqref{amplioo} has to be
considered in detail.  Due to the presence of the linear term
$l_{t}(q_{sx}+q_{ix})$, that function is, in general, much narrower in the
$x$ direction than it is in the $y$ direction.  This fact causes the
two-photon coincidence detection amplitude to not emulate the pump beam
angular spectrum completely\cite{molina-terriza05,fedorov07}. For example, in the collinear
$(\mu_{oe}=\mu_{eo}=0)$ degenerate monochromatic
($\omega_{s}=\omega_{i}=\omega_{p}/2$) case, when the detectors are
scanned in the same direction in the $\rmvect{q}$ space 
($\rmvect{q}_{s}=\rmvect{q}_{i}=\rmvect{q}$), the amplitude
$\Phi_{oo}$ is proportional to
\begin{equation}
v(2\rmvect{q})\,\sinc\,(2l_{t}q_{x})\exp(-2il_{t}q_{x}).
\end{equation}
It is clear that the angular spectrum will be clipped by the sinc 
function and subjected to transverse walk-off.

In type II phase matching, one of the down-converted fields has 
ordinary polarization and the other one has extraordinary 
polarization ($e\rightarrow oe$). In this case, one has two 
amplitudes:
\begin{eqnarray}
\label{amplioe}
\Phi_{oe}&\approx& C_{oe}\ G_{s}(\omega_{s})G_{i}(\omega_{i})\
v(\rmvect{q}_{s}+\rmvect{q}_{i})\delta(\omega_{s}+\omega_{i}-\omega_{p})\nonumber\\
&&\times\,\sinc\left(\mu_{oe}+ l_{t} q_{sx} +
l_{t}^{\prime} q_{ix}+
\frac{L}{4K}\left|\frac{\rmvect{q}_{s}}{r_{s}} -
\frac{\rmvect{q}_{i}}{r_{i}}\right|^2\right)\nonumber\\
&&\times\, \exp[-i(\mu_{oe}+ l_{t} q_{sx} +
l_{t}^{\prime} q_{ix})],
\end{eqnarray}
and
\begin{eqnarray}
\label{amplieo}
\Phi_{eo}&\approx& C_{eo}\ G_{s}(\omega_{s})G_{i}(\omega_{i})\
v(\rmvect{q}_{s}+\rmvect{q}_{i})\delta(\omega_{s}+\omega_{i}-\omega_{p})\nonumber\\
&&\times\,\sinc\left(\mu_{eo}+l_{t}^{\prime}q_{sx} +
l_{t}q_{ix}+
\frac{L}{4K}\left|\frac{\rmvect{q}_{s}}{r_{s}} -
\frac{\rmvect{q}_{i}}{r_{i}}\right|^2\right)\nonumber\\
&&\times\, \exp[-i(\mu_{eo}+ l_{t}^{\prime} q_{sx} +
l_{t} q_{ix})],
\end{eqnarray}
where
\begin{eqnarray}
\mu_{oe}&=&\left[\frac{\bar{n}+\bar{\eta}}{2}-\eta_{p}+ \frac{\omega_{s}-\omega_{i}}{2\omega_{p}}
(\bar{n}-\bar{\eta})\right]\frac{KL}{2\eta_{p}},\\
\mu_{eo}&=&\left[\frac{\bar{n}+\bar{\eta}}{2}-\eta_{p}-
\frac{\omega_{s}-\omega_{i}}{2\omega_{p}}
(\bar{n}-\bar{\eta})\right]\frac{KL}{2\eta_{p}},
\end{eqnarray}
$l_{t}=\alpha_{p}L/2$, and
$l_{t}^{\prime}=\left(2\alpha_{p}-\bar{\alpha}\right){L}/{4}\approx 
l_{t}/2$.  The walk-off
parameter $\bar{\alpha}$ and the refractive indices $\bar{n}$
(ordinary) and $\bar{\eta}$ (extraordinary) are calculated at the
frequency $\bar{\omega}=\omega_{p}/2$.  Note the linear dependence of
$\mu_{oe}$ and $\mu_{eo}$ on $\omega_{s}-\omega_{i}$. For the crossed cone configuration\cite{kwiat95}, 
the state generated by type II SPDC is, according to Eq. \eqref{amplieo},
\begin{equation}
\ket{\psi}_{\textsc{SPDC}}\approx 
C_{\text{v}}\ket{vac}+C_{2}\ket{\psi}_{\textsc{II}},
\end{equation}
where
\begin{eqnarray}
\ket{\psi}_{\textsc{II}}&\approx&\int d\omega_{s}\int d\omega_{i}\int 
d\rmvect{q}_{s}\int 
d\rmvect{q}_{i}\ (\Phi_{oe}\ket{\rmvect{q}_{s},\omega_{s},\rmvect{e}_{o}}
\ket{\rmvect{q}_{i},\omega_{i},\rmvect{e}_{e}}\nonumber\\
&&+
\Phi_{eo}\ket{\rmvect{q}_{s},\omega_{s},\rmvect{e}_{e}}
\ket{\rmvect{q}_{i},\omega_{i},\rmvect{e}_{o}}).
\end{eqnarray}
Again, if the crystal is thin enough and narrow-band filters are used,
the state $\ket{\psi}_{\textsc{II}}$ in collinear frequency-degenerate
phase matching reduces to
\begin{equation}
\ket{\psi}_{\textsc{II}}\approx\iint\limits_{D}
d\rmvect{q}_{s}
d\rmvect{q}_{i}\ 
v(\rmvect{q}_{s}+\rmvect{q}_{i})(\ket{\rmvect{q}_{s},\rmvect{e}_{o}}
\ket{\rmvect{q}_{i},\rmvect{e}_{e}}+
\ket{\rmvect{q}_{s},\rmvect{e}_{e}}
\ket{\rmvect{q}_{i},\rmvect{e}_{o}}).
\end{equation}
\par 
Eqs.  (\ref{amplioe}) and (\ref{amplieo}) predict different
transverse and longitudinal walk-off for ordinary and extraordinary
down-converted photons.  In type II phase matching, these effects can
be considerable, and may decrease the quality of the entanglement
between the photons.  
\subsection{Coincidence detection probability and probability amplitude}
The coincidence count rate is proportional to the
fourth-order correlation function
\begin{equation}
C(\rmvect{\rho}_{s},\rmvect{\rho}_{i}) = \bra{\psi}\oper{E}^{(-)}({\rmvect{\rho}_{s}})\oper{E}^{(-)}({\rmvect{\rho}_{i}})
\oper{E}^{(+)}({\rmvect{\rho}_{s}})\oper{E}^{(+)}({\rmvect{\rho}_{i}})  \ket{\psi},
\label{eq:angspec2}
\end{equation}
where $\oper{E}$ is the electric field operator and $\rmvect{\rho}_{s}$ and $\rmvect{\rho}_{i}$
are the transverse coordinates on the signal and idler detection planes respectively.  Since  $\ket{\psi}$ is a two-photon state, the correlation function (\ref{eq:angspec4}) can be put in the form \cite{rubin96}
\begin{equation}
C(\rmvect{\rho}_s,\rmvect{\rho}_i)=|\Psi(\rmvect{\rho}_s,\rmvect{\rho}_i)|^2, 
\label{eq:Pprob}
\end{equation}
where $C(\rmvect{\rho}_s,\rmvect{\rho}_i)$ can be interpreted as the two-photon detection probability and   
\begin{equation}
\Psi(\rmvect{\rho}_s,\rmvect{\rho}_i)=\bra{vac}\oper{E}^{(+)}(\rmvect{\rho}_s)\oper{E}^{(+)}(\rmvect{\rho}_i)\ket{\psi}, 
\label{eq:wavefunction}
\end{equation}
as the two-photon detection amplitude.  Eq. (\ref{eq:wavefunction}) can be viewed as the two-photon, or {\emph {bi-photon}} wavefunction.  The electric
field operator is given by
\begin{equation}
\oper{E}^{(+)}(\rmvect{\rho})= C \int d\rmvect{q}\ 
\oper{a}(\rmvect{q})\,
\exp{[i(\rmvect{q}. \rmvect{\rho} + \sqrt{k^2-q^2}z)]},
\label{eq:angspec3}
\end{equation}
where C is a constant and $z$ is the propagation distance.

\subsection{The role of the spatial properties of the pump field}
\label{pump}
The quantum states derived in section \ref{sec:state} show a
dependence on the angular spectrum of the pump beam.  This dependence
indicates that the photon pairs can be prepared in a variety of quantum
states through the manipulation of the pump beam.  This preparation has measurable effects in
the coincidence counting rate \cite{monken98a,pittman96}.  In order to illustrate this idea, we
will describe one experiment in which the spatial shape of the
coincidence distribution in the detection plane was prepared through
manipulation of the pump beam \cite{monken98a}.

\begin{figure}
\begin{center}
\includegraphics[width=0.5\linewidth]{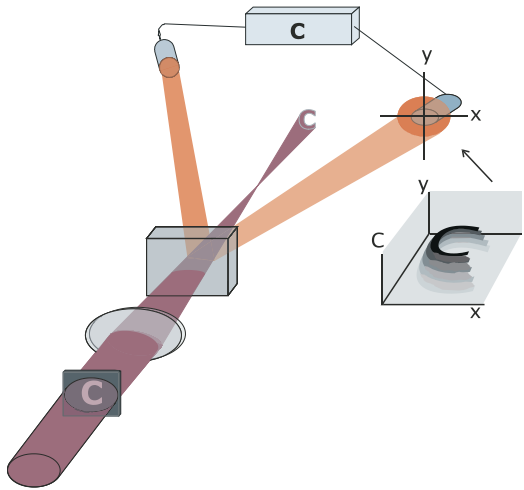}
\caption{Illustration of an experiment to show the transfer of the angular spectrum from the pump beam to the spatial correlations of down converted photons \cite{monken98a}.}
\label{angspec}
\end{center}
\end{figure}

The set-up is shown in Fig.\ref{angspec}. The pump laser passes
through a mask and a lens, before pumping the non-linear crystal.
The mask has an aperture with the shape of the letter ``C". The
lens is used to form the image of the mask in a plane situated
after the crystal. The
down-converted beams are detected on the plane
where the image of the mask is formed.  In this condition, it is possible
to recover the shape of the image ``C", in the transverse
coincidence distribution. In order to image the coincidence distribution, the idler (or
signal) detector is kept fixed, while the signal (or the
idler) detector is scanned in two dimensions on the
detection plane. A typical result is displayed in the inset in
Fig. \ref{angspec}. The angular spectrum of the
pump laser is transferred to the correlations between signal and
idler photons, while the local intensities of each photon field are not affected \cite{monken98a}.
\par
The coincidence count rate can be calculated using the state
given in Eq. \eqref{eq:spdc2} together with several approximations
simplifying the function $\Phi(\rmvect{q}_{s},\rmvect{q}_{i})$ given
by Eq. \eqref{amplioo}. Assuming the thin crystal approximation, the SPDC state can be described by
\begin{equation}
\ket{\psi}= C_2 \int\hspace{-2mm}\int\limits_{D}\hspace{-1mm}
d\rmvect{q}_{s} d\rmvect{q}_{i}\  v(\rmvect{q}_{s}+\rmvect{q}_{i})
\ket{\rmvect{q}_{s},\sigma_{s}}_{s} \ket{\rmvect{q}_{i},\sigma_{i}}_{i},
\label{eq:angspec1}
\end{equation}
where $C_2$ is a constant. We neglect $\ket{vac}$ in
Eq. \eqref{eq:angspec1}, since  it does not contribute to coincidence
counts.

Using Eqs.  \eqref{eq:angspec1}, \eqref{eq:angspec2} and
\eqref{eq:angspec3}, the coincidence rate can be calculated:
\begin{align}
C(\rmvect{\rho}_s,\rmvect{\rho}_i)&= |C_2|^2 \left| \int d\vect{\rho}\, {\cal W}(\vect{\rho})
\exp{\left(-i \frac{K}{2Z_0}|\rmvect{R} - 
\rmvect{\rho}|^2\right)}\right|^2 
\nonumber\\
&= |C_2|^2 \left|{\cal W}(\rmvect{R};Z_0)\right|^2,
\label{eq:angspec4}
\end{align}
where ${\cal W}(\rmvect{R};Z_0)$ is the transverse spatial profile of the pump beam propagated to $Z=Z_0$, and
\begin{align}
\frac{1}{Z_0} &= 
\frac{\omega_s}{\omega_p}\frac{1}{z_s}+\frac{\omega_i}{\omega_p}\frac{1}{z_i},
\\ \nonumber
\rmvect{R} &= 
\frac{\omega_s}{\omega_p}\frac{Z_0}{z_s}\rmvect{\rho}_s+\frac{\omega_i}{\omega_p}\frac{Z_0}{z_i}\rmvect{\rho}_s,
\label{eq:angspec5}
\end{align}
where $z_s$ and $z_i$ are distances between the crystal and signal and
idler detection planes, respectively.

Eq. \eqref{eq:angspec4} shows that the coincidence distribution
depends on the Fresnel propagator of the transverse profile of the pump
field. This structure coincides with the image
prepared in the pump, depending on the relationship between the
wavelengths of the three fields and the free propagation
distances.   This allows the engineering of spatial correlations by manipulating the pump laser beam.  Another important aspect is that the coincidence
distribution depends on the sum of the signal and idler detector
coordinates, and not either detector position alone, which is a signature of the spatial entanglement between the down-converted photons.

%% file: section4.tex
\section{Double slit experiments with twin photons}
\label{sec:dslitexperiments}

Spatial properties of twin photons were experimentally investigated already at the time 
of the first experimental observation of parametric down-conversion by Burnham and Weinberg \cite{BurnhamWeinberg}.
They observed that the intensity correlations were stronger for certain
combinations of detection angles.  A more comprehensive study of these
far field correlations were made by Grayson and Barbosa \cite{GraysonBarbosa}.
Experiments exploring interference at a double slit experiment using twin photons and coincidence counting
came in 1994 \cite{Ribeiro94a} and 1995 \cite{strekalov94} and was further
investigated in later work \cite{burlakov97,fonseca99a,nogueira02,fonseca99b,fonseca99c,fonseca00,kim00,nogueira01,walborn02,peeters09}.  In this section,  we review some of the key experiments exploring aspects of interference of photon pairs using a double slit.  
\subsection{Nonlocal dependence of spatial coherence}
\begin{figure}
\begin{center}
\includegraphics[width=5cm]{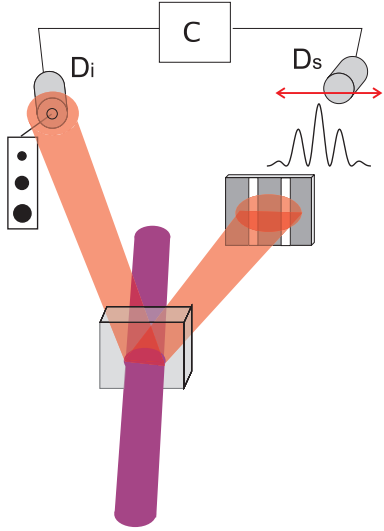}
\caption{Double slit experiment with twin photons.  In Ref. \cite{Ribeiro94a},  the signal detector $D_s$ was scanned to observe interference fringes.}
\label{dslit1}
\end{center}
\end{figure}

\begin{figure}
\begin{center}
\includegraphics[width=5cm]{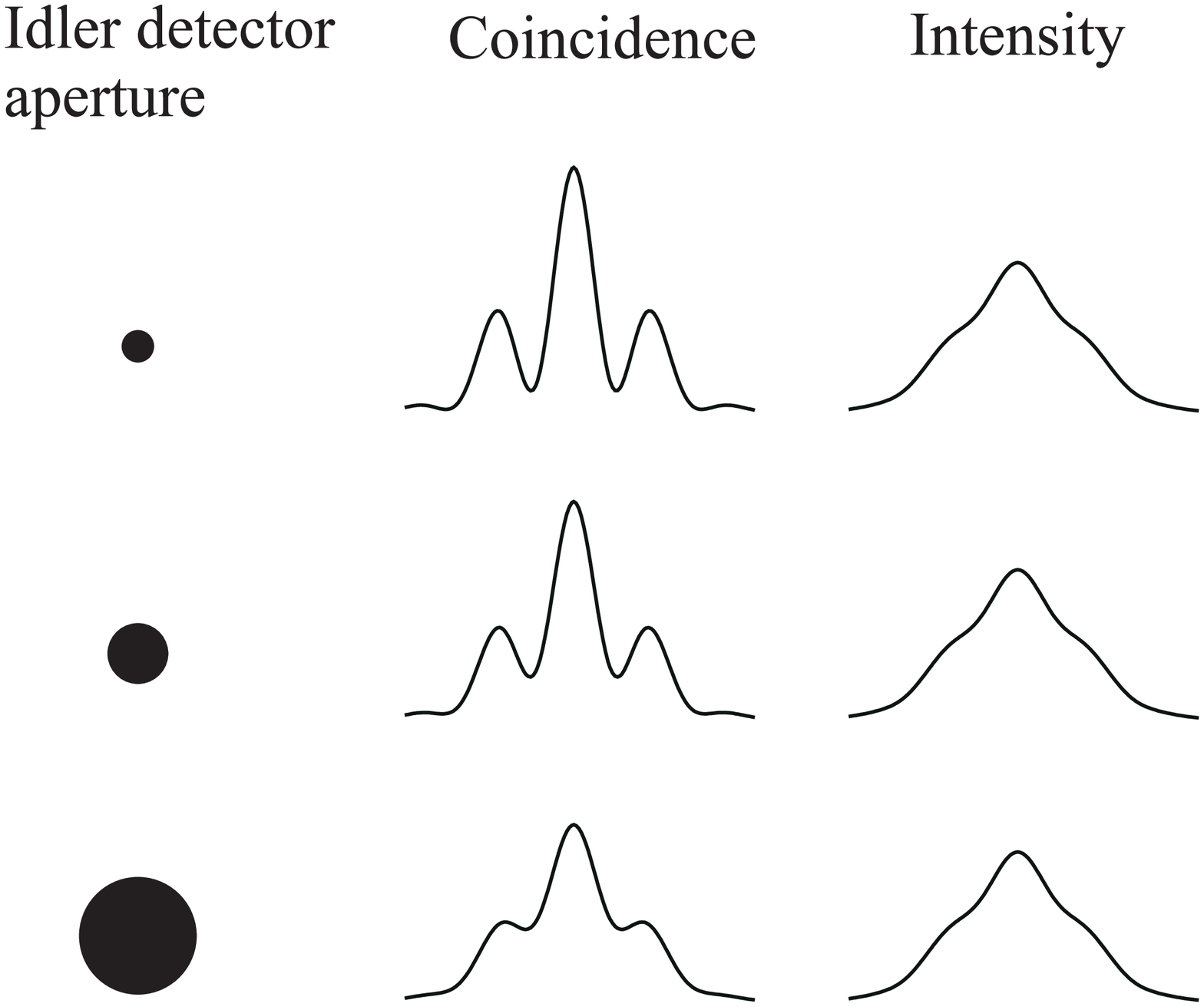}
\caption{Qualitative comparison between intensity (single-photon) and coincidence fringes obtained by scanning the signal detector $D_S$ as a function of the aperture of the idler detector $D_i$.}
\label{fringes}
\end{center}
\end{figure}

There are several possible ways to send one or two photons through a Young's double slit.  Historically, the first choice is to send one of the
photons--say the signal photon--through the double slit aperture, and let the idler propagate freely to the detection plane \cite{Ribeiro94a}, as illustrated in Fig. \ref{dslit1}.   The 
two-photon coincidence distribution can be obtained in the same way as in Section \ref{pump}. In order to take into account the presence of the slits in the signal beam, we include the diffraction integral in the electric field operator \cite{monken02}:
\begin{equation}
\oper{E}^{(+)}(\rmvect{\rho})= \int d\rmvect{q}\, d\rmvect{q}^{\prime}\, \oper{a}(\rmvect{q}^{\prime})
{\cal T}(\rmvect{q}-\rmvect{q}^{\prime})\,  
\exp {\left\{i\left[\rmvect{q}\cdot \rmvect{\rho} - \frac{q^2}{2k}Z - \frac{q^{\prime 2}}{2k}z_A \right]\right\}},
\label{dslitoperator}
\end{equation}
where ${\cal T}(\rmvect{q})$ is the Fourier transform of the double slit aperture $A(\vect{\rho})$, $z_A$ is the longitudinal coordinate of the slits, and $Z=z-z_A$. Using this operator and the quantum state (\ref{eq:angspec1}) in Eq. (\ref{eq:wavefunction}), we arrive at:
\begin{equation}
\Psi(\vect{\rho}_s,\vect{\rho}_i) \propto \iint d\vect{\rho}_s^\prime d\vect{\rho}_i^\prime A(\vect{\rho}_s^\prime)\mathcal{W}\left(\frac{1}{2}\vect{\rho}_s^\prime+\frac{1}{2}\vect{\rho}_i^\prime,z_A\right) 
e^{-i\frac{K}{4Z}\left(\left|\vect{\rho}_s - \vect{\rho}_s^\prime\right|^2+\left|\vect{\rho}_i - \vect{\rho}_i^\prime \right|^2\right)},
\label{eq:ghost1}
\end{equation}
where $\mathcal{W}(\vect{\rho},z_A)$ is the transverse profile of the pump beam propagated to $z=z_A$ and we have assumed that $k_s=k_i=K/2$ and $z_s=z_i\equiv z$.  Assuming that the distance $Z$ is such that the Fraunhofer approximation is valid, the detection probability is
  \begin{equation}
C(\vect{\rho}_s,\vect{\rho}_i) \propto \left| v\left(\frac{k}{Z}\vect{\rho}_i\right)  \mathcal{T}\left(\frac{k}{Z}\vect{\rho}_s-\frac{k}{Z}\vect{\rho}_i\right) \right|^2.
\label{eq:ghost2}
\end{equation}
Approximating the aperture function by $A(\vect{\rho}_s)=\delta(x_s+d)+\delta(x_s-d)$ gives $\mathcal{T}(\rmvect{q})\propto\cos(q_x d)$.  In this case the detection probability is 
 \begin{equation}
C(\vect{\rho}_s,\vect{\rho}_i) \propto \left|v\left(\frac{k}{Z}\vect{\rho}_i\right)\right|^2 \cos^2\left[\frac{kd}{2Z}(x_s-x_i)\right].
\label{eq:ghost3}
\end{equation}
From Eq. (\ref{eq:ghost3}) we see that the typical oscillations of a double slit interference
pattern is found, as a function of the difference between the transverse coordinates of the signal and idler
detectors. In Ref. \cite{Ribeiro94a}, the coincidence patterns were obtained by scanning the signal detector $D_s$ after the slits.   A qualitative comparison between the intensity fringes and the coincidence fringes observed as a function of the width of the idler detector aperture
is shown in Fig. \ref{fringes}. 
These results show that the visibility
of the coincidence fringes depends on the diameter of the idler detector, while
the visibility of the intensity (single-photon) fringes depends only on the source properties.
The visibility is high for a narrow aperture and low for a large aperture, implying a nonlocal dependence of the spatial coherence.   The detection probability in this case is obtained from Eq. (\ref{eq:ghost3}) by integrating over the size of the idler detector
\begin{equation}
C_D(\vect{\rho}_s,\vect{\rho}_i) \propto \int d\vect{\rho}_i D(\vect{\rho}_i-\vect{\sigma})\left|v\left(\frac{k}{Z}\vect{\rho}_i\right)\right|^2 \cos^2\left(\frac{kd}{2Z}x_s-\frac{kd}{2Z}x_i\right),
\label{eq:ghost4}
\end{equation}
where $D(\vect{\rho}_i)$ is the function describing the idler detector centered at position $\vect{\sigma}$ in the detection plane.  For a very narrow detection aperture, $D(\vect{\rho}_i) \sim \delta(\vect{\rho}_i-\vect{\sigma})$, and
$C_D(\vect{\rho}_s,\vect{\rho}_i) \propto\cos^2[{kd}/2Z(x_s-\sigma_x)]$, resulting in maximum visibility.  In other words, the post-selection by the narrow aperture 
in the idler mode projects the signal beam in a state with a narrower angular
bandwidth.  For a very broad detector, so that the width of $D$ is much larger than the width of the angular spectrum $v$, we can approximate $D\sim1$.  In this case Eq. (\ref{eq:ghost4}) gives a convolution, which washes out the oscillations of the cosine function, giving low visibility.    
A simple way of relating the idler detector diameter and the visibility
of the coincidence fringes, following the lines of the classical van Cittert-Zernike theorem, was developed in Ref. \cite{Ribeiro95a}. 
\subsection{Ghost interference}
\label{ghostinterference}
\begin{figure}
\begin{center}
\includegraphics[width=5cm]{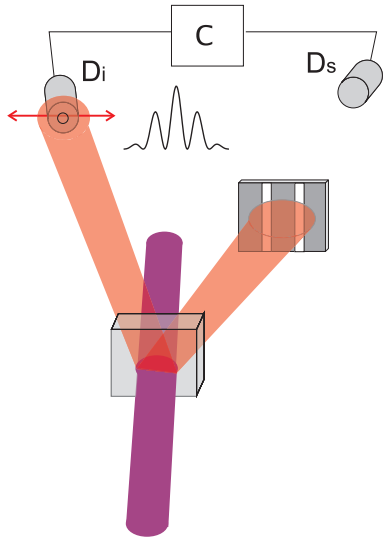}
\caption{Double slit experiment used to observe ghost interference fringes.}
\label{ghost}
\end{center}
\end{figure}

Another possibility in terms of double slit experiments using twin photons
and coincidence detection is the scanning of the idler detector instead of the
signal detector \cite{strekalov94}, as shown in Fig. \ref{ghost}. The coincidence distribution displays interference fringes, while the intensity distribution of the idler field would never present interference fringes, 
since the double slit aperture is in the other beam. These fringes can be understood
in terms of Eq. (\ref{eq:ghost3}), which shows that the parameters of the
coincidence interference pattern depend on the difference between the signal and
idler detectors coordinates.  Thus, by fixing the signal detector, and scanning the idler,  the interference fringes appear as a function of the position of the idler detector, even though the idler photon never passes through the double slit aperture.  This type of experiment was named ``ghost" interference, in reference to
the ``spooky" action at a distance attributed by Einstein to the non-local nature of entangled particles.
\subsubsection{The Advanced Wave Picture}
\begin{figure}
\begin{center}
\includegraphics[width=5cm]{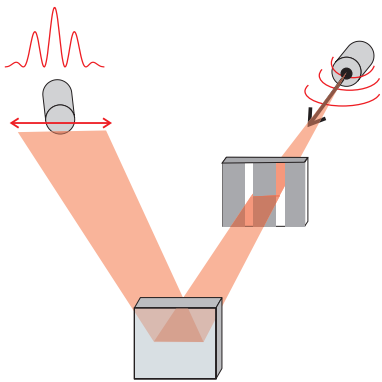}
\caption{Advanced wave picture.}
\label{awp}
\end{center}
\end{figure}

A method that aids in the understanding and design of experiments exploring the spatial
correlations between twin photons was introduced by A. V. Belinski and D. N. Klyshko \cite{belinski94b}. This method, known as the ``Advanced Wave Picture" (AWP), consists of associating
the coincidence detection and post selection to a temporal inversion of the propagation
of one of the beams. In the case
of the experiment of Fig.\ref{ghost}, the AWP considers the system as if the
fixed detector (signal) behind the slits was an incoherent light source and the
scanning detector (idler) was registering the resulting intensity. The crystal
plays the role of a mirror, as shown in Fig. \ref{awp}.
The coincidence pattern obtained when the idler detector is
scanned will be equal to the hypothetical intensity pattern obtained according
to the AWP scheme described above. The AWP works independently of the optical
components placed in the signal and idler beams. It is important to notice that
the crystal is treated as a mirror and the kind of mirror depends on the shape of the pump
beam. This dependency was discussed in section \ref{pump}.  For example, in Ref. \cite{pittman96} it was shown that the curvature of the ``mirror" is directly related to the curvature of the pump beam.

\subsection{The de Broglie wavelength of a two-photon wave packet}
\label{sec:debroglie}
Interesting possibilities arise when one sends both photons through the double slit aperture. Using the spatial correlations between the twin photons, it is possible to perform a double slit experiment in which
the two photons always pass through the same slit \cite{fonseca99b}.
\begin{figure}
\begin{center}
\includegraphics[width=5cm]{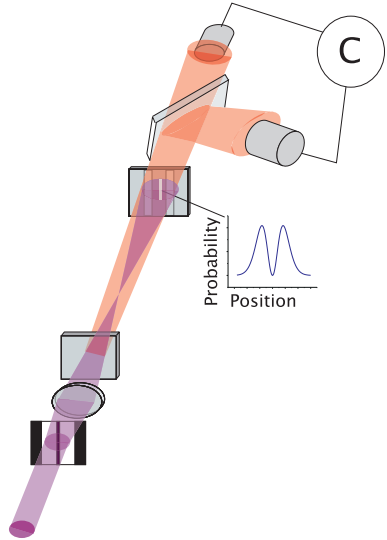}
\caption{Experimental setup for two-photon interference at a double slit.  The resulting interference pattern has oscillations corresponding to the spatial frequency of a two-photon wave packet.}
\label{debrogl}
\end{center}
\end{figure}
In this way, the interference for a wave packet containing two photons
can be observed.  The signature of the two-photon interference is the 
spatial frequency of the fringes, which is twice the 
frequency of the fringes that arise in single photon interference.  The spatial frequency is directly related to the de Broglie wavelength $\lambda_{db}=\lambda/N$ of the $N$-photon wave packet \cite{jacobson95}.
An experiment which observed the de Broglie wavelength of a two-photon wave packet was reported in Ref. \cite{fonseca99b}, and is sketched in Fig. \ref{debrogl}. A twin photon pair is produced in collinear SPDC. The beam containing the down-converted photons is sent through a double slit aperture with slit separation of $2d$, after which the photons are
split on a 50/50 beam splitter and detected in coincidence. In order to
control the spatial correlations between the pair of photons in the plane
of the slits, the pump beam is sent through a wire and a lens, placed before the crystal. The angular
spectrum of the pump is produced so that the image of the wire is projected
onto the double slit plane. The pump beam is blocked after the crystal, and does not actually pass through the slits. However the image of the wire is 
transferred to the correlations between the twin photons, such that
there is a relative spatial separation $2d$ between the photons. Therefore, if one photon
passes through one of the slits, the probability of finding its twin in the
other slit is practically zero. Then, almost all photon pairs that pass through
the slits must have passed through the same slit. 
\begin{figure}
\begin{center}
\includegraphics[height=5cm,width=9cm]{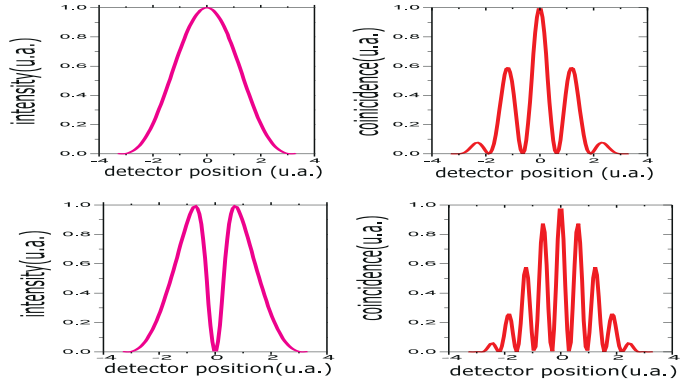}
\caption{Typical experimental results for two-photon interference at a double slit.  The plots on the left show the profile of the pump beam at the double slit.  The plots on the right show the two-photon interference pattern.  In the lower pattern, the spatial frequency is doubled.}
\label{debrog2}
\end{center}
\end{figure}
The detection of the coincidence distribution is performed by scanning both detectors
together, in order to mimic a two photon detector. Again, we can use 
the two-photon state (\ref{eq:angspec1}) and detection operator (\ref{dslitoperator})
for both the signal and idler fields to calculate the coincidence count distribution.  We obtain \cite{fonseca99b}:
\begin{eqnarray}
C(x) \propto |B_d(x)|^2 + 4|B_0(x)|^2 + |B_{-d}(x)|^2 + \\ \nonumber 
4B_d(x)B_0(x) \cos\left(\frac{kd^2}{z_A} + \frac{k x 2d^2}{z_1}\right) + \\ \nonumber
4B_0(x)B_{-d}(x) \cos\left(\frac{kd^2}{z_A} - \frac{ k x 2d^2}{z_1}\right) + \\ \nonumber
2B_d(x)B_{-d}(x) \cos\left(\frac{2 k x 2d^2}{z_1}\right),
\label{eq:debroglie}
\end{eqnarray}
where $x$ is the position of the detectors, $B_d$, $B_0$ and $B_{-d}$
are envelope functions which depend on the spatial amplitude distribution of the
pump beam and the dimensions of the double slit aperture, $z_A$ is the distance from the crystal to the double slit and $z_1$ is
the distance from the slits to the detectors.  In particular, $B_j$ is proportional to the square root of the transverse profile of the pump beam at position $x=j$.   
We can see in Eq. (\ref{eq:debroglie}) that two oscillating terms depend on
$kx$ and one term depends upon $2kx$, which is the two-photon interference term.
Therefore, if $B_0(x)=0$, then only the two-photon interference terms will
survive. This condition is achieved through the preparation of the pump profile,
using the wire and the lens to obtain a zero intensity profile at $x=0$. Then the coincidence rate will be given by:
\begin{eqnarray}
C(x) \propto |B_d(x)|^2 + |B_{-d}(x)|^2 + 2B_d(x)B_{-d}(x)\cos\left(\frac{2 k x 2d^2}{z_1}\right).
\label{eq:debroglie2}
\end{eqnarray}
Fig. \ref{debrog2} shows a computer simulation of typical experimental results, similar to those obtained in Ref. \cite{fonseca99b}.  The type of two-photon interference fringes depends strongly upon the transverse profile of the pump beam at the double slit aperture \cite{fonseca00}.    When the pump beam is a broad Gaussian distribution,  the interference pattern depends only on the wavelength and not on the number of photons \cite{walls}, and an interference pattern, shown in the upper right-side of
Fig. \ref{debrog2}, is obtained. However, if one manipulates the pump beam so that the photons always pass through the same slit, the spatial frequency of the interference pattern changes to twice the original frequency, as illustrated in Fig. \ref{debrog2}b.   Additional experiments reporting the observation of the de Broglie wavelength of two or more photons have been reported \cite{fonseca01,soutoribeiro01a,keiichi02,walther04,vidal08,luo09}.  The increase in spatial frequency in these experiments is equivalent to an increase in resolution, which may lead to interesting applications in optical lithography. 

\subsubsection{Quantum Lithography}
\label{sec:quantumlithography}
\par 
We can conclude from the experiment discussed in section \ref{sec:debroglie}, that there are instances in which electromagnetic
radiation with a given wavelength behaves as if it had a smaller wavelength,
depending on the photon number.  This idea has motivated the proposition of
applications such as lithography processes with increased resolution.  
It was called \emph{quantum lithography} \cite{yablonovitch99,boto00,kok01,dangelo01} and is 
based on the possibility of creating two-photon wavepackets \cite{fonseca99b}.  In brief, using
entangled pairs of photons and coincidence techniques, it is possible to obtain
two-photon wavepackets which interfere and diffract in the same fashion as light with twice its
wavelength. In principle, this idea can be extended to $N$-photon wavepackets and in this
case the increase in the resolution increases with $N$.
\par
The term \emph{quantum lithography} comes from the fact that the concept of photon
is required to distinguish the behavior of one photon wavepackets and
$N$ photon wavepackets. However, there are propositions for the implementation
of quantum lithography using classical light sources \cite{wang04,hemmer06,sun07,peer,boyd06}.  In this case it is the interaction between the light and some material which would only absorb $N$-photon wavepackets,   and not $J$-photon wavepackets ($J \neq N$).  Detection of only $N$-photon wavepackets projects the field onto the $N$-photon component. 
\par
Optical lithography is a process used to fabricate integrated
circuits. It consists in the removal of a thin film from a substrate through
the exposure to light. This results in drawings of very small circuits
and electronic components. The future of integrated circuits in large
scale depends on the possibility of reducing the size of these drawings.
A limitation is given by the diffraction limit of the light used in the
process, which depends upon the wavelength. Quantum lithography would allow one to obtain, for instance, the optical resolution
of a UV beam, using near infra-red light.  
The main challenge towards quantum lithography seems to be finding proper
materials which selectively absorb only $N$-photon wavepackets \cite{tsang07}.
\par
An experiment demonstrating this idea, similar to the one described in section \ref{sec:debroglie}, was later reported in Ref. \cite{dangelo01}.  

\subsection{Non-local double slit}
\begin{figure}
\begin{center}
\includegraphics[width=5cm]{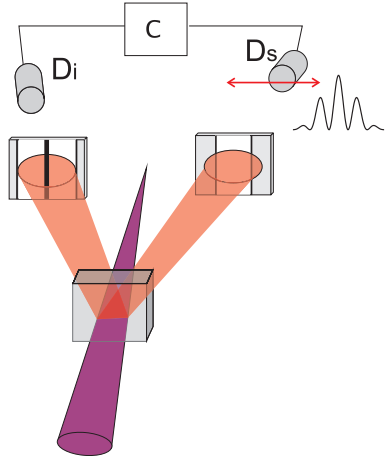}
\caption{Experimental set-up for the non-local double slit.}
\label{nldslit}
\end{center}
\end{figure}

Another possibility in terms of variations of a double slit experiment using twin
photons and coincidence detection is a non-local double slit \cite{fonseca99c},  illustrated in Fig. \ref{nldslit}.  In this experiment double-slit interference is observed, even though neither photon passes through a double-slit aperture. 
The non-local double slit is built from a single slit placed in the signal beam
and a thin wire placed in the idler beam. The combined effect of these
two components results in double-slit type coincidence fringes. In order
to calculate the coincidence count distribution, we can again use the two photon state
given by Eq. (\ref{eq:angspec1}) and the electric field operator given by Eq. (\ref{dslitoperator}), where the transfer function ${\cal T} = {\cal T}_1$ for 
the signal field corresponds to a single slit and the transfer function 
${\cal T} = {\cal T}_2$ for the idler corresponds to a thin wire.  In order to observe the coincidence fringes for the non local double slit, it
is necessary to focus the pump beam on the $z_A$ plane. This allows one to approximate the pump beam amplitude transverse distribution by a delta function. The resulting
coincidence count distribution is given by \cite{fonseca99c}:
\begin{equation}
C(\vect{\rho}_s,\vect{\rho_i}) \propto \left| \int d\vect{\xi} {A}_1(\vect{\xi})\, {A}_2(-\vect{\xi})\, 
\exp{\left(ik\frac{\xi^2}{z_A}\right)}
\exp {\left[ ik \frac{\left| \frac{1}{2}(\rmvect{\rho}_1 - \rmvect{\rho}_2) - \vect{\xi} 
\right|^2}{z_D - z_A} \right]}\right|^2,
\label{nldslitcc}
\end{equation}
where ${A}_1$ and ${A}_2$ are the transmission functions of the
single slit and the thin wire respectively, $z_A$ is the longitudinal
coordinate of the single slit and thin wire, considered to be at equal distances
from the crystal ($z=0$), and $z_D$ is the position of the detection plane for
both signal and idler detectors. The integral above is a Fresnel
integral and the argument is given by the product of the functions ${A}_1$ and ${A}_2$ which 
results in the transmission function of a usual double slit aperture.  Another characteristic
which is present in this result, as well as many other SPDC experiments, is the dependence on the
difference between the detector coordinates appearing in the Fresnel propagator in Eq. \ref{nldslitcc}.   
\subsection{The double-slit quantum eraser}
A double slit experiment with twin photons
can be used to implement a quantum eraser \cite{scully82,scully91}, which shows the complementarity between wave-like and particle-like behavior.  Exploiting the polarization correlation between down-converted photons, the idler photon can be used to herald either interference fringes (wave-like) in the signal field or to determine which slit the signal photon passed through (particle-like). Furthermore, one can delay the decision about erasing this information or not.
\par
Typically, if one cannot determine through which slit a photon passes, then interference fringes will be observed in the intensity distribution after the double slit aperture.  However, if one can mark the photon's path, so that the two paths are in principle distinguishable, then no interference fringes appear.  It was previously argued that the disappearance of interference fringes was due to the perturbation caused by the which-slit marker \cite{feynman65v3}.  However, Scully, Englert and Walther showed that the interference fringes can reappear if one erases the which-slit information in the marker \cite{scully91}.  Hence the name \emph{quantum erasure}.  Let us begin by describing a simple double-slit quantum eraser with single photons.
\begin{figure}
\begin{center}
\includegraphics[width=5cm]{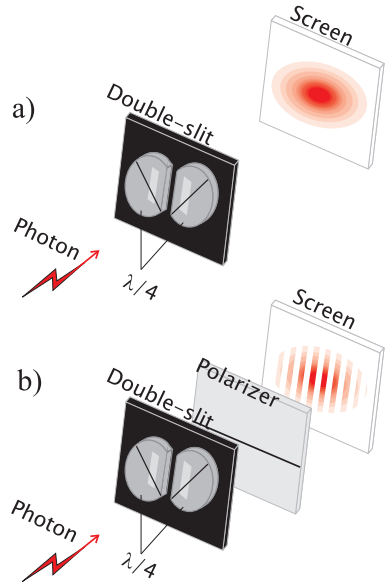}
\caption{The double slit quantum eraser for one photon. $\lambda/4$ represents
a quarter wave plate.}
\label{qeraser1}
\end{center}
\end{figure}
Let us suppose that a source emits single photons which propagate to a 
double slit. If interference fringes are visible on the detection screen after many photons have emitted, then one cannot 
determine through which slit each photon has passed. 
The state of the photons after the slits can be given by: 
\begin{equation}
\label{eq1:qeraser}
|\psi\rangle_{1photon} = \frac{1}{\sqrt{2}}[|\psi_1\rangle+ |\psi_2\rangle],
\end{equation}
where $\psi_1$ and $\psi_2$ represent the passage through slit 1 and 2 respectively.
In order to mark the path of each photon, let us place quarter wave-plates in front of each slit, so that if the horizontally polarized photon crosses one of the slits, it leaves circularly polarized to the right and if it crosses the other it leaves
circularly polarized to the left, as in Fig.\ref{qeraser1}a. Then, a measurement of the photon polarization is enough to determine through which slit it has passed. This causes the interference pattern to disappear, even if the polarization
is not measured.  Even the possibility of obtaining which-path information
is enough to destroy the interference, since the presence of the wave plates in each slit entangles states
$\psi_1$ and $\psi_2$ with the polarization degree of freedom, so that the state becomes:
\begin{equation}
\label{eq2:qeraser}
|\psi\rangle_{1photon} = \frac{1}{\sqrt{2}}[||R\rangle |\psi_1\rangle+|L\rangle |\psi_2\rangle],
\end{equation}
where $|R\rangle$ and $|L\rangle$ are right and left circular polarization
states respectively. 
The which-path information is available due to the entanglement between the state
describing the passage through each slit and the output polarization
state. It can be erased by projecting the two orthogonal polarization states $|R\rangle$ and $|L\rangle$
onto a--say--linear polarization state, $|H\rangle$(horizontal) or $|V\rangle$(vertical)
for instance, as in Fig. \ref{qeraser1}b. Therefore, after the polarizer, the which-path information is
no longer available, and the interference fringes are recovered.
\par
\begin{figure}
\begin{center}
\includegraphics[width=6cm]{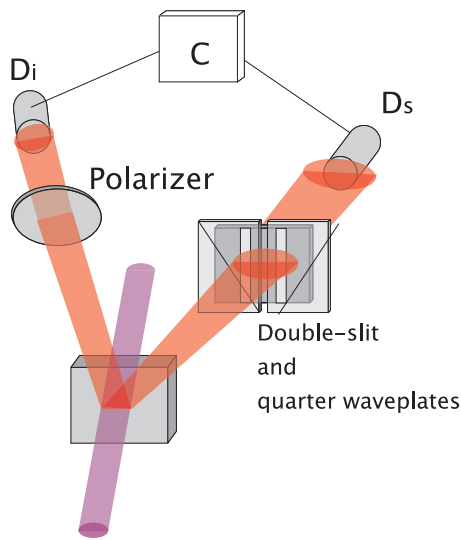}
\caption{The delayed choice double slit quantum eraser. Quarter wave-plates are used to mark the path of the signal photon.  Which-path information can be recovered or erased by projecting the idler photon onto the appropriate polarization state.}
\label{qeraser2}
\end{center}
\end{figure}
\begin{figure}
\begin{center}
\includegraphics[width=4cm]{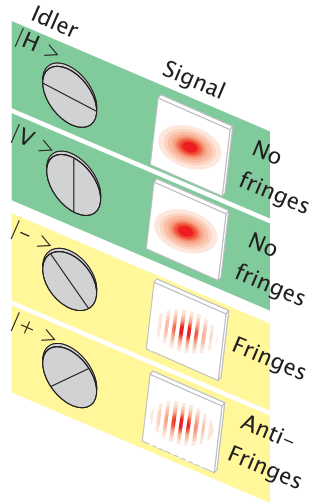}
\caption{The delayed choice double slit quantum eraser. Each polarization
projection of the idler photon corresponds to a state preparation of the signal photon.
Projections onto vertical and horizontal polarizations results in which-path
information. Projections onto diagonal polarizations results in erasure of
which-path information and interference.}
\label{qeraser3}
\end{center}
\end{figure}

Using pairs of photons allows for a delayed choice quantum eraser, as reported in Refs. 
\cite{kim00,walborn02,walborn03d}.  A sketch of an experiment is shown in Fig. \ref{qeraser2}.
A pair of photons is prepared in a polarization-entangled state:
\begin{equation}
\label{eq3:qeraser}
|\psi\rangle_{QE} = \frac{1}{\sqrt{2}}[|H\rangle_i |V\rangle_s+|V\rangle_i |H\rangle_s].
\end{equation}
The signal photon is sent through the double slit aperture and quarter-wave plates. The state of the photons after passage through the slits is
given by:
\begin{equation}
\label{eq4:qeraser}
|\psi\rangle_{QE} = \frac{1}{2}[|H\rangle_i(|L\rangle_s|\psi_1\rangle + |R\rangle_s|\psi_2\rangle) + 
                                      i |V\rangle_i(|R\rangle_s|\psi_1\rangle + |L\rangle_s|\psi_2\rangle)].
\end{equation}
The polarization entanglement allows the idler photon to be used as a which-path marker or eraser. If it is projected onto polarization
$H$ or $V$ the signal photon will be projected onto state $|L\rangle_s|\psi_1\rangle + |R\rangle_s|\psi_2\rangle$ or
$|R\rangle_s|\psi_1\rangle + |L\rangle_s|\psi_2\rangle$ respectively. Therefore, no interference is observed. 
However, if one projects the idler photon onto $|+\rangle$(+45 degrees) or 
$|-\rangle$(-45 degrees) linear polarizations, the which-path information is erased and
the interference fringes return, as illustrated in Fig. \ref{qeraser3}.  
In order to see this, we write the state (\ref{eq4:qeraser}) in the $+/-$ basis:
\begin{equation}
\label{eq5:qeraser}
|\psi\rangle_{QE} = \frac{1}{\sqrt{2}}
[|+\rangle_i|+\rangle_s(|\psi_1\rangle - i|\psi_2\rangle) + i
|-\rangle_i|-\rangle_s(|\psi_1\rangle + i |\psi_2\rangle)].
\end{equation}
We see in Eq. (\ref{eq5:qeraser}) that projections of the idler photon
onto $|+\rangle_i$ or $|-\rangle_i$ projects the signal photon onto a
superposition of $|\psi_1\rangle$ and $|\psi_2\rangle$ states, which gives interference fringes. This quantum eraser allows for 
``delayed choice", meaning that the decision about the measurement of
the idler photon can be delayed until the signal has already been detected.

%% file: section5.tex
\section{Quantum imaging}
\label{sec:qimages}
In sections \ref{pump} and \ref{sec:debroglie} we discussed experiments in which the image or interference pattern of an object appears in the two-photon coincidence counts.  These types of experiments have led to the idea of ``quantum imaging", which in some cases may present advantages over classical imaging techniques, such as increased spatial resolution \cite{santos03,santos05,santos08,giovannetti09}, as discussed in section \ref{sec:debroglie}. 
The term quantum images generally refers to images that appear in two-photon coincidence distributions. They
are called {\it quantum}, because they were first produced with the quantum-correlated twin photons produced in parametric down-conversion. However,
it is also possible to produce correlated images with classically correlated
light sources. In this section we will discuss quantum and classically
correlated images and some proposed applications.
\subsection{Quantum images}
The image of a double-slit aperture can be understood in terms of the diffraction theory of classical optics as the near-field distribution of the double slit, whereas the interference pattern is related to the far-field distribution.  The amplitude distributions of the electric field
in these two cases are related by a Fourier transform. This is also true for the correlated
images and interference patterns which may appear in coincidence distributions.   However, the propagation and detection of both signal and idler fields may influence the resulting spatial distribution. 
\par
\begin{figure}
\begin{center}
\includegraphics[width=8cm]{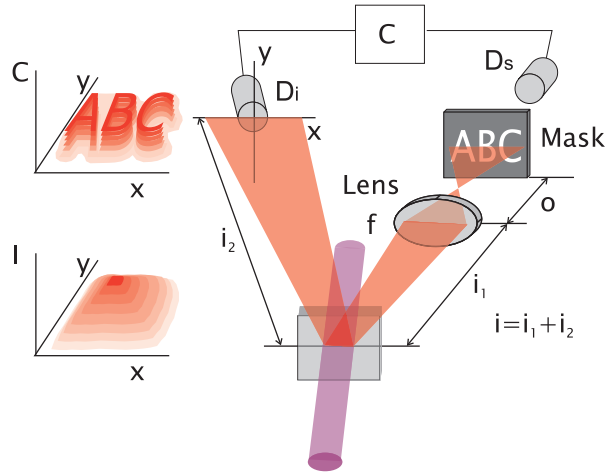}
\caption{Observation of a quantum image, similar to the experiment of Pittman et al. \cite{pittman95}.  The image of the ABC mask appears in the coincidence counts when the detector D1 is scanned.}
\label{image1}
\end{center}
\end{figure}

In order to introduce the idea of a quantum image, let us consider the
experiment reported by Pittman et al.\cite{pittman95} and illustrated
in Fig.\ref{image1}. Parametric down-conversion is produced as usual. An
aperture mask is placed in the signal beam close to the detector and
a lens also placed in the signal beam. 
Coincidence detection is performed
such that the signal detector is a fixed large-aperture {\em bucket} detector, while the idler detector is
scanned in the transverse plane. A 3-D plot of the coincidence counts
as a function of the transverse coordinates $x$ and $y$ reproduces the
image of the aperture {\it ABC}. See the inset in the upper part of 
Fig. \ref{image1}.
The lens in the signal beam is placed
according to the advanced wave picture (see subsection \ref{ghostinterference} 
for details). The imaging condition given by the thin lens formula 
$\frac{1}{f}=\frac{1}{i}+\frac{1}{o}$ is applied, where $f$ is the
focal length, $o$ is the distance between the lens and the aperture mask
(the image plane) and $i$ is the distance between the lens and the crystal
plus the distance between the crystal and the idler detector. According to 
the advanced wave picture, the crystal plays the role of a mirror in the
imaging process.

There are some particular aspects which distinguish this image pattern
from ordinary intensity images. One feature is that the image is observed in coincidence as a function of the position of the idler detector, even though the idler did not pass through the aperture mask.  The image does not appear in the intensity distribution of the idler alone, but only when it is counted in coincidence with a detection of the signal photon. The intensity distribution
depends only on the source shape as usual, as illustrated in the inset in the lower part of Fig. \ref{image1}.  Another aspect is that the dimensions and 
the resolution of the quantum image depend on the pump beam wavelength and spatial properties.  

Initially, it was widely believed that the quantum correlation of the down-converted photons was necessary to observe quantum images \cite{abouraddy01}.  Thus, it was quite surprising when similar types of correlated images were produced with classical sources. This is the subject of the next section.

\subsection{Classical two-photon images}
\begin{figure}
\begin{center}
\includegraphics[width=8cm]{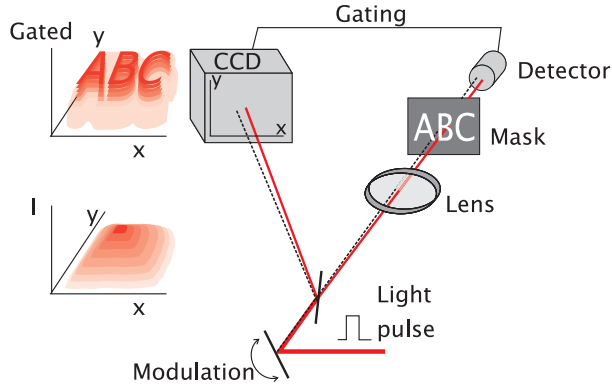}
\caption{Observation of a classically correlated image using a laser pulse as reported in Ref. \cite{bennik02}.}
\label{image2}
\end{center}
\end{figure}

A great deal of discussion was triggered by the work of Bennink et al.
\cite{bennik02,bennik04},
in which it was demonstrated that it is possible to obtain correlated images using classical light, in the
same fashion as it is done with entangled photons. Bennink et al.'s  experiment
is sketched in Fig.\ref{image2}.
A laser pulse is sent to a mirror and then to a 50-50 beam splitter. The transmitted
beam propagates through an aperture mask and is detected. Let us call this the signal
beam. The idler beam is reflected at the beam splitter and sent directly to a CCD camera, 
which records the spatial intensity distribution. It is not necessary to use a pulsed laser, the pulse-like scheme can be obtained for instance with a chopper in a c.w. laser. However, it is important to have pulses, so that there will be a temporal correlation between signal and idler pulses. 
\par
The angle of the mirror before the beam splitter is
randomly modulated in both axial and azimuthal directions. This artificially broadens the angular
spectrum of the light sent to the beam splitter, and the
temporal correlation results in an angular correlation between the signal and idler beams. The signal detector after the
aperture mask is kept fixed, while the idler is detected with the CCD camera
which only records when a trigger pulse is sent by the signal. The triggered CCD image reproduces the {\it ABC} aperture, just as in the case of correlated photon pairs from SPDC.  Almost all
ingredients naturally present in the pair of photons produced in parametric
down-conversion, are artificially introduced here: {\it i)} temporal correlation and
{\it ii)} angular correlation.  However, since the correlated pulses are produced classically, there is no entanglement. 
This experiment demonstrated that it is possible to obtain a correlated
image using classical light, and opened up the discussion about the role of entanglement in quantum imaging experiments.   
\par
\begin{figure}
\begin{center}
\includegraphics[width=8cm]{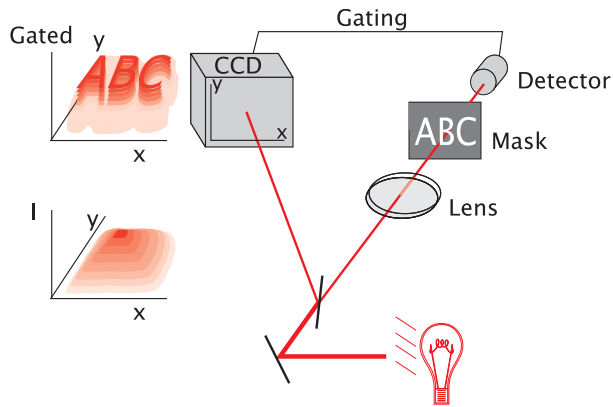}
\caption{Observation of a classically correlated image using a thermal light source.}
\label{image3}
\end{center}
\end{figure}
\begin{figure}
\begin{center}
\includegraphics[width=8cm]{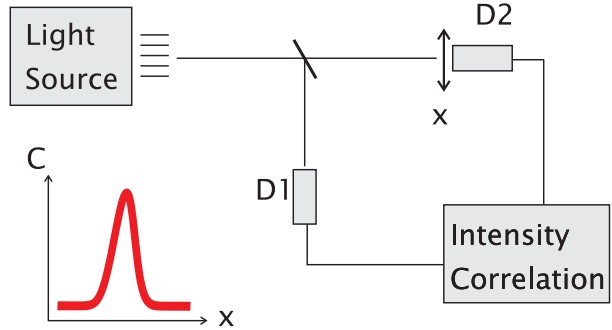}
\caption{Sketch of the Hanbury Brown and Twiss experiment. C is the intensity
correlation. It is proportional to $\mu_{12}$, the normalized spatial degree of coherence.}
\label{image4}
\end{center}
\end{figure}

The production of correlated images with classical light was extended to
thermal/chaotic light sources \cite{wang04,lugiato04a,lugiato04b,ferri05,Scarcelli}, supporting 
the idea that correlated images were not actually quantum.  A typical experimental set-up
for the observation of correlated images with thermal light is shown In Fig. 
\ref{image3} and is similar to Fig. \ref{image2}. However, the pulsing to obtain temporal correlation and spatial modulation to obtain spatial correlation is not necessary anymore. Due to the natural bunching of photons in thermal light, it is possible to obtain temporal and spatial correlation directly from the thermal light. 
This kind of correlation was already observed in the famous experiment by 
Hanbury Brown and Twiss (HBT)\cite{HBT}. 
In the HBT experiment, sketched in Fig. \ref{image4}, the light is sent to a beam splitter and intensity correlations
are measured. One detector is kept fixed, while the other is scanned along the
detection plane. Results show that the width of the coincidence distribution is proportional to the spatial coherence of the input field. The larger the coincidence
distribution, the smaller the spatial correlation and the larger the spatial coherence.
For incoherent sources like thermal sources, the spatial coherence depends only
on the source dimensions, as discussed in section \ref{sec:vCZ}. 
In this way the HBT interferometer was used to measure the diameter of stars.
Correlated imaging with a laser source required a modulation
to induce spatial correlations since the spatial correlation
is inversely proportional to the spatial coherence, which in the case of a laser source can be quite large.
\par
While classical light can also be used to produce correlated images, there is a price 
to be paid: the signal to noise ratio is generally higher than for the output state of a down-converter \cite{lugiato04a,erkmen09}.  Quantum imaging has been analyzed in the general case of light fields with Gaussian photon statistics \cite{tan08,erkmen08}, and this formalism has been applied to two-photon imaging \cite{erkmen08b}. 
\subsection{Phase objects}
\begin{figure}
\begin{center}
\includegraphics[width=8cm]{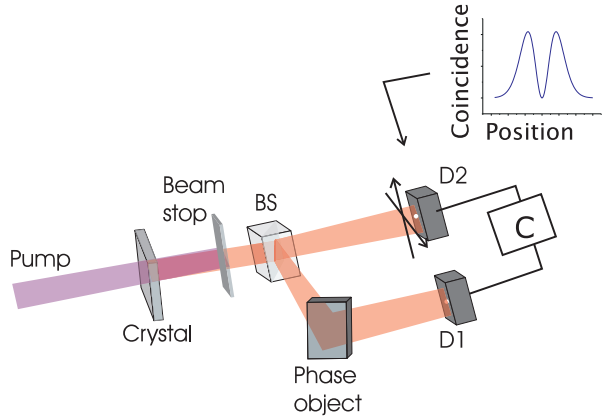}
\caption{Observation of a quantum image with a pure phase object.}
\label{image5}
\end{center}
\end{figure}

The discussion about the differences between quantum and classically
correlated images has led to an important test, proposed and realized 
experimentally by Abouraddy et al.\cite{boston}.
The authors observed correlated images using phase objects. Fig. \ref{image5} shows a sketch of a typical experimental set-up.
The transmission configuration using an aperture is replaced by a reflection
configuration using a micro electromechanical system (MEMS) micro-mirror. The 
spatial amplitude distribution of the reflected beam is not changed, while
the spatial phase distribution is modulated. 
The signal photon propagates through this
plate and is detected afterwards by a fixed detector D1, outfitted with a small pinhole. 
The idler photon propagates directly to detector D2, which is scanned in the transverse plane. 
The spatial coincidence distribution will be determined by the diffraction pattern 
of the signal beam. A typical coincidence profile 
is illustrated in the inset of Fig. \ref{image5}. This kind of two peak profile, is
obtained when a double-slit object with phase $\pi$ separated by a line of phase 0
is prepared in the MEMS. It was shown in  Ref.\cite{bache06}, that coherent imaging can also be achieved with pseudo-thermal sources.
The observation of quantum images of phase objects is a further demonstration
of the differences between the correlated images with classical and quantum light.

\subsection{Spatial resolution of magnified images: quantum versus classical}

It is well know that the spatial resolution of an image increases
when the wavelength of light used for illuminating the object is
decreased. On the other hand, short wavelengths in the UV range or
smaller usually damage an illuminated object. Needless to say, it would be
advantageous to achieve the spatial resolution of short wavelengths while illuminating
the object with a light beam with longer wavelength. Spatially entangled two-photon
light beams generated by SPDC are good candidates for this
special light source because it has two wavelengths associated
with the two-photon beams: the central wavelengths of the
individual photons $\lambda$ and the de Broglie wavelength
associated with the biphoton wavepacket equal to $\lambda/2$, as discussed in section \ref{sec:debroglie}.  
It was demonstrated experimentally
in Ref. \cite{santos05} that magnified images of objects
illuminated by two-photon wavepackets, as in a microscope \cite{salehpatent},
can be generated with spatial resolution better than the
diffraction limit and are self-apodized \cite{klein86}.
\par
For a comparison with a classical imaging experiment suppose an
object is illuminated by a classical coherent light source.  The
image is produced with a lens with a focal length $f$, separated
from the object by a distance $z_L$. The image plane is located at
a distance $z_D$ far from the lens. If it is assumed that the light
arriving at the object is a plane wave, the electric field at the
image plane is
\begin{eqnarray}   \label{classica}
E(\vect{\rho}) &=& \int O(\vect{\xi})T_{L}\left( \frac{k\vect{\xi}}{z_L}+\frac{k\vect{\rho}}{z_D}%
\right)d\vect{\xi},
\end{eqnarray}
where $O(\vect{\xi})$ is the transmission function of the object,
$\vect{\rho}$ is the transverse position in the image plane, $k =
2\pi/\lambda$ is the wave vector of the
incident plane wave and $T_L$ is the Fourier transform of the
magnitude of the lens transmission function. This last term limits
the spatial resolution of the image due to the finite size of the
lens aperture \cite{klein86}.

In the imaging experiment with photon pairs generated by SPDC, the object information appears in the two-photon probability amplitude at the image
plane. It was shown theoretically in Ref. \cite{santos03} that the
 probability amplitude of simultaneously detecting two down-converted photons in
the image plane of an object illuminated by the signal photon is
\begin{eqnarray}   \label{quantica}
\Psi(\vect{\rho}) &=& \int O(\vect{\xi})T_{F}\left( \frac{2k\vect{\xi}}{z_L}+\frac{2k\vect{\rho}}{z_D}%
\right)d\vect{\xi},
\end{eqnarray}
where it is assumed that
$z_L = z_{Li} = z_{Ls}$, $z_D = z_{Di} = z_{Ds}$. Vector $\vect{\rho}$ describes the detection position at both signal and idler
detection planes, so that $\vect\rho_{i}= - \vect\rho_{s} = \vect\rho$. This last
assumption means that image is measured by displacing the signal
and idler detectors simultaneously in opposite directions. It has
also been assumed that the pump laser beam at the object plane
has a transverse profile which is sufficiently narrow to be described
by a Dirac delta function. $T_F$ is the Fourier transform of the function $F$ defined as
\begin{equation}   \label{lente}
F(\rmvect{v}) = \int |A_{Li}(\rmvect{u}+\rmvect{v})||A_{Ls}(\rmvect{u}-\rmvect{v})|d\rmvect{u},
\end{equation}
where $|A_{Ls}|$ and $|A_{Li}|$ are the magnitude of the aperture functions of the lenses placed in signal and idler paths, respectively.
$F(\rmvect{v})$ is the correlation of the lens transmission function
magnitudes \cite{santos03}.
In comparing Eqs. (\ref{classica}) and (\ref{quantica}), one
notices the presence of the factor $2k$ in Eq. (\ref{quantica})
instead of $k$ as in Eq. (\ref{classica}).  Eq. (\ref{quantica}) is equivalent to an image generated by single photons, however with
wavelength equal to the De Broglie wavelength
\cite{jacobson95,fonseca99b} of the biphoton
$\lambda/2$, and thus shows improvement in the spatial resolution.
On the other hand, the function $F$ present in Eq. (\ref{quantica})
describes an effective lens with aperture transmission function that has a magnitude that is equal to the
correlation of the magnitudes of the aperture transmission
functions of the actual lenses, as it can be seen in Eq.
(\ref{lente}). The function $F$ describes an effective apodized
lens \cite{klein86}.  
\par
The image generated by the parametric down-converted photons is
better resolved than a similar one generated by an infrared
classical light source with the same wavelength, and is resolved
better than the diffraction limit in this case. In spite of this, the quantum
image resolution is not as good as the resolution of the image
produced by the pump laser source ($\lambda_p=\lambda/2$), since the
transmission function of the effective lens seen by the photon
pairs is not equal to the transmission function of the original
lenses.  Apodization effects are also observed in the twin photon image without physically apodizing the lenses used, which could lead to interesting applications.
 \par
The spatial resolution of images obtained in quantum fourth-order
imaging has also been compared with that obtained in a classical
second-order incoherent imaging method \cite{santos08}. The intensity at the image plane of an object
illuminated by a classical incoherent light source is \cite{goodman96}:
\begin{equation}
I(\vect{\rho}) = \int \left|O(\vect{\xi})\right|^{2}T_{\wp}\left(
\frac{2k\vect{\xi}}{z_L}+\frac{2k\vect{\rho}}{z_D}
\right)d\vect{\xi},
\label{e:f}
\end{equation}
  where $T_{\wp}$ is the Fourier transform of the function $\wp$,
defined as
\begin{equation}
    \wp(\rmvect{u}) = \int A_{L}(\rmvect{u} + \rmvect{v})A_{L}(\rmvect{u} - \rmvect{v})d\rmvect{v},
    \label{e:g}
\end{equation}
$k = 2\pi/\lambda$, in which $\lambda$ is the central wavelength
of the incoherent light beam and $A_{L}$ is the aperture
transmission function of the lens. The function $\wp(\rmvect{u})$ is
the auto-correlation function of the aperture
function of the lens used for imaging, and  
describes the aperture function of an effective lens.  Note that in
the two-photon image the effective lens transmission function (\ref{lente}) is not an auto-correlation function but a cross correlation function
of the signal and idler aperture lens transmission functions.  Another difference with respect to the object
function is that in the incoherent imaging only the square modulus
of the object function appears in the optical intensity
expression, i.e., this method is not sensible to the object phase.
In quantum imaging the complete object function
is present in the coincidence distribution at the image plane,
i.e., the image carries the information about the object spatial
phase.  A situation where the object function has no phase
variation was analyzed in Ref. \cite{santos08}, but the two imaging
methods still produced completely different results. The
transmission function of the lens pair in the two-photon imaging
was modified such that the cross-correlation function between them
was different from the auto-correlation of one of the lenses. When
the object was illuminated by an incoherent classical light
source, the image was not resolved spatially. On the other hand,
an image with very high spatial resolution was measured when the
object was illuminated by one down-converted photon, and both photons were detected in coincidence at the image
plane.

%% file: section6.tex
\section{Spatial correlations: quantum versus classical}
\label{sec:spatialcorr}

In the 1990's, many novel and interesting features of spatial correlations were observed experimentally and well described by the quantum theory.
Many of these results suggested that the spatial correlations were indeed quantum correlations, even though
the non-existence of a classical analog was not discussed in most
of these contributions. For example, the non-local conditionality of two-photon interference fringes \cite{strekalov94,fonseca01} was well described by quantum mechanics.  However, there was never any formal proof that the conditional interference fringes could not be exactly reproduced by some special classical light source.   In Ref. \cite{zou91},
it was shown that the simple detection of coincidence counts above the accidental coincidence rate is a proof of non-classical 
behavior.  However, this refers to the temporal correlations and cannot
be directly extended to spatial variables.  
\par
Perhaps the first experimental demonstration of the quantum nature of light via a fourth-order correlation function was photon anti-bunching,  first observed in the time domain by Kimble et al. in 1977
\cite{kimble77}.  Demonstration of the non-classical nature of photon anti-bunching was based on the violation of a classical inequality.  In
the same spirit, the anti-bunching of photons was later demonstrated in the transverse spatial domain in Ref. \cite{nogueira01}, and was probably the first formal demonstration of the non-classical behavior of spatial correlations.  In 2004, Howell et al. \cite{howell04} and D'Angelo et al. \cite{dangelo04} demonstrated that the pairs of down-converted photons are indeed entangled by experimentally violating classical separability criteria.    
\par
In this section we review several of these important results concerning the non-classical nature of transverse correlations.  The observation, detection and quantification of spatial entanglement will be discussed in sections \ref{sec:spatialentanglement} and \ref{sec:transversemodes}.  

\subsection{Conditional interference and complementarity in one and
two-particle interference patterns} 
\begin{figure}
\begin{center}
\includegraphics[width=0.3\linewidth]{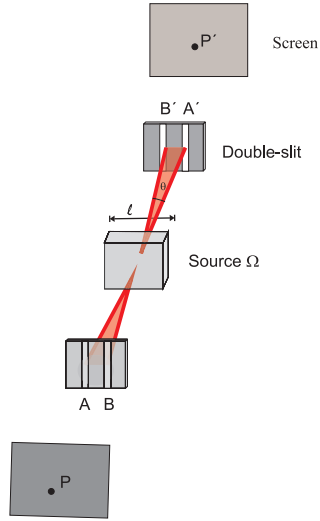}
\caption{Double-slit interference with two photons.}
\label{fig:condit1}
\end{center}
\end{figure}
An interesting idealized
two-particle double-slit interference experiment was discussed by
Greenberger, Horne and Zeilinger \cite{ghz93} in 1993. In
this gedanken experiment, two identical double-slits were placed in opposite
sides of a linear source $\Omega$ with extension $l$ that always
emits two particles with momenta approximately equal and
opposite, as sketched in Fig. \ref{fig:condit1}. The angle between the slits and the source is
$\theta$ and, due to the momentum correlation, the particles  pass either through the slits
$A$ and $A'$ or $B$ and $B'$. The arrival of the particles is
registered at two opposite screens after each double-slit. For the case where the
daughter particles are photons and $l >> \lambda /\theta$, the probability amplitude 
for one of the particles to arrive at $P$ and the other to arrive at $P'$ is \cite{ghz93}
\begin{equation}
\Psi_{2}(x_i,x_s) \propto \,\cos\left[\,k\,\theta^\prime(x_{i}-x_{s})\right]
\label{nc1}
\end{equation}
where $x_{s}$ and $x_{i}$ are the transverse coordinates, $k = 2\pi / \lambda$, $\lambda$ is the wavelength (de Broglie or optical) related to the emitted
particles, and $\theta^\prime$ is the angle that is subtended by the hole pairs and  the
detecting screens.
On the other hand, if $l <<\lambda /\theta$, the probability
amplitude is
\begin{equation}
\Psi_{2}(x_i,x_s) \propto \,\cos(\,k\,\theta'\,x_{i}) \times
\cos(\,k\,\theta'\,x_{s}). \label{nc2}
\end{equation}
Expression (\ref{nc1}) shows that the coincidence interference
pattern detected when $l >> \lambda /\theta$  presents
``conditional fringes".    If the interference pattern is
recorded by two movable detectors $D_i$ and $D_s$, the term ``Conditionality" means
that the position of the interference fringes will vary depending on the
position of both detectors.  Suppose the interference pattern is
recorded by keeping one of the detectors ($D_i$, for example)
fixed at some position $x$ while $D_s$ is scanned. For two
different positions $x$, two displaced interference pattern will
be recorded. Expression (\ref{nc2}) shows that for $l <<\lambda
/\theta$ a product of independent single-particle interference
patterns (``independent fringes") is measured and conditional
fringes are not detected. This condition is the requirement to
detect the usual Young single-particle interference pattern.

\begin{figure}
\begin{center}
\includegraphics[width=0.4\linewidth]{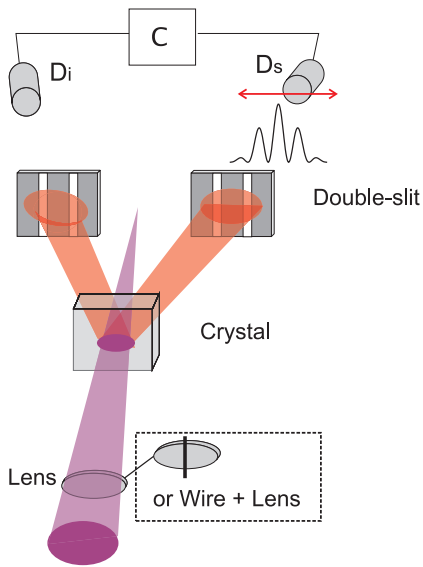}
\caption{Double-slit interference with two photons and parametric down-conversion. Focusing the pump beam with a lens induces an anti-correlation (dependence on the sum of coordinates) and using a a lens plus a wire induces correlation (dependence on the difference of coordinates).}
\label{fig:condit2}
\end{center}
\end{figure}

This idealized experiment was tested experimentally in reference
\cite{fonseca00} using photons obtained from SPDC.  The experiment is illustrated in Fig. \ref{fig:condit2}. Two-photon interference
experiments with spatial interference patterns were also observed
in \cite{ghosh87,Ribeiro95b,strekalov94,Ribeiro95a,hong98,fonseca99b} but as discussed above 
\cite{ghz93} 
the detection of a fourth-order interference pattern does not guarantee the presence
of conditional fringes.  Using the quantum multimode theory presented in section \ref{sec:fundamentals} to calculate the
fourth-order correlation function as a function of the detector positions, the number of coincident photons
at positions $x_{i}$ and $x_{s}$ is \cite{fonseca00}:
\begin{eqnarray}
N_{c}(x_{i},x_{s}) \propto  A(x_{i},x_{s}) +
2\,B_{1}(x_{i},x_{s})\,
B_{2}(x_{i},x_{s})\,\cos\left(\frac{k\,d^{2}}{z_{A}}+
\frac{k\,x_{i}\,2\,d}{z_{1}}\right)\,+\nonumber \\
+
2\,B_{1}(x_{i},x_{s})\,B_{3}(x_{i},x_{s})\,\cos\left(\frac{k\,d^{2}}{z_{A}}+
\frac{k\,x_{s}\,2\,d}{z_{1}}\right)\,+\nonumber \\ +
2\,B_{2}(x_{i},x_{s})\,B_{4}(x_{i},x_{s})\,\cos\left(\frac{k\,d^{2}}{z_{A}}-
\frac{k\,x_{s}\,2\,d}{z_{1}}\right)\,+\nonumber \\
+ 2\,B_{3}(x_{i},x_{s})\,
B_{4}(x_{i},x_{s})\,\cos\left(\frac{k\,d^{2}}{z_{A}} -
\frac{k\,x_{i}\,2\,d}{z_{1}}\right)\, +\nonumber \\ + 2\,B_{2}(x_{i},x_{s})\,
B_{3}(x_{i},x_{s})\,\cos\left[\frac{2\,d\,k\,(x_{i}-x_{s})}{z_{1}}\right]\,
+\nonumber \\
+
2\,B_{1}(x_{i},x_{s})\,B_{4}(x_{i},x_{s})\,\cos\left[\frac{2\,d\,k\,(x_{i}+x_{s})}{z_{1}}\right]\, \label{nc6}
\end{eqnarray}
with
\begin{eqnarray}
A(x_{i},x_{s}) = \vert\,B_{1}( x_{i},x_{s})\,\vert^{2} +
\vert\,B_{2}( x_{i},x_{s})\, \vert^{2} +  \vert\,B_{3}(
x_{i},x_{s})\,\vert^{2} + \vert\,B_{4}( x_{i},x_{s})\,\vert^{2}.
\label{nc7a} 
\end{eqnarray}
Here $2d$ is the separation of the double-slits, $ 2a $ is
the width of each slit. The first four cross terms in (\ref{nc7a}) give independent fringes that depend only on $x_s$ or $x_i$, while the last two terms give conditional fringes which depend upon $x_s \pm x_i$. 
 The interesting
fact that appears from this calculation is that the diffraction
terms $B_{j}( x_{i},x_{s})$, $j = 1, 2, 3, 4$, are proportional to
the transverse profile of the pump laser at different transverse positions in the plane of the double slit aperture:  $B_{j}( x_{i},x_{s}) \propto \mathcal{W}(0,z_{A})$ with $j =
2,3$, $B_{1}( x_{i},x_{s}) \propto \mathcal{W}(d,z_{A})$ and $B_{4}(
x_{i},x_{s}) \propto \mathcal{W}(-d,z_{A})$. Generating the photon pairs
using pump beams with different transverse profiles at the double-slit
plane,  the authors in Ref. \cite{fonseca00} were able to
measure two-particle independent fringes, as well as conditional fringes
that depend on the difference or the sum of the detector position
coordinates.  Conditional fringes which depend on the sum of the
position coordinates have been predicted theoretically in
references \cite{horne94,horne98}.
\par
Greenberger, Horne and Zeilinger \cite{ghz93}, analyzing
the two extreme cases discussed above (independent versus
conditional fringes) affirmed: ``... there is a sort of
complementarity between one- and two-particle fringes: the
condition for seeing one precludes the possibility of seeing the
other''. When the source dimension or the distance from the double-slits  to the source is such that the idealized experiment is analyzed in an
intermediate geometry when compared with the 
extreme cases mentioned above, both single-particle fringes and conditional
two-particle fringes are present
\cite{Greenberger92,ghz93,horne98,horne89}. Horne
\cite{horne98} derived a complementarity relation for the one-
($v_{1}$) and the two-particle ($v_{12}$) fringe visibility for a
general state of a two-particle system \cite{jaeger95}:
\begin{eqnarray}
v_{1}^{2} + v_{12}^2 = 1. 
\label{complementarity1}
\end{eqnarray}
The single-particle fringe visibility is derived from the
normalized single-particle probability density, which is obtained  by integrating the
joint probability density with respect to the position of one of the particles. The two-particle visibility is
obtained from the two-particle ``corrected'' probability density
\cite{jaeger93,jaeger95}. This definition prevents the visibility
$v_{12}$ from being unity when particles $1$ and $2$ are prepared in
a product state. Abouraddy and collaborators
\cite{abouraddy01b,saleh00} studied the two-particle double-slit
interferometer for the case where the particles are generated by
means of SPDC. Photon
pairs generated collinearly by the crystal are passed through a
double-slit aperture and are imaged by a $2f$ lens system at the
coincidence detector camera. For this geometry, the authors verified the above complementarity
relation experimentally \cite{abouraddy01b}.  Calculation of $v_{1}$ and $v_{12}$ from
the measured data follows the strategy used by Horne
\cite{horne98}. The single-photon coincidence probability
 associated with the
two-photon pair transmitted by the double-slit is calculated by integrating the fourth-order
correlation functions with respect to one of the position
variables.
\subsection{Spatial antibunching}
\label{subsec:antibunch}     
The experimental observation of the spatial anti-bunching of 
photons consists in the production of a homogeneous optical field for which it is more probable to detect coincident photons at spatially-separated positions in the transverse plane. The original idea of anti-bunching of photons\cite{kimble77} was connected to the time interval between
the emission of two photons by a light source. Spatial anti-bunching means that the probability of finding two photons together at the same position 
in the transverse propagation plane is smaller than the probability of finding them at different positions. For a classical light
beam, the best that one can do is to have equal probability of detecting photons
together or apart. A thermal source produces light that is bunched in time and in space, while 
an ideal laser produces light which is neither bunched nor anti-bunched.
It is also important to note that both bunching and anti-bunching require that the fields be stationary (depend only on $\tau = t_1-t_2$) and homogenous (depend only on $\vect{\delta}=\vect{\rho}_1- \vect{\rho}_2$). If this were not the case, it would be possible to simulate temporal anti-bunching using a low intensity pulsed laser, for instance.
\par
From a mathematical point of view, the inequality
\begin{equation}
\label{eq:antibunch}
\Gamma^{(2,2)}(\vect{\delta},\tau) \leq \Gamma^{(2,2)}(0,0),
\end{equation}
must be satisfied by any classical field.  $\Gamma^{(2,2)}$ is the fourth-order coherence function, which is proportional to the coincidence count probability. 
The violation of this inequality implies anti-bunching if the condition of stationarity and homogeneity is fulfilled. This means that the temporal and spatial arguments of all coherence functions $(t,t+\tau,\vect{\rho},\vect{\rho}+\vect{\delta})$
can be replaced by $(\tau,\vect{\delta})$.
\begin{figure}
\begin{center}
\includegraphics[width=0.4\linewidth]{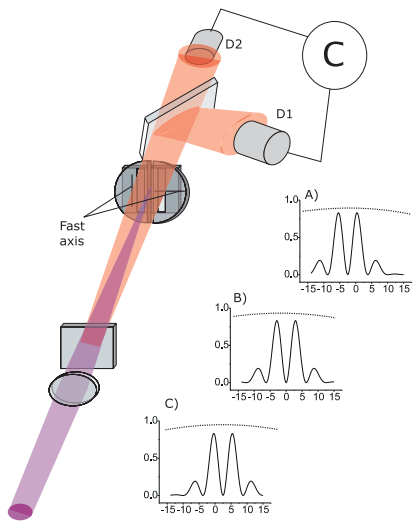}
\caption{Experiment for the observation of the spatial anti-bunching. The vertical axes in the plots are normalized coincidence rates for the solid lines and single count rates for the dashed lines. The horizontal axes are D2
detector positions in arbitrary units. The position of detector D1 is fixed
in A) -2.5, B) 0 and C) +2.5 and D2 is scanned.}
\label{fig:anti}
\end{center}
\end{figure}
The experiment performed in Ref. \cite{nogueira01} is sketched in Fig. \ref{fig:anti}. The non-linear crystal is pumped by a Gaussian profile laser beam and collinear type II down-conversion is produced.  Signal and idler photons are sent to a double-slit aperture equipped with quarter wave-plates (see below). After propagation through the slits, the photons are sent to a 50/50 beam splitter (BS) and single-photon detectors are placed in each output port.  The detectors are carefully calibrated, so that the transverse coordinates in both detection planes are equivalent. 
This arrangement is essentially a two-photon detector.  Detector D1 is kept fixed while D2 is scanned in the transverse plane. For an ordinary double-slit, the resulting coincidence pattern would display the usual interference fringes, equivalent to the intensity fringes in a double-slit experiment. To produce a spatially anti-bunched field, a phase shift of $\pi$ is produced in the coincidence fringes, so that the interference maxima are now minima and vice versa. In this case, the probability of finding two photons together in the same point  ($\vect{\delta}=0$) is smaller than the probability of finding them apart. This is the signature of spatial anti-bunching of photons, provided the condition of homogeneity is fulfilled.
\par
The $\pi$ phase shift is obtained using a double-slit aperture with a zero-order quarter-wave plate placed in front of each slit. The fast axis of one wave plate is oriented along the vertical direction and the fast axis of the other is oriented along the horizontal direction. The pump beam is focused in the plane of the slits, which causes an anti-correlation in the positions of the photons in this plane \cite{monken98a}, so that if one photon crosses one slit, its twin will pass through the other slit. The type-II SPDC produces photons in the polarization state $|H\rangle \otimes |V\rangle$. The position anti-correlation at the plane of the double-slit guarantees that the photons pass through different slits, resulting in the state
 \begin{equation}
|\Psi\rangle_{AB}= \frac{1}{\sqrt{2}}( |\psi_1,H\rangle|\psi_2,V\rangle + e^{i\pi/2} |\psi_2,H\rangle e^{i\pi/2}|\psi_1,V\rangle),
\end{equation}
where $\psi_1$ and $\psi_2$ are the quantum states describing the spatial variables when the photons pass through slit 1 or slit 2, respectively.  The $\pi/2$ phase factors are due to the delay introduced by the zero-order quarter-wave plates. The coincidence probability is \cite{nogueira01,nogueira02}
\begin{equation}
C(x_s,x_i) \propto 1- \cos\left[ \frac{kd}{z}(x_1-x_2) \right],
\end{equation}
where $k$ is the wave number of the down-converted fields, $d$ is the separation between the slits, $z$ is the propagation distance, and $x_1$ and $x_2$ are the positions of detectors D1 and D2.  The coincidence probability is zero when $x_1=x_2$, corresponding to a spatially anti-bunched field.   
 Typical coincidence and intensity measurements observed in Ref. \cite{nogueira01} are shown in the insets of Fig. \ref{fig:anti}. The coincidence distribution is near zero at when both detectors observe the same point, and the single-photon intensity is nearly constant for all positions. The coincidence distributions violate inequality (\ref{eq:antibunch}), as $\Gamma^{(2,2)}(\vect{\delta}) > \Gamma^{(2,2)}(0)$, for many values of $\vect{\delta}$. The measurements are performed for several positions in the transverse planes of both detectors, demonstrating homogeneity. It was also demonstrated that the same effect can be observed without the use of a double-slit, with free propagating signal and idler beams \cite{caetano03}.  Spatial anti-bunching has also been produced using Hong-Ou-Mandel interference of photons in an anti-symmetric polarization state \cite{nogueira04a}.  This arrangement allows for the production of a beam of photons in a ``singlet" polarization state.    

%% file: section7.tex
\section{Spatial Entanglement}
\label{sec:spatialentanglement}
Quantum entanglement is a correlation that can exist between two or more quantum systems.  In the case of pure states, entanglement implies that the quantum state $\ket{\Psi}_{12}$ describing the--say--two systems cannot be separated into a product of quantum states $\ket{\psi}_1\otimes\ket{\phi}_2$.  If the bipartite state can be written in this form it is said to be separable.  In the more general case of mixed states, a separable bipartite state can be written as a convex sum of tensor products of local density matrices \cite{horodecki09,guhne09}:
\begin{equation}
\hat{\varrho}_{\mathrm{sep}}=\sum\limits_{i} p_i \hat{\varrho}_{1i}\otimes  \hat{\varrho}_{2i},
\label{eq:sepstate}
\end{equation}
where $\sum_i p_i =1$ and $p_i \geq 0$.  If a bipartite quantum state cannot be written in the form (\ref{eq:sepstate}), it is entangled.  For a more in depth discussion of quantum entanglement in general, we refer the reader to two recent review articles \cite{horodecki09,guhne09}.    
\par
As discussed in sections \ref{sec:dslitexperiments}, \ref{sec:qimages}, \ref{sec:spatialcorr}, although many experiments performed in the 1990's  strongly suggested that photon pairs produced by SPDC were indeed entangled in the spatial degrees of freedom, it wasn't until 2004 that experimental tests confirmed this fact \cite{howell04,dangelo04}.  
These experiments made use of entanglement criteria involving measurements of the sum and difference of the position and momentum variables \cite{duan00,mancini02}, or EPR criteria involving conditional measurements \cite{reid88,reid89}.  

\subsection{Continuous Variable Entanglement Criteria}
\label{sec:CVentcrit}
There has been much work concerning the detection and characterization of entanglement in continuous variable degrees of freedom.    An overview of the continuous variable formalism, and entanglement criteria for continuous variables, can be found in a number of review articles \cite{braunstein05,adesso07}.   Though the physical system of interest is usually the field quadratures of intense fields, continuous variable entanglement conditions are also applicable to spatial entanglement of photon pairs.       
\par
For the spatial degrees of freedom of photon pairs, it is most useful to consider the global operators  
\begin{subequations}
\begin{equation}
\oper{X}_{\pm} = \oper{x}_1 \pm \oper{x}_2, 
\end{equation}
\begin{equation}
\oper{P}_{\pm} = \oper{p}_1 \pm \oper{p}_2, 
\end{equation}
\end{subequations}
where in the present case $\oper{x}$ and $\oper{p}$ are the position and momentum operators for photons 1 and 2, and we assume that $[\oper{x}_j,\oper{p}_k]=i\delta_{j,k}$ and $j,k=1,2$.  
In the case of spatially entangled photons, the position and momentum observables correspond to measurements in the near and in the far-field respectively, relative to the source plane, namely the SPDC crystal. The near field is associated with a position measurement at the source plane, while the far field is associated with a momentum measurement at the source plane.  
\par
In most cases, entanglement can be identified through violation of one of a number of inequalities, which thus serve as entanglement witnesses.  For instance, 
the Mancini-Giovannetti-Vitali-Tombesi (MGVT)  criteria \cite{mancini02} reads
\begin{equation}
\langle\Delta^2\oper{X}_{\pm}\rangle \langle\Delta^2\oper{P}_{\mp}\rangle \geq 1,   
\label{eq:mancini}
\end{equation}
where $\Delta^2$ stands for the variance, while the separability criteria of Duan, Giedke, Cirac and Zoller (DGCZ) \cite{duan00} is
\begin{equation}
a^2\langle\Delta^2\oper{X}_{\pm}\rangle + \frac{1}{a^2} \langle\Delta^2\oper{P}_{\mp}\rangle \geq 2, 
\label{eq:duan}
\end{equation}
where $a$ is a local scaling parameter which guarantees that the quantities in the sum are dimensionless.  One can optimize condition \eqref{eq:duan} over $a$, in which case one arrives at condition \eqref{eq:mancini} \cite{hyllus06}.  
All separable states of the form \eqref{eq:sepstate} will obey  inequalities \eqref{eq:mancini} and \eqref{eq:duan}, and thus violation of either of them is a sufficient condition for identifying quantum entanglement. 
Inequalities \eqref{eq:mancini}, \eqref{eq:duan} are  examples of a more general class of entanglement witnesses involving second-order moments \cite{giovannetti03}, for which Hyllus and Eisert have provided an optimization procedure \cite{hyllus06}.     
\par
The Simon criterion  \cite{simon00} is a necessary and sufficient condition for $1\times N$ mode Gaussian states \cite{braunstein95}.   To evaluate this criterion, it is necessary to reconstruct the covariance matrix \cite{braunstein05,adesso07}.  Violation of the Simon, MGVT or DGCZ criteria are sufficient for identifying quantum states with negative partial transpose (NPT)\cite{peres96,horodecki96,nha08}.  In the general case, violation of these criteria is not a necessary condition for entanglement even in the case of Gaussian states, due to the existence of bound entangled states which have a positive partial transpose.    
 \par
For non-Gaussian states, these second-order criteria are sufficient, but not necessary, and there exist NPT entangled states which do not violate any second-order criteria.  In this case one can use the NPT criteria of Shchukin and Vogel, which provides a hierarchy of inequalities involving combinations of second and higher-order moments \cite{shchukin05}.  In fact, they have shown that inequalities \eqref{eq:mancini}, \eqref{eq:duan} and the Simon criterion are special cases of this general NPT criteria. Violation of any inequality in the Shchukin-Vogel hierarchy indicates that the quantum state is NPT, and is thus a sufficient condition for entanglement.  Higher-order inequalities obtained from the Shchukin-Vogel criterion have been violated recently for a spatially non-Gaussian two-photon state which does not violate any second-order criteria \cite{gomes09b}.  We note that there exist several other CV entanglement criteria involving higher order moments \cite{agarwal05,hillery05,hillery06b,chen07,rodo08,walborn09,miranowicz09,hillery09,sperling09b,sperling09,adesso09}.  
  
\subsection{Correlations in the near and far field}
\par
The CV entanglement criteria discussed in section \ref{sec:CVentcrit} can be easily applied to the spatial correlations of SPDC photon pairs.  In this case, the position variable $x$ corresponds to the transverse position $\rho$ and the momentum variable $p$ is related to the transverse wave vector via $p= \hbar q$. The position and wave vector distributions can be observed through intensity measurements in the near and far field of the non-linear crystal, respectively.  
Using simple optical systems composed of lenses and free space, it is possible to measure the spatial distributions of the down-converted photons in the near and far field.  In the near field, one generally observes a strong correlation between the positions of the signal and idler photons.  An intuitive way to understand this correlation is through the fact that the two down-converted photons are ``born" from the same pump photon simultaneously. Thus, the photons are produced at approximately the same transverse position inside the crystal. If the crystal were infinitesimally thin in the longitudinal direction, this would produce a point-like correlation between their positions in the crystal plane.  
\par
  The field operator which describes detection in the near-field is given by 
\begin{equation}
\oper{E}_{\mathrm{nf}}(\vect{\rho}) \propto \int d \vect{\rho}\, \oper{a}(\vect{q}) e^{-i\vect{q}\vect{\rho}}.  
\label{eq:Enf}
\end{equation} 
Assuming that the quantum state of SPDC photon pairs is given by Eq. \eqref{eq:quantumstate}, the near-field correlations are governed by the profile of the pump beam $\mathcal{W}(\vect{\rho})$ and the Fourier transform of the phase matching function: $\Gamma(\vect{\rho})=\mathcal{F}\{\gamma(\vect{q})\}$. Typically, $\Gamma(\vect{\rho})$ is much narrower than $\mathcal{W}(\vect{\rho})$ at the crystal face.  The detection amplitude is defined as $\Psi_{\mathrm{nf}}(\vect{\rho}_s,\vect{\rho}_i) = \langle 0 | \vect{E}^{+}(\vect{\rho}_i)\vect{E}^{+}(\vect{\rho}_s)|\Psi\rangle$ and its square modulus is proportional to the coincidence rate.   

If $\mathcal{W}(\vect{\rho})$ is approximately constant in the region where $\Gamma(\vect{\rho})$ varies appreciably, then the near-field detection amplitude is given by 
\begin{equation}
\Psi_{\mathrm{nf}}(\vect{\rho}_s,\vect{\rho}_i) \propto \Gamma(\vect{\rho}_s-\vect{\rho}_i).
\label{eq:Psinf}
\end{equation} 
$\Gamma(\vect{\rho})$ is generally a narrow function peaked at $\vect{\rho}=0$, which guarantees that the positions of the down-converted photons are correlated:  $\vect{\rho}_i = \vect{\rho}_s$.  We note that this is a much simplified picture of near-field correlations.  It has very recently been shown that the near-field distribution of the biphoton may present a complex structure due to fourth-order interference of photon pairs that originate from different transverse planes of the crystal, and is strongly dependent upon the collinear phase mismatch \cite{pires09b,pires09c}.   
\par
In the far-field region, the down-converted photons exhibit anti-correlation.  This is understood intuitively via momentum conservation in the SPDC process.  If the pump beam is centered around $\vect{q}=0$, momentum conservation guarantees that the down-converted photons are produced such that $\vect{q}_s \approx -\vect{q}_i$.  Since the transverse position in the far-field is associated with the momentum distribution at the crystal, this translates into anti-correlation in the detection positions in the far-field.    
Equivalently, if $\mathcal{W}(\vect{\rho})$ is much larger than $\Gamma(\vect{\rho})$, then this implies that for the Fourier transforms of these functions the inverse is true: $v(\vect{q})$ is much narrower  than $\gamma(\vect{q})$. The field operator corresponding to detection in the far-field as 
\begin{equation}
\oper{E}_{\mathrm{ff}}(\vect{\rho}) \propto \oper{a}\left(\frac{\vect{\rho}}{b}\right),  
\label{eq:Eff}
\end{equation} 
so that the detection amplitude can be approximated by
\begin{equation}
\Psi_{\mathrm{ff}}(\vect{\rho}_s,\vect{\rho}_i) \propto v\left(\frac{\vect{\rho}_s}{b_s}+\frac{\vect{\rho}_i}{b_i} \right), 
\label{eq:Psiff}
\end{equation} 
where $b_s$ and $b_i$ are scaling factors with dimension of length$^{-2}$, as the argument of the function $v$ has units of a transverse wave vector (spatial frequency).  Equation \eqref{eq:Psiff} shows that position measurements in the far-field are associated with the momentum profile $v(\vect{q})$ in the near-field.  Thus, through position measurements in the near and far-field, one can test the entanglement criteria \eqref{eq:mancini} or \eqref{eq:duan}.  For example, in this case the MGVT criterion \eqref{eq:mancini} gives
\begin{equation}
\langle\Delta^2\oper{X}_{-}\rangle \langle\Delta^2\oper{P}_{+}\rangle = \Delta^2 v  \Delta^2 \Gamma  \geq 1, 
\end{equation}  
where $\Delta^2 v $ and  $\Delta^2 \Gamma $ are the variances of the functions $v$ and $\Gamma$, respectively.  For a Gaussian pump beam with width $w$ at the crystal face, $\Delta^2 v = 1/w^2$, and $\Delta^2 \Gamma \approx \lambda_p L/(2 \pi)$ \cite{chan07}, where $L$ is the crystal length and $\lambda_p$ is the wavelength of the pump beam.  Since $v$ and $\Gamma$ are independent functions, they need not respect the limit on the right-hand side.  In fact, for a $\lambda = 400$nm Gaussian pump beam with width $w=1$mm, and $L=1$mm crystal, the left-hand side is on the order of $10^{-4}$.    
\par
 The strong spatial correlation depends explicitly on the widths of the pump beam and phase matching function. For long crystals and strongly focused pump beams, this correlation can be decreased so that the two-photon state is almost separable \cite{law04,walborn07c}.  In this regime the shape and form of the pump beam and phase matching function become very important.  In the approximation where both functions are described by Gaussians, the two-photon state is separable when $\Delta^2 v  = \Delta^2 \Gamma $.
 \par
 \begin{figure}
\begin{center}
\includegraphics[width=7cm]{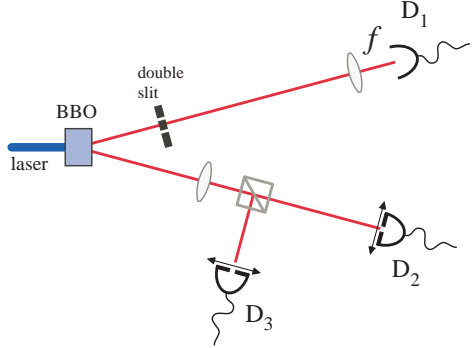}
\caption{Experimental setup for test of spatial entanglement performed by D'angelo et al. \cite{dangelo04}. Measurements in the near field are performed by scanning detector D$_2$ in the image plane of an imaging lens system.   Far-field measurements are realized by scanning detector D$_3$ in the Fourier plane of the lens.  Detector D$_1$ is a bucket detector, placed in the focal plane of a lens.}
\label{fig:dangelo}
\end{center}
\end{figure}
A condition  similar to inequalities \eqref{eq:mancini} and \eqref{eq:duan} was  tested by D'Angelo {\it et al.} for down-converted photons using a ghost imaging setup \cite{dangelo04}, as illustrated in figure \ref{fig:dangelo}.  Here a double slit was placed in the path of the signal photon, which is detected by a stationary ``bucket" detector.  The idler photon is detected by scanning a point-like detector in either the near-field or far-field, resulting in a ghost imaging or ghost interference scenario.    The non-classicality conditions tested were 
 \begin{align}
\label{eq:dangelo}
 \Delta^2(x_1-x_2) & < \min(\Delta^2 x_1,\Delta^2 x_2); \nonumber \\
  \Delta^2(p_1+p_2) & < \min(\Delta^2 p_1,\Delta^2 p_2). 
 \end{align} 
 Violation of both inequalities by the same quantum state implies entanglement.  In the experiment, $\Delta^2(p_1+p_2)$ was evaluated through the visibility of the ghost interference setup, while  $ \Delta^2(x_1-x_2)$ is determined through the ghost imaging of the double slit.  By further considering the divergence of the down-converted beams and the localization of photons by the double slit, both inequalities in \ref{eq:dangelo} are violated.  
          
\subsection{Einstein-Podolsky-Rosen Non-locality}
\label{sec:EPR}
In 1935, Einstein, Podolsky and Rosen (EPR) began the ongoing discussion and study of quantum entanglement in their seminal paper ``Can Quantum-Mechanical Description of Physical Reality Be Considered Complete?".  EPR argued that quantum mechanics was inconsistent with seemingly reasonable notions of locality and ``elements of reality".  Through analysis of a gedanken experiment involving  two particles with perfectly correlated position and momentum, they argued that one should be able to attribute well-defined values to complementary physical quantities of one of the particles.  This conflicted with the predictions of  quantum mechanics, in particular with Bohr's complementarity principle, quantified by the Heisenberg uncertainty relation.  EPR then argued that quantum mechanics must be incomplete.  The EPR paradox illustrates the incompatibility between what is now known as \emph{local realism} and the standard quantum theory.   
 \par   
 Though EPR considered perfect position and momentum correlations, it is possible to apply the EPR argument to a more realistic setting.  Reid and collaborators have shown that EPR-like correlations can be identified by 
the violation of the inequality \cite{reid88,reid89,reid08}
\begin{equation}\label{eq:reid}
\Delta^2_{\mathrm{min}}(x_{1}{|x_{2}})\Delta^2_{\mathrm{min}}
(p_{1}|p_{2})>\frac{1}{4}.  
\end{equation}
 Here $\Delta^2_{\mathrm{min}}(r_{1}|r_{2})$ is the minimum inferred variance, which represents the minimum uncertainty in inferring variable $r_1$ of system 1 conditioned upon measurement of variable $r_2$ of system 2.  
 Explicitly, 
 \begin{equation}
 \Delta^2_{\mathrm{min}}(r_{1}|r_{2}) = \int d\xi_2 \mathcal{P}(\xi_2) \Delta^2(r_1|\xi_2),
 \end{equation}
 where $\Delta^2(r_1|\xi_2)$ is the variance of the conditional probability distribution $\mathcal{P}(r_1|\xi_2)$, which gives the probability of $r_1$ given that the measurement of system 2 gave \emph{result} $\xi_2$, and $\mathcal{P}(\xi_2)$ is the probability that result $\xi_2$ is obtained.      
Inequality \eqref{eq:reid} was first violated with quadrature measurements of two intense beams \cite{ou92}.  
  \par
\begin{figure}
\begin{center}
\includegraphics[width=12cm]{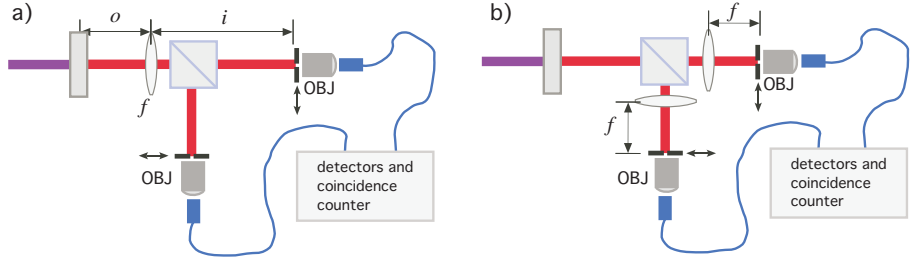}
\caption{Experimental setup for test of the EPR criteria performed by Howell et al. \cite{howell04}. a) Position measurement are performed by scanning a slit aperture in the image plane of an imaging lens system.  b) Momentum measurements are realized by scanning the slit in the Fourier plane of a lens.}
\label{fig:howell}
\end{center}
\end{figure}

   
  Using SPDC, it is possible to produce a two-photon quantum state that is similar to the original EPR state.  The SPDC state (see section \ref{sec:fundamentals}) in the limit $w\longrightarrow \infty$ and $L \longrightarrow 0$ becomes 
\begin{equation}
\ket{\psi}_{\mathrm{EPR}} = \int d\vect{q} \ket{\vect{q}}_1 \ket{-\vect{q}}_2 = \int d\vect{\rho} \ket{\vect{\rho}}_1 \ket{\vect{\rho}}_2,
\label{eq:eprstate}
\end{equation}
where the equality on the right side is obtained by Fourier transform.   
Inequality \eqref{eq:reid} for spatial correlations was first tested experimentally by Howell {\it et al.} \cite{howell04}, using the experimental setup illustrated in figure \ref{fig:howell}.  The near ($x$) and far-field ($p$) distributions were measured using an imaging lens system and a Fourier transform lens system, respectively.  A value of $\Delta^2_{\mathrm{min}}(x_{1}{|x_{2}})\Delta^2_{\mathrm{min}}(p_{1}{|p_{2}})= 0.01  << 0.25$ was obtained, demonstrating the high correlation of the spatial degrees of freedom of the photon pairs.  The separability condition \eqref{eq:mancini} was also tested and violated.  
\par
 A thorough discussion of the EPR paradox, including the experimental results to date can be found in a recent review article \cite{reid08}.
 As inequality \eqref{eq:reid} deals with EPR non-locality, it is generally more restrictive than those which identify non-separability of continuous variable systems \cite{duan00,mancini02}.  Recently, it has been shown that violation of the EPR inequality  \eqref{eq:reid} is sufficient to discard ``local hidden state" models, which have been related to Schr\"odinger's ``steering" phenomenon \cite{wiseman07, jones07, cavalcanti09}.  An EPR inequality based on an entropic uncertainty relation was introduced and tested experimentally in Ref. \cite{walbornEPR}.    

\subsection{Propagation of spatial entanglement}
The above experiments show quantum entanglement and EPR non-locality by identifying correlations in the near and far field of the SPDC crystal.  In general, after free propagation, the transverse spatial correlations switch from a correlation in the near-field, given by  equation \eqref{eq:Psinf} to an anti-correlation in the far-field, given by equation \eqref{eq:Psiff}.  Chan {\it et al.} have shown that this change from correlation to anti-correlation implies that at some intermediate plane there must be very little correlation present in the spatial intensity distributions \cite{chan07}.   Following Chan {\it et al.}, let us consider the case where the cardinal sine function in the momentum configuration is approximated with a Gaussian:  $\sinc(b q^2)\propto \exp(-\alpha b q^2)$, with $\alpha =0.455$($\alpha =0.455$ optimizes the approximation), and the pump is a Gaussian beam. After propagation, the wave function in $q$-space is   
\begin{eqnarray}
\Phi(\vect{q}_{s},\vect{q}_{i},z) = C \exp\biggr[ -\frac{a_+ + ib_+(z)}{4}(\vect{q}_s + \vect{q}_i)^2 - \\ \nonumber \frac{a_- + ib_-(z)}{4} (\vect{q}_s - \vect{q}_i)^2 \biggr],
\label{eq:psiq}
\end{eqnarray} 
where $C$ is a normalization constant and $a_+ = w_o^2/(1+z_R^2/R^2)$, $a_- = \alpha L/ K$, $b_+(z) = 2(z+L)/K + 2R/K(1+z_R^2/R^2)$, and $b_-(z)=(2z+L)/K$.  Here $R$ and $z_R$ are the radius of curvature  and Rayleigh range of the Gaussian pump beam in the source plane, respectively. The $\rho$-space wave-function is the Fourier transform of \eqref{eq:psiq}, and is given by  
 \begin{eqnarray}
\Psi(\vect{\rho}_{s},\vect{\rho}_{i},z) = D \exp \biggr [ -\frac{a_+ - ib_+(z)}{4(a_+^2 + b_+^2(z))}(\vect{\rho}_s + 
\vect{\rho}_i)^2 - \\ \nonumber  \frac{a_- - ib_-(z)}{4(a_-^2 + b_-^2(z))} (\vect{\rho}_s - \vect{\rho}_i)^2 \biggr].  
\label{eq:psirho}
\end{eqnarray}   
One can see that both wave functions \eqref{eq:psiq} and \eqref{eq:psirho} are completely separable (un-entangled) when $a_+=a_-$ and $b_+(z)=b_-(z)$, and thus $\Psi_{\mathrm{sep}}(\vect{\rho}_{s},\vect{\rho}_{i},z)=\psi_s(\vect{\rho}_s,z)\psi_i(\vect{\rho}_i,z)$.  This condition depends upon the initial parameters of the pump beam, as well as the length of the non-linear crystal, and cannot be met by propagation alone.     
However, at a certain distance $z=z_0$ from the non-linear crystal, such that  $a_+[a_-^2+b_-^2(z_0)]=a_-[a_+^2+b_+^2(z_0)]$, the modulus squared of the wave function is separable:  
\begin{equation}
|\Psi(\vect{\rho}_{s},\vect{\rho}_{i},z_{0})|^2=|\psi_s(\vect{\rho}_s,z_0)|^2 |\psi_i(\vect{\rho}_i,z_0)|^2.
\label{eq:psiintsep}
\end{equation}     
Thus, in the transverse plane located a distance $z=z_0$ from the crystal, there is no correlation present in the intensity distribution.  Since at $z_0$ we have $b_+(z_0)(a_-^2+b_-^2(z_0)) \neq b_-(z_0)(a_+^2+b_+^2(z_0))$, $\Psi(\vect{\rho}_{s},\vect{\rho}_{i},z_{0})$ is not separable, and we conclude all initial spatial correlations have ``migrated" to the phase of the two-photon state \cite{chan07}.  Thus, to properly identify the entanglement present at $z=z_0$, some type of interferometric measurement capable of accessing phase information must be performed \cite{chan07}. 
\par
In the above arguments, the phase matching function $\gamma(\vect{q})$ is approximated by a Gaussian function, which is necessary for the separability of the modulus of the wave function \eqref{eq:psiintsep}.  When considering the phase matching function in the form of a sinc function(in the momentum space), the correlations in principle never completely disappear.  Nevertheless, one observes a strong decrease in spatial correlations at $z=z_0$.         
\par
Since the correlations present in the two-photon state migrate from the amplitude to the phase due to local unitary evolution in the form of free propagation, it should be possible to retrieve the correlations by undoing the unitary evolution.  In this vein, it is convenient to parametrize the evolution in a more general framework.   

\subsection{Fractional Fourier Transform}
\label{sec:frft}
\begin{figure}
\begin{center}
\includegraphics[width=10cm]{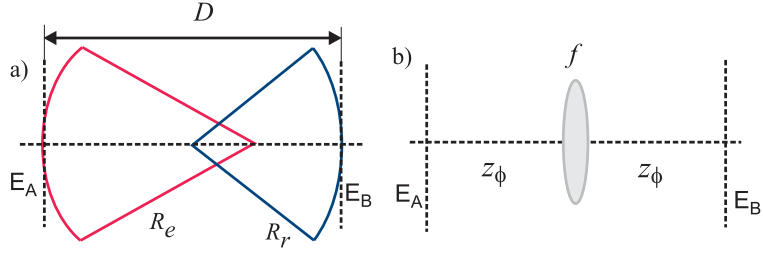}
\caption{a) Free space propagation as an FRFT.  b) Optical lens system used to implement an FRFT.}
\label{fig:FRFT}
\end{center}
\end{figure}
The analysis of the propagation of spatial correlations can be cast in a general setting using the concept of the fractional Fourier transform (FRFT) \cite{mendlovich93,pellat-finet94,lohmann95,ozaktas01}.  Since its first appearance in 1929 \cite{wiener29}, the FRFT has found widespread use in quantum mechanics \cite{namias80,mcbride87,chountasis99} as well as signal processing and optics \cite{ozaktas01}. 
The FRFT of order $\phi$ of a function $E(\vect{\zeta})$ can be defined by the integral transform \cite{pellat-finet94}
 \begin{align}
\mathcal{F}_{\phi}\left[E\right](\vect{\xi}) = &   \iint F_{\phi}(\vect{\xi},\vect{\zeta})
  E(\vect{\zeta}) d \vect{\zeta},
  \label{eq:FRFTmath}
\end{align}  
where the FRFT kernel is given by
 \begin{equation}
F_{\phi}(\vect{\xi},\vect{\zeta}) =  \frac{i \exp{(-i\phi)}}{2 \pi \sin \phi}   \exp\left({-i\frac{ \xi^2 \cot \phi}{2}} 
{-i \frac{\zeta^2 \cot \phi}{2}}  
  + {i\frac{\vect{\zeta}\cdot\vect{\xi}}{\sin \phi}} \right),
  \label{eq:FRFTmathkernel}
\end{equation}  
and $\vect{\zeta}$ and $\vect{\xi}$ are two-dimensional variables.  Note that $\phi=\pi/2$ corresponds to the usual Fourier transform.  Furthermore, in the limit $\phi \longrightarrow 0$,   $ F_{0} \longrightarrow \delta(\vect{\zeta}-\vect{\xi})$, and when $\phi \longrightarrow \pi$ the kernel $F_{\pi} \longrightarrow \delta(\vect{\zeta}+\vect{\xi})$.  Just as the Fourier transform appears naturally in the context of Fraunhofer diffraction, it has been shown that the FRFT appears in the Fresnel diffraction regime \cite{pellat-finet94}.  To identify ordinary free space propagation with the fractional Fourier transform it is necessary to choose properly scaled coordinates \cite{pellat-finet94}.  Consider the usual Fresnel diffraction integral corresponding to the propagation of a field from the plane $A$ to plane $B$, as shown in Fig. \ref{fig:FRFT} a).  Choosing dimensionless coordinates $\vect{\zeta} = \sqrt{k \tan \phi /D}\vect{\rho}$ and $\vect{\xi} = \sqrt{k \sin \phi \cos \phi /D}\vect{\rho}^\prime$, where $D$ the distance between planes, leads to     
\begin{equation}
E_B(\vect{\xi}) = \exp(i \phi) \cos \phi \exp(-i \xi^2 \tan \phi /2)\mathcal{F}_{\phi}\left[E_A\right](\vect{\xi}),  
\end{equation}
which expresses $E_B$ as a FRFT of $E_A$ multiplied by a phase term \cite{pellat-finet94}.  It is possible to eliminate the phase term if one observes the output field on a spherical surface of radius $R_\phi = -D/ \sin^2 \phi$, as shown in Fig. \ref{fig:FRFT} a).  This spherical surface can also be mapped to a plane using a lens with focal length equal to $-R_\phi$.  
\par  
It has also been shown that one can implement a FRFT using lenses \cite{lohmann93,lohmann95}.  For example, consider the symmetrical optical system shown in Fig. \ref{fig:FRFT} b).  Here $l$ is the focal length of the lens and $z$ is the propagation distance before and after the lens.  Using $z=2 l \sin^2(\phi/2)$ and $f=l\sin\phi$, it has been shown that this optical system implements a FRFT operation given by 
\begin{equation}
\mathcal{F}_{\phi}\left[E\right](\vect{\rho}^\prime)  =  
   \iint F_{\phi} \left (\frac{k}{f}\vect{\rho}^\prime,\frac{k}{f}\vect{\rho} \right)
  E(\vect{\rho}) d \vect{\rho},
  \label{eq:FRFT}
\end{equation}    
which is equivalent to the transform given in Eq. \eqref{eq:FRFTmath} when one chooses scaled variables $\vect{\zeta} = \sqrt{k/f}\vect{\rho}$ and $\vect{\xi} = \sqrt{k/f}\vect{\rho}^\prime$.  
\par
A generic optical system consisting of  lenses, mirrors and sections of free space can be described in terms of FRFTs by scaling the transverse spatial coordinates appropriately.    
The convenience of the FRFT comes from the fact that it can be viewed as a simple rotation in phase space.  This can been seen most easily using geometrical optics.  Consider the symmetrical lens system shown in Fig. \ref{fig:FRFT} b).  Free propagation of an optical ray $\vect{r}=(r,\theta)$ can be represented by the ABCD matrix \cite{fowles89}
\begin{equation}
{S}_z = \left(\begin{array}{cc}
1 & z \\
0  & 1
\end{array}\right),
\label{eq:Smatrix}
\end{equation}
where $z$ is the propagation distance.
Passage through a thin lens is described by the matrix 
\begin{equation}
{L}_l = \left(\begin{array}{cc}
1 & 0 \\
-1/l  & 1
\end{array}\right),
\label{eq:Lmatrix}
\end{equation}
where $l$ is the focal length of the lens.
Choosing $z=2l\sin^2(\phi/2)$ and defining $f=l\sin\phi$ as a scaled focal length,  the complete optical FRFT system is given by the matrix  
\begin{equation}
{F}_{\phi} = {S}_z{L}_l{S}_z=\left(\begin{array}{cc}
\cos\phi & f \sin\phi  \\
-\frac{1}{f} \sin\phi & \cos\phi
\end{array}\right). 
\label{eq:FRFTmatrix}
\end{equation}
Matrix
(\ref{eq:FRFTmatrix}) represents a $\phi$-order FRFT, and is recognized as a rotation matrix scaled
by $f$. This scaling is necessary since $r$ has dimension of length and 
 $\theta$ is adimensional, and thus the scaled focal
length $f$ acts on $\theta$ so that $f\theta$ has dimension of length.
If the scaled focal lengths $f$ of two FRFTs are equal, it
is easy to see from the ray matrix in Eq. (\ref{eq:FRFTmatrix})
that the FRFT is additive, that is
${F}_{\phi_1}{F}_{\phi_2}={F}_{\phi_1+\phi_2}$,
such that the order of the combined optical FRFT is
$\phi_1+\phi_2$.  By introducing the FRFT with properly scaled coordinates, and using the additivity property, one can describe propagation through any first-order optical system consisting of lenses, mirrors and free space.  Propagation is now parametrized solely by the order of the overall FRFT, given by the angle $\phi$.  

\par
In quantum-mechanical operator formalism, the FRFT operator is defined as \cite{ozaktas01}
\begin{equation}
\label{EQ:FRFToperator}
\oper{{F}}_{\phi} \equiv e^{i \phi/2}
\exp{\left [ -i \frac{\phi}{2}(\oper{x}^2+\oper{p}^2) \right ]}\;\;,
\end{equation}
where $\oper{x}$ and $\oper{p}$ are the dimensionless position and momentum operators satisfying $[\oper{x},\oper{p}]=i$.  
This operator is equivalent to the evolution operator of the quantum harmonic oscillator, which has the hamiltonian  $\oper{H}={(\oper{x}^2+\oper{p}^2)}/{2}$.  
Application of the FRFT operator $\oper{{F}}_{\phi}$  to the $\oper{x}$ and $\oper{p}$ corresponds to rotation of angle $\phi$ in phase space \cite{lohmann93,ozaktas01}:  
\begin{equation}
\label{eq:RhoqEvolved}
\oper{{F}}^{\dagger}_{\phi}
   \left(\begin{array}{c}
 \oper{x}\\
  \oper{p}
  \end{array}\right)
  \oper{{F}}_{\phi}
  =
  \left(\begin{array}{cc}
  \cos\phi& \sin\phi\\
  -\sin\phi & \cos\phi
  \end{array}\right)
  \left(\begin{array}{c}
 \oper{x}\\
  \oper{p}
  \end{array}\right)
  \equiv 
   \left(  \begin{array}{c}
    \oper{x}_{\phi}\\
  \oper{p}_{\phi}
  \end{array}\right),
\end{equation}
where $ \oper{x}_{\phi}$ and $\oper{p}_{\phi}$ are the rotated operators.  
\par
The FRFT is a rotation in the $x$,$p$ phase space.  In this respect, it can be used to measure the marginal probability distribution along any axis in phase space, and to perform tomography of the spatial Wigner function \cite{mcalister95,kang10}.  We note that spatial tomography of a field can also be performed  using interference of rotated and displaced halves of the field \cite{mukamel03,smith05}. 

\subsection{Correlations at Intermediate Planes}
 To describe the spatial correlations of photon pairs at some propagation distances $z_s$ and $z_i$, through arbitrary first-order optical systems, it suffices to consider FRFT's of arbitrary order angle $\alpha$ and $\beta$.
 A two-photon state
$\ket{\Psi}$ after arbitrary propagation is given by
$
\ket{\Psi_{\alpha,\beta}}=\oper{{F}}^{(1)}_{\alpha} \otimes
\oper{{F}}^{(2)}_{\beta}\ket{\Psi}
$.  The two-photon wave function then becomes  $\Psi_{\alpha,\beta}(x_1,x_2)=\langle x_1,x_2|\Psi_{\alpha,\beta}\rangle $, where  
 \begin{align}
\Psi_{\alpha,\beta}(x_1,x_2) 
& =  \iint dx_1^{\prime}dx_2^{\prime}
 \bra{x_1}\oper{F}_{\alpha}\ket{x_1^{\prime}}  \bra{x_2}\oper{F}_{\alpha}\ket{x_2^{\prime}}\Psi(x_1^\prime,x_2^\prime) \\ \nonumber 
& =  \iint dx_1^{\prime}dx_2^{\prime}
 {{F}}_{\alpha}(x_1,{x_1^{\prime}}) {{F}}_{\beta}(x_2,{x_2^{\prime}})
\Psi(x_1^\prime,x_2^\prime),
 \end{align}  
and the kernels are defined in Eq. \eqref{eq:FRFTmathkernel}.   For simplicity, we consider only one spatial dimension. As an example, let us consider the propagation of the EPR state in one dimension  
\begin{equation}
\ket{\Psi^{\mathrm{EPR}}} = \iint dx_1 dx_2 \delta(x_1-x_2) \ket{x_1}_1\ket{x_2}_2,   
\label{eq:EPR}
\end{equation} 
which presents a perfect correlation, since detection of photon $2$ at position $x$ projects photon $1$ onto a position eigenstate $\ket{x}$.  This situation is approximated by the state produced by SPDC when the pump beam can be treated as a plane wave.   This state evolves to \cite{tasca09}   
\begin{align}\label{eq:EPRevolved}
  \ket{\Psi_{\alpha,\beta}^{\mathrm{EPR}}}= A_{\alpha+\beta} &\iint dx_1 dx_2 \exp\left[ i\frac{\cot(\alpha+\beta)}{2}\left(\rho_1^2 +\rho_2^2\right) \right ] \times \nonumber \\ 
  & \exp \left[-i \frac{x_1\cdot x_2}{\sin(\alpha+\beta)} \right] \ket{x_1}_1\ket{x_2}_2.   
\end{align}
\par
We can now analyze the type of correlations present.  Whenever $\alpha+\beta = 0 \;(\mmod 2\pi)$, the original state \eqref{eq:EPR} is recovered.  That is, the EPR state \eqref{eq:EPR} is an eigenstate of operators of the type $\oper{{F}_{\alpha}}\oper{{F}}_{2\pi-\alpha}$, $\oper{{F}_{\alpha}}\oper{{F}}_{4\pi-\alpha}$, etc.   
When  $\alpha+\beta  = \pi \;(\mmod 2\pi)$, the correlated EPR state \eqref{eq:EPR} evolves to an anticorrelated EPR state
 \begin{equation}
\ket{\Phi^{\mathrm{EPR}}} = \iint dx_1 dx_2 \delta(x_1+x_2) \ket{x_1}_1\ket{x_2}_2.   
\label{eq:EPRanti}
\end{equation} 
In this case the detection of photon $2$ at $x$ projects photon $1$ onto the state $\ket{-x}$.   When  $\alpha+\beta  = \pi/2 \;(\mmod 2\pi)$, this state becomes  
 \begin{equation}
\ket{\Omega} = \int dx  \ket{x}_1\ket{p(x)}_2,   
\label{eq:EPR2}
\end{equation} 
which presents no intensity correlation.  Here $\ket{p(x)}\propto \int \exp(i p(x)\cdot x)\ket{x}$ is the momentum eigenstate conjugate to $\ket{x}$.   An equivalent result is found for $\alpha+\beta  = 3\pi/2 \;(\mmod 2\pi)$.  We note that the conditions for correlation, anti-correlation, and no-correlation depend on the sum of the FRFT angles of the down-converted fields, and not the individual angles $\alpha$ and $\beta$.  Thus, for any propagation characterized by an FRFT $\mathcal{F}_{\alpha}$ on photon 1, one can find a suitable transformation $\mathcal{F}_{\beta}$ on photon 2 such that a correlation, anti-correlation or no intensity correlation is recovered.  These conditions were tested experimentally in Ref. \cite{tasca09}.      
\par
This simple picture drawn for the ideal EPR-state is followed approximately by the two-photon state in Eq. (\ref{eq:psirho}).  Let us consider two simple cases.  First, let us assume that the FRFT angles are the same for both down-converted photons:  $\alpha=\beta$.  Then, no intensity correlation is observed whenever 
\begin{equation}
\cot^2 \alpha = \frac{a_-(a_+^2+b_+^2)-a_+(a_-^2+b_-^2)}{a_+-a_-},
\end{equation}  
where parameters $a_\pm$ and $b_\pm$ are defined in (\ref{eq:psiq}) and $b_\pm$ refers to $b_\pm(z)$ at the crystal plane.   
Consider another simple case in which  $b_+=b_-=0$ at the crystal face, with no restriction on $\alpha$ and $\beta$.  
Analytical calculation shows that, in order to have no intensity correlation,
 \begin{equation}
\cot\alpha\cot\beta =a_{-}a_{+}. 
\label{eq:no-intensity-correlation-condition}
\end{equation} 
Eq. \eqref{eq:no-intensity-correlation-condition} is satisfied by FRFT orders such that $\alpha+\beta=\pi/2 \;(\mmod 2\pi)$ or $\alpha+\beta=3\pi/2 \;(\mmod 2\pi)$
only when $a_{-}=1/a_{+}$. 
Nevertheless, the intensity correlations present in the state Eq. (\ref{eq:psirho}) 
propagate in a fashion similar to idealized case of the EPR state.  
\par 
The above arguments apply to intensity correlations, the absence of which does not necessarily imply that a quantum state is separable.  Let us now discuss the identification of entanglement via intensity correlations of transverse spatial variables.  The DGCZ
inequality \eqref{eq:duan} for transverse variables rotated using the FRFT operator (\ref{eq:RhoqEvolved}) is  \cite{tasca08}:
\begin{eqnarray}\label{EQ:VariancesPRIME}
    \langle (\Delta\oper{X}^{\prime }_-)^2 \rangle + \langle
(\Delta\oper{P}^{\prime}_+)^2 \rangle = 
&\frac{1+ \cos(\alpha+ \beta)}{2} \left[
\langle(\Delta\oper{X}_-)^2 \rangle + \langle
(\Delta\oper{P}_+)^2 \rangle \right]  + \nonumber \\
&\frac{1- \cos(\alpha + \beta)}{2} \left[
\langle(\Delta\oper{X}_+)^2 \rangle + \langle
(\Delta\oper{P}_-)^2 \rangle \right]  - \nonumber  \\ &\frac{\sin(\alpha + \beta)}{2}\left[ \langle\{ \oper{X}_+,
\oper{P}_+ \}\rangle - 2\langle \oper{X}_+\rangle \langle
\oper{P}_+\rangle  \right] + \nonumber \\ &\frac{\sin(\alpha + \beta)}{2}\left[ \langle \{\oper{X}_-
,\oper{P}_-\} \rangle - 2\langle \oper{X}_-\rangle \langle
\oper{P}_-\rangle \right], 
\end{eqnarray}
where  
$\oper{X}^{\prime}_- \equiv \oper{x}_{\alpha} -
\oper{x}_{\beta}$ and $\oper{P}^{\prime}_+ \equiv
\oper{p}_{\alpha} + \oper{p}_{\beta}$,
are defined in terms of the rotated variables, and  $\oper{X}_- = \oper{x}_1 - \oper{x}_2$
and $\oper{P}_+ = \oper{p}_1 + \oper{p}_2$ in terms of the variables at the source. The sum of variances for the
rotated variables coincides with the sum of variances for the
variables at the source when $\alpha +
\beta (\mbox{mod}\,2\pi) = 0$.  If the two-photon state is entangled, with intensity correlation in the $\oper{x}$ and $\oper{p}$ variables at some initial plane, then, for any propagation of the signal photon,  characterized by $\alpha$, it is possible to
find a propagation of the idler field, characterized by $\beta$, so that an intensity correlation is recovered and entanglement can be identified. It is important to note that
Eq. (\ref{EQ:VariancesPRIME}) does not depend on the state
and is applicable to any
bipartite continuous variable systems.  Eq. (\ref{EQ:VariancesPRIME}) was experimentally tested for intermediate propagation planes of SPDC photons in Ref. \cite{tasca08}. 

%% file: section8.tex
\section{Transverse Modes}
\label{sec:transversemodes}
\begin{figure}
\begin{center}
\includegraphics[width=12cm]{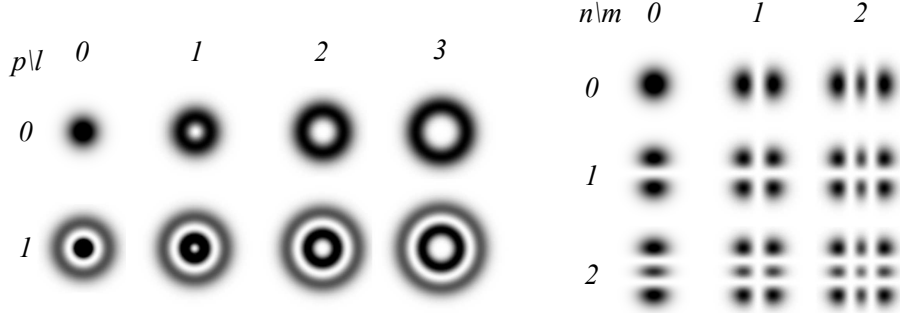}
\caption{Laguerre-Gaussian and Hermite-Gaussian modes.}
\label{fig:modes}
\end{center}
\end{figure}
A simple way to describe spatial correlations in the paraxial regime is through the use of transverse spatial modes described by discrete indices, such as the Laguerre Gaussian (LG) modes or  the Hermite Gaussian (HG) modes.  Like the well known Gaussian beam, the Hermite-Gaussian and Laguerre-Gaussian beams are also solutions to the
paraxial Helmholtz equation (\ref{eq:paraxhelm}) \cite{saleh91}. In section \ref{sec:lg}, we will introduce the Laguerre Gaussian modes, which have well defined orbital angular momentum (OAM).  In section \ref{sec:lgspdc} we review the conservation of OAM in SPDC and the generation of entangled OAM states.  In sections \ref{sec:hg} and \ref{sec:hgspdc} we present the Hermite-Gaussian modes and their application to SPDC.    

\subsection{Laguerre Gaussian modes:  orbital angular momentum of light}
\label{sec:lg} 
Since the 1990's it has been known that any electromagnetic paraxial beam with an azimuthal phase dependence of the form $e^{il\phi}$ carries an orbital angular momentum $l\hbar$ per photon \cite{allen92}. The Laguerre-Gaussian beams are probably the most well-known and studied examples of beams carrying orbital angular momentum. They are given by \cite{beijersbergen93}
\begin{align}
U_{p}^{\ell}(\rho,\phi,z)=&
D_{p}^{\ell}\frac{1}{w(z)}\left(\frac{\sqrt{2}\rho}{w(z)}\right)^{\ell}L_{p}^{\ell}\left(\frac{2\rho^2}{w(z)^2}\right)
\exp\left(-\frac{\rho^2}{w(z)^{2}}\right)\nonumber \\
& \exp\left\{-i\left[\frac{k \rho^2}{2R}
-(n+m+1)\gamma(z)\right]-(p-\ell)\phi\right\},
\label{eq:lg}
\end{align}
where $(\rho,\phi,z)$ are the usual cylindrical coordinates, $D_{p}^{\ell}$ is a constant which depends on the azimuthal index $\ell$ and radial index $p$.  $L_{p}^{\ell}$ are the Laguerre polynomials.  Here $z$ is the longitudinal propagation direction, $R(z)$ is the radius of curvature, $w(z)$ is the beam waist,  and $\gamma(z)$ is the phase retardation or \emph{Gouy phase}.  The parameter $z_{R}$ is the
\emph{Rayleigh range}.  The order of the $LG$ beam is  defined as $\mathcal{N} = |\ell|+2p$.  The usual Gaussian beam is the zeroth-order $U_{0}^{0}$ beam.
See plots of LG modes in Fig. \ref{fig:modes}(left).
\par
  In addition to interesting implications in classical and quantum optics, the orbital angular momentum of a light field raises possibilities for technical applications. It has been shown that the orbital angular momentum of light can be used to rotate micro-particles in optical traps \cite{grier03}.  In terms of quantum optical applications, the orbital angular momentum of single photons in LG modes provides a possible $d$-dimensional qudit encoding scheme \cite{arnaut01,vaziri02,vaziri03a,molina04,molina05,molina07}, which allows for the creation of multi-dimensional entanglement in discrete bases.  Devices that discriminate the orbital angular momentum of Laguerre-Gaussian  beams have been proposed \cite{xue01_OL} and experimentally tested \cite{leach02,wei03}.  
\subsection{Entanglement and Conservation of Orbital Angular Momentum in SPDC}
\label{sec:lgspdc}
\par
The first experimental investigation of OAM conservation in SPDC suggested that OAM is not conserved \cite{arlt99}, in contrast to previous experimental investigations of other non-linear optical processes such as second-harmonic generation (SHG).   Using an LG mode as the fundamental beam in SHG showed a conservation relation, such that the OAM of the second-harmonic beam is double that of the fundamental beam:  $\ell_{\mathrm{SH}} = 2 \ell_{\mathrm{fund}}$ \cite{dholakia99,courtial99b}.  In these experiments, the intensity and phase structure of the output beam was observed using a CCD camera.  The same measurement method was used in the inverse experiment, investigation of OAM conservation in SPDC by Arlt et al. \cite{arlt99}.  However, by imaging the near and far field of the collinear down-converted fields using a CCD camera, correlations in second-order are observed.  Theoretical investigations by Arnaut and Barbosa showed that, under suitable experimental conditions, OAM conservation appears only in fourth-order \cite{arnaut01}.  The two-photon state has well-defined OAM equal to that of the pump beam, and $\ell_{\mathrm{signal}}+\ell_{\mathrm{idler}} = \ell_{\mathrm{pump}}$.  Moreover, they showed that the two-photon state is entangled in OAM, and thus the individual signal and idler beams do not have well defined OAM.  For this reason, the experiment by Arlt {\it et al.} did not observe OAM conservation.  Shortly thereafter, a clever experiment by Mair {\it et al.} confirmed the conservation and entanglement of OAM in SPDC \cite{mair01}.      
\par
\begin{figure}
\begin{center}
\includegraphics[width=6cm]{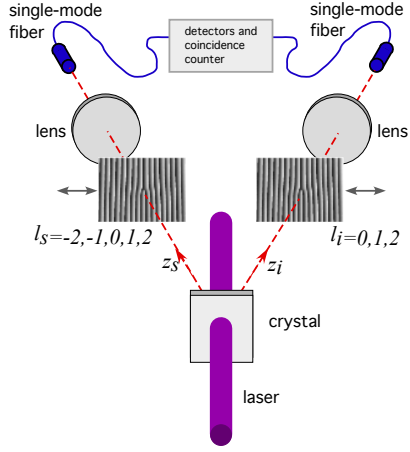}
\caption{First measurement of OAM entanglement in SPDC by Mair et al. \cite{mair01}.  Forked holographic masks of OAM $l_s$ and $l_i$ are used in conjunction with lenses and single mode-optical fibers to project onto modes with well-defined OAM.  To project onto superpositions of OAM, the masks are shifted in the transverse direction.}
\label{fig:mair}
\end{center}
\end{figure}

 The experiment by Mair et al. \cite{mair01}  is illustrated in figure \ref{fig:mair}.   The OAM correlation was measured using forked holographic masks and single-mode optical fibers to project onto individual OAM values of $m$.  The forked hologram mask functions in much the same way as a diffraction grating.  A Gaussian mode passing through a mask with $m$ dislocations and diffracting into the $n^\mathrm{th}$-order becomes an LG mode with $l=m n$.  The detection scheme of Mair et al. exploits this process in the reverse.  A mode with OAM $m$ that diffracts into the first-order of a forked mask with $m$ dislocations will now be in a Gaussian mode.  Single mode fibers are used to select only those photons in the Gaussian mode, which are then registered by single photon detectors.  The pump laser beam with $l=-1,0,1$ was shown to produce entangled photons with $m_{signal}=l-m_{idler}$. However, this confirms only a classical correlation between the OAM of the down-converted photons.  To prove that the photons are indeed entangled in OAM, Mair et al. projected the photons onto superpositions of OAM states.  This was done by shifting the position of the holographic mask.  A shifted mask with $m$ dislocations and the single-mode fiber detection scheme described above projects the single-photons onto superposition states of the form $\alpha\ket{0} + \beta \ket{m}$, where $\alpha$ and $\beta$ depend on the position of the mask. 
\par
The two-photon state for frequency degenerate down-converted fields ($\lambda_{s}=\lambda_{i}=2\lambda_{p}$) in a quasi-collinear geometry can be decomposed in terms of LG modes in a relatively simple form \cite{arnaut01,franke-arnold02,walborn04a,ren04}.  The wave function in this case can be written as 
\begin{equation}
\Psi(\vect{\rho}_{s},\vect{\rho}_{i}) = {U}_p^l\left(\frac{\vect{\rho}_{s}+\vect{\rho}_{i}}{\sqrt{2}}\right)\ \mathcal{F}\left(\frac{\vect{\rho}_{s}-\vect{\rho}_{i}}{\sqrt{2}}\right),
\label{eq:psif2}
\end{equation}
where ${U}_p^l$ describes the pump beam profile and $\mathcal{F}$ is the phase-matching function.
\par
Decomposing this wave function into signal and idler LG modes results in an alternative form:
\begin{equation}
\Psi(\vect{\rho}_{s},\vect{\rho}_{i})=\sum\limits_{l_{s},p_{s}}\sum\limits_{l_{i},p_{i}}N_{p_{s}p_{i}}^{l_{s}l_{i}}{U}_{p_{s}}^{l_{s}}(\vect{\rho}_{s})\,{U}_{p_{i}}^{l_{i}}(\vect{\rho}_{i}), 
\end{equation}
where the coefficient $N_{p_{s}p_{i}}^{l_{s}l_{i}}$ is given by \cite{walborn04a}
\begin{align}
N_{p_{s}p_{i}}^{l_{s}l_{i}} =&\int\hspace{-2mm}\int\hspace{-1mm} d\vect{\rho}_{s}
d\vect{\rho}_{i}\  {U}_{p}^{l}\left(\frac{\vect{\rho}_{s}+\vect{\rho}_{i}}{\sqrt{2}}\right)\,\mathcal{F}\left(\frac{\vect{\rho}_{s}-\vect{\rho}_{i}}{2}\right)\nonumber \\ &\times {U}_{p_{s}}^{*l_{s}}(\vect{\rho}_{s})\,{U}_{p_{i}}^{*l_{i}}(\vect{\rho}_{i}).
\label{eq:cdef}
\end{align} 
Assuming that the crystal length $L$ is much smaller than the Rayleigh range $z_R$ of the pump beam, and the function $\mathcal{F} \sim 1$, the coefficient $N_{p_{s}p_{i}}^{l_{s}l_{i}}$ becomes
\begin{equation}
N_{p_{s}p_{i}}^{l_{s}l_{i}}\propto \delta_{l_{s}+l_{i},l}\int\hspace{-1mm} q\,dq\   v_{p}^{l}(\sqrt{2}\,q)\,v_{p_{s}}^{*l_{s}}(q)\,v_{p_{i}}^{*l_{i}}(q), 
\label{eq:cfour2}
\end{equation} 
where $v_{p}^{l}(q)$ is the Fourier transform of the radial component of $U_p^l(\vect{\rho})$. 
The presence of the Kronecker delta function guarantees that OAM is conserved:  $l_i + l_s = l$, and the down-converted signal and idler photons are emitted with correlated OAM.  
 For simplicity, let $\ket{m}$ denote a single photon state carrying an orbital angular momentum $m\hbar$ and $l$ be the azimuthal index of the LG pump beam.  Coherence of the SPDC process guarantees that the two-photon state can be written as a Schmidt decomposition of the form \cite{law04}
\begin{equation}
\ket{\psi}=\sum_{m=-\infty}^{+\infty}P_{m}\ket{l-m}\ket{m}.
\label{eq:OAMstate}
\end{equation}
Here the states $\ket{m}$ represent spatial modes with well-defined OAM $m$.  
\par
\par
The derivation of the two-photon state \eqref{eq:OAMstate} is valid for quasi-collinear paraxial fields in the thin-crystal approximation.  Other effects, such as the birefringence and anisotropy of the non-linear medium, as well as a non-collinear geometry, were ignored.     
Some authors have studied SPDC and OAM conservation in more general contexts \cite{molina03,torres05,osorio08}.  A comprehensive discussion has been provided recently by Osorio {\it et al.} \cite{osorio08}.  They have shown that when one considers all the photon pairs produced by the crystal, distributed along the entire down-conversion cone, OAM is conserved.  In a more realistic setting, in which signal and idler photons are detected in small regions of the down-conversion cone, the two-photon wave function may be spatially asymmetric and not reproduce the angular spectrum of the pump beam.  The deformation generally corresponds to an ellipticity that depends upon the non-collinear angle $\varphi$, the crystal length, and  the width of the pump beam $w$ , and can be quantified by the quantity $L_d = w_/\sin \varphi$.  When the crystal length $L \ll L_d$, the two-photon wave function reproduces the transverse field of the pump beam, and OAM is conserved.  If this condition is not fulfilled, the two-photon wave function is not given by $\Psi(\vect{\rho}_{s},\vect{\rho}_{i}) = {U}_p^l\left(\frac{\vect{\rho}_{s}+\vect{\rho}_{i}}{\sqrt{2}}\right)$, and OAM will not be conserved.    Ellipticity of the two-photon spatial wave function caused by non-collinear geometry and focusing of the pump beam has been observed experimentally \cite{molina-terriza05}.  Torres, Osorio and Torner \cite{torres04} have considered the extreme situation in which the down-converted fields are either counter-propagating and perpendicular to the pump field, or co-propagating with respect to each other but counter-propagating to the pump field.  In these cases the usual OAM selection rule (as in the collinear geometry above) does not apply.     
\par
Another consideration is the spatial walk-off caused by the birefringence of the non-linear SPDC crystal, which occurs for either the pump and/or one of the down-converted fields.  In the case of type-I down-conversion, in which only the pump field is polarized in the extraordinary direction, the critical parameter is the ``walk-off length" $L_w = w/ \tan \rho_0$ \cite{torres05,osorio08}.  Here $\rho_0$ is the spatial walk-off angle, which depends upon the extraordinary refractive index and the propagation direction in the crystal.  When the length of the non-linear crystal $L \ll L_w$, the effect of walk-off on the shape of the two-photon wave function is negligible, and OAM is conserved.    
Fedorov \cite{fedorov07}.   
\par
\subsection{Probing the phase structure of the two photon state: two-photon interference}
\label{sec:mmhom}
Two-photon interference at a beam
splitter was first demonstrated by Hong, Ou and
Mandel (HOM) \cite{hom87}. It has subsequently become the key process for two-photon quantum logic, and is used in quantum information protocols such as Bell-state measurements
\cite{mattle96,bouwmeester97} and to construct two-photon logic gates
\cite{pittman02,obrien03,langford05,kiesel05,okamoto05}. On the one hand, these applications are designed for perfect interference with a single spatial mode.   
On the other hand, multimode interference is a useful method to probe the transverse phase structure of the two-photon state \cite{walborn03a,walborn05a}, and has been used to show the conservation of orbital angular momentum in SPDC  \cite{walborn04a}, as well as to quantify spatial entanglement \cite{peeters07}.  It has also been shown to be useful in Bell-state analysis of polarization entangled photons \cite{walborn03b}, and to produce two-photon fields exhibiting spatial antibunching \cite{nogueira04a}.       
\begin{figure}
\begin{center}
\includegraphics[width=10cm]{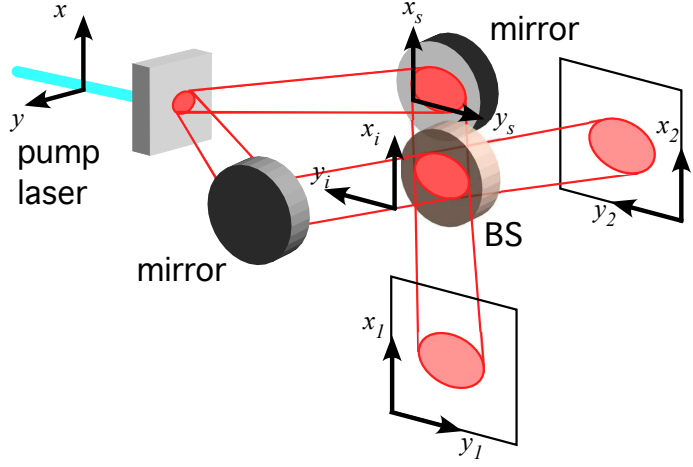}
\caption{Illustration of Hong-Ou-Mandel interference with multiple spatial modes.}
\label{fig:MultimodeHOM}
\end{center}
\end{figure}
\par
Let us illustrate the probing of the phase structure. 
Consider the HOM interferometer shown in Fig. \ref{fig:MultimodeHOM}, in which two photons are generated by SPDC and then reflected onto opposite sides of a beam
splitter (BS).  If the path length difference is greater than the coherence length of the down-converted photons, then there is no interference and the photons leave either side of the beam splitter randomly.  Here we will assume that lengths of paths $s$ and $i$ are equal. Using the reference frames illustrated in Fig. \ref{fig:MultimodeHOM}, the annihilation operators 
in exit modes $1$
and $2$ \emph{after} the beam splitter can be expressed in terms of the operators in input modes $s$ and $i$:
\begin{subequations}
\label{eq:ab}
\begin{align}
\oper{a}_{1}(\rmvect{q},\sigma)& = t\oper{a}_{s}(q_{x},q_{y},\sigma) + i r
\oper{a}_{i}(q_{x},-q_{y},\sigma) \\
\oper{a}_{2}(\rmvect{q},\sigma)& = t\oper{a}_{i}(q_{x},q_{y},\sigma) + i r
\oper{a}_{s}(q_{x},-q_{y},\sigma),
\end{align}
\end{subequations}
where $\sigma$ is the polarization, $q_x$ and $q_y$ are the transverse components of the momentum and $t$ and $r$ are respectively the transmission and reflection coefficients of the beam splitter ($|t|^{2}+|r|^{2}=1$).  A field reflected from the beam splitter undergoes a spatial inversion in the horizontal ($y$) direction, while a transmitted field does not suffer any inversion, as illustrated in Fig. \ref{fig:MultimodeHOM}. The negative sign that appears in the $q_{y}$ components of the reflected field is due to this
spatial inversion.  The two-photon wave function is split into 
four components, according to the four possibilities of transmission 
and reflection of the two photons \cite{walborn03a,walborn05a}: 
\begin{align}
\vect{\Psi}_{tr}(\rmvect{r}_{+},\rmvect{r}_{-})=\vect{\Psi}_{rt}(\rmvect{r}_{+},\rmvect{r}_{-})=&
itr \exp{\left\{\frac{2iK}{Z}(x_-^2+y_+^2)\right\}}
\nonumber \times \label{eq:psirt}\\
& \left[\mathcal{W}\left(x_+,-y_-,Z\right)
+\mathcal{W}\left(x_+,y_-,Z\right)\right],
\end{align}
\begin{equation}
\vect{\Psi}_{tt}(\rmvect{r}_{+},\rmvect{r}_{-})=
t^2 \exp{\left\{\frac{2iK}{Z}(x_-^2+y_-^2)\right\}}
\mathcal{W}\left(x_+,y_+,Z\right),\label{eq:psitt}
\end{equation}
\begin{equation}
\vect{\Psi}_{rr}(\rmvect{r}_{+},\rmvect{r}_{-})=
-r^2 \exp{\left\{\frac{2iK}{Z}(x_-^2+y_-^2)\right\}}
\mathcal{W}\left(x_+,-y_+,Z\right),\label{eq:psirr}
\end{equation}
 \par
where $\rmvect{r}_\pm=(x_\pm,y_\pm)=(x_s/2\pm x_i/2,y_s/2\pm y_i/2)$.  Let us assume that $t=r=1/\sqrt{2}$.  If the pump beam profile $\mathcal{W}$ is symmetric with respect to the $y$ coordinate ($\mathcal{W}[y]=\mathcal{W}[-y]$), then the two photons always leave in the same output port of the BS, since the amplitude for coincidence counts in different output ports is $\vect{\Psi}_{tt}+\vect{\Psi}_{rr}=0$. This results in the usual HOM interference dip.  If $\mathcal{W}$ is antisymmetric with respect to $y$ ($\mathcal{W}[y]= - \mathcal{W}[-y]$), then $\vect{\Psi}_{tr}=\vect{\Psi}_{rt}=0$, and the photons always leave the BS in different output ports, giving an interference peak.  
\par
The analysis above considered that all other degrees of freedom are in symmetric states.  Combining the spatial degrees of freedom with the polarization degree of freedom, for instance, provides a way to control two-photon interference of different polarization states.  This allows for partial Bell-state analysis in the coincidence basis \cite{walborn03b}, as well as for the creation of a two-photon ``singlet" beam, in which the photon pairs in the same beam are in an anti-symmetric polarization state, and consequently are spatially antibunched \cite{nogueira04a}.  
\par
Other experiments have used two-photon interference to probe the phase structure of the two-photon state, and have shown that the down-converted photon pairs are entangled in OAM \cite{walborn04a,walborn05a}.  For a pump beam in a LG mode, the two-photon coincidence detection probability at the output of the HOM interferometer is
\begin{equation}
P(\rmvect{r}_{1},\rmvect{r}_{2})\propto |u_{p}^{\ell}(R)|^{2}\sin^{2} l\theta,
\label{eq:prob1}
\end{equation}   
where $u_{p}^{\ell}(R)$ is the radial component of the LG mode, $\ell$ is the OAM of the pump beam which is transferred to the two-photon state, $R=|\vect{\rho}_1+\vect{\rho}_2|$, and  $\sin \theta = (\rho_1 \sin \phi_1 + \rho_2 \sin \phi_2)/R$.  When $\rmvect{r}_2=0$, the coincidence profile is  $P(\rmvect{r}_{1},0)\propto |u_{p}^{\ell}(\rho_1/\sqrt{2})|^{2}\sin^{2} \ell\phi_1$.  Thus, fixing one detector and scanning the other, oscillations with period $\ell$ will appear in the azimuthal coordinate  $\phi_1$.  In Ref. \cite{walborn04a}, the phase structure was observed for LG pump beams with $\ell=1,2$.    
   
\par
\subsection{Quantifying Spatial Entanglement}
In section \ref{sec:spatialentanglement}, the detection of spatial entanglement using entanglement witnesses was discussed.  In many cases, these witnesses involve only a few measurements of the transverse spatial variables.  Quantifying spatial entanglement, on the other hand, presents a considerable challenge, due to the large dimensionality of the involved Hilbert spaces.  Nonetheless, theoretical and experimental methods exist which allow for the quantification of bipartite spatial entanglement of photon pairs.    
\subsubsection{Schmidt decomposition of SPDC}
The Schmidt decomposition \cite{chuang00} of the two-photon state produced by SPDC has been derived by Law and Eberly \cite{law04}.    Absorbing any phase factors into the states $\ket{m}$ allows one to write the two-photon amplitude \eqref{eq:OAMstate} in the form \cite{law04}
\begin{equation}
\Psi(\vect{\rho}_s,\vect{\rho}_i) = \sum\limits_{m=-\infty}^{+\infty} \sum\limits_{n=0}^{\infty}\sqrt{\lambda_{n,m}}u_{n,l-m}(\vect{\rho}_s)u_{n,m}(\vect{\rho}_i),
\label{eq:sd}
\end{equation}    
where the real numbers $\lambda_{n,m}$ are the eigenvalues of the reduced density operators.  The functions $u_{n,m}$  are normalized transverse mode functions with well-defined OAM \cite{law04}.   The entanglement depends upon the number of non-zero Schmidt coefficients $\lambda_{n,m}$.  Each $\lambda_{n,m}$ depends on the OAM $\ell$, wavenumber and width of the pump laser beam, and the length of the non-linear crystal, and can be calculated  by determining the overlap integral of the product of mode functions with the two-photon amplitude given in equation \eqref{eq:psif2}:
\begin{equation}
\sqrt{\lambda_{n,m}} = \int \vect{\rho}_s \vect{\rho}_i u^*_{n,l-m}(\vect{\rho}_s)u^*_{n,m}(\vect{\rho}_i) {U}_p^l\left(\frac{\vect{\rho}_{s}+\vect{\rho}_{i}}{\sqrt{2}}\right)\ \mathcal{F}\left(\frac{\vect{\rho}_{s}-\vect{\rho}_{i}}{\sqrt{2}}\right).  
\end{equation}
Again, effects due to crystal anisotropy and non-collinear geometry have been ignored.  Law and Eberly \cite{law04} have used the decomposition \eqref{eq:sd} to calculate the entanglement of the two-photon state generated by a Gaussian pump beam using the so-called Schmidt number, $S = 1/ \sum_{n,m} \lambda^2_{n,m}$ \cite{law04}.  They have shown that the entanglement depends upon the parameter $L/2 z_R$, where $z_R$ is the Rayleigh range of the pump beam and $L$ is the crystal length.  If one approximates the phase matching function by a Gaussian function with the same variance, $\gamma(\vect{q}) \approx  \exp(-L\vect{q}^2/8K)$ \cite{chan07}, the Schmidt number in this case can be put in the analytical form $S=(\sqrt{L/4z_R} + \sqrt{4z_R/L})^2/4$.  The entanglement in this Gaussian approximation serves as a lower bound, since Gaussian states are always less entangled than non-Gaussian states with the same covariance matrix \cite{wolf06}.             
\par
For typical values of the experimental parameters, the predicted Schmidt number can be very large.  For example, for $L = 1$mm and $z_R=1$m, the predicted Schmidt number is on the order of $10^3$. However, as has been pointed out by van Exter {\it et al.} \cite{vanexter06}, and Osorio {\it et al.} \cite{osorio08}, even in the case of collinear propagation and negligible crystal anisotropy, this does not necessarily correspond to the usable entanglement present in a realistic experimental scenario.  The effect of spatial filtering due to the finite spatial bandwidth of an optical setup is to decrease the Schmidt number, and hence the entanglement.   This can be a rather large effect, of several orders of magnitude.    
\par
In Ref. \cite{pires09a}, Di Lorenzo Pires {\it et al.} demostrated that for the case of a two-photon pure state such as \eqref{eq:quantumstate}, the overall Schmidt number (including contributions from radial and azimuthal parameters) is given by the inverse of the overall degree of coherence of either down-converted field.  Assuming that--say--the signal field is quasihomogeneous, the Schmidt number takes the form 
\begin{equation}
S \approx \frac{1}{(2 \pi)^2}\frac{\left[\int I_{nf}(\vect{\rho}_s) d\vect{\rho}_s\right ]^2}{\int I_{nf}^2(\vect{\rho}_s) d\vect{\rho}_s}\frac{\left[\int I_{ff}(\vect{q}_s) d\vect{q}_s\right ]^2}{\int I_{ff}^2(\vect{q}_s) d\vect{q}_s},
\end{equation}    
where $I_{nf}$ and $I_{ff}$ are the near-field and far-field intensity distributions of the signal photon field, respectively.  Thus, through second-order intensity measurements, the spatial entanglement of the two-photon state can be determined.  It was shown that the entanglement  depends critically on the collinear phase mismatch \cite{pires09a}.  Schmidt numbers above 400 were obtained experimentally.   
\par
An experimental technique to measure the Schmidt number associated with the OAM decomposition \eqref{eq:sd} was developed by Peeters {\it et al.} \cite{peeters07}, who investigated multimode HOM interference in the presence of image rotation of the signal beam before the beam splitter.  They showed that the amount of spatial entanglement can be inferred from the visibility of the HOM dip:
\begin{equation}
V(\vartheta) = \sum\limits_{m=-\infty}^{\infty} P_m \cos(2 m \vartheta),
\label{eq:vispeeters}
\end{equation}
where $\vartheta$ is the angle of image rotation of the signal field.  Here $P_m$ are the coefficients of the Schmidt decomposition  \eqref{eq:OAMstate}.  To quantify spatial entanglement, an azimuthal Schmidt number can be defined as $S_{az} = (\sum\limits_{m=-\infty}^{\infty} P_m^2)^{-1}$.  
The Schmidt coefficients can be determined by measuring $V(\vartheta)$ for a number of values of $\vartheta$ and inverting the Fourier series \eqref{eq:vispeeters}.  In Ref. \cite{peeters07}, this was done for a number of different types of spatial filtering systems.  Schmidt numbers as high as $7.3\pm0.3$ were determined experimentally.  More recently,  Di Lorenzo Pires {\it et al.} performed a similar experiment in a collinear geometry using a bucket detector system \cite{pires10}.  This impressive experiment allowed for the measurement of the complete OAM distribution of the correlated down-converted photons, and  Schmidt numbers above 35 were measured.  This was in excellent agreement with theoretical predictions based on the experimental parameters.  It was also shown that the phase mismatch could be used to increase the entanglement. In this case, the mismatch could be controlled through the temperature of the periodically poled crystal.     
\par
Another method to quantify entanglement using multimode HOM interference has been proposed in \cite{walborn07c}. In this scheme, it is possible to measure the concurrence directly using two copies of the quantum state \cite{walborn06b,walborn07a}.    
     \par
An interesting series of experiments have been performed employing  phase plates and single-mode fibers to detect and analyze spatial entanglement \cite{oemrawsingh05,pors08,pors08b}. Two main types of phase plates have been investigated:  spiral phase plates and Heaviside step plates. When a Gaussian mode passes through a half-integer spiral phase plate, it is transformed into a superposition of LG modes, given by \cite{oemrawsingh06}
\begin{equation}
\sum\limits_{p,\ell} |C_{p,\ell}|e^{i f_{p,\ell}(\alpha)}U_{p}^{\ell}(\rho,\phi), 
\label{eq:spp}
\end{equation}    
where $f_{p,\ell}(\alpha)$ is a function which depends upon the orientation angle $\alpha$ of the phase plate.  By considering the inverse scenario, it is apparent that the spiral phase plate transforms the superposition (\ref{eq:spp}) into a Gaussian mode, which can be selected with a single mode optical fiber.  By changing the orientation angle $\alpha$ of the plate, one can project onto a number of superposition states of the form (\ref{eq:spp}).  In Ref. \cite{oemrawsingh06} it was shown that the number of non-zero terms in the superposition (\ref{eq:spp}) can be quite large ($\sim 10^2-10^3$), and depends upon the half-integer value of the plate.
 \par
 Half-integer spiral phase plates were used to observe fractional OAM entanglement in down-converted photons \cite{oemrawsingh05}.  The experimental setups are similar to that of Figure \ref{fig:mair}, with the forked holographic masks replaced by the phase plates.    The coincidence detection probability, given by \cite{oemrawsingh05}
 \begin{equation}
 P(\alpha_s,\alpha_i) \propto \left( \pi - |\alpha_s-\alpha_i| \right)^2, 
 \label{eq:Pspp} 
 \end{equation}
 depends only upon the relative angle $|\alpha_s-\alpha_i|$ between the analyzers in the signal and idler fields, which is typical of a quantum correlation.      
 \par  
 Another type of phase plate which has been studied  is a Heaviside phase plate, in which there is relative phase difference of $\pi$ between the phase imparted by each section of the plate \cite{oemrawsingh06,pors08}.  The transmission function is given by $T(\phi,\alpha)=1-2[\Theta(\phi-\alpha)-\Theta(\phi-\alpha-\beta)]$, where $\Theta$ is the Heaviside step function, $\alpha$ is the orientation angle of the plate and $\beta$ is the angular width of the $\pi$-step section.  The simplest case corresponds to two half-circle sections ($\beta=\pi$).  These step phase plates can also be used to probe multidimensional entanglement.     Pors {\it et al.} have shown that the amount of usable entanglement that is available is given by the dimensionality of the measurement apparatus \cite{pors08,pors08b}.    The authors have dubbed this the ``Shannon dimensionality", in analogy with the Shannon number of a classical communication channel \cite{shannon}.  The Shannon dimensionality is an effective Schmidt number, corresponding to the amount of entanglement which can be accessed by the measurement system.  
 In the case of the phase plate analyzers, the dimensionality has a simple experimental interpretation.  It is given by $2 \pi$ divided by the area under the normalized coincidence curve, plotted as a function of the relative angle of the phase plates.  Experimental and theoretical results show that the half-circle plate corresponds to dimensionality of three, while a quarter section plate gives dimensionality of six.  The authors also considered step plates with multiple sections, and have shown that it should be possible to probe spaces with dimension on the order of about 50 \cite{pors08b}.      
 \par      
 There is thus an inconsistency between the amount of entanglement which is theoretically available and the amount of entanglement actually accessible in a given experimental scenario, due in particular to the geometry of the down-conversion setup and the spatial bandwidth of the optical systems used in manipulation and detection of the fields involved.  
  
Another experiment investigating this kind of entanglement quantification, used entangled states prepared placing
apertures in the path of the beams generated by SPDC.
In this case the aperture photon path defines the qubit state. In reference \cite{neves07}, two
double-slits placed at the path of the idler and signal photons were used to generate
two-qubit entangled states. States with different degrees of entanglement were prepared
by modifying the pump beam profile that generates the twin photons. Measurements of either
two-photon conditional interference or marginal probabilities were
 used for characterizing entanglement in two-qubit spatial pure states.

The concurrence ${\cal C}$ is an entanglement quantifier for bipartite systems \cite{wootters98}.  For a general pure entangled state of two qubits,
\begin{equation}\label{pure}
|\psi\rangle=c_{11}|11\rangle+c_{10}|10\rangle+c_{01}|01\rangle+c_{00}|00\rangle\,,
\end{equation}
the concurrence is given by
\begin{equation}\label{wootters}
{\cal C}(\psi)=2|c_{11}c_{00}-c_{10}c_{01}|\,.
\end{equation}
State (\ref{pure}) may be written as  a Schmidt decomposition \cite{ekert94}
\begin{equation}\label{schmidt}
|\psi\rangle=c^\prime_{11}|11\rangle^\prime+c^\prime_{00}|00\rangle^\prime\,,
\end{equation}
where $|ii\rangle^\prime$ ($i=0,1$)  is the {\it Schmidt basis}, and the coefficients $c^\prime_{ii}$ 
are real and positive. In terms of these coefficients,  the concurrence is given by the simpler expression
\begin{equation}\label{csimple}
{\cal C}(\psi)=2c^\prime_{11}c^\prime_{00}\,.
\end{equation}
In Ref. \cite{neves07}, states $|0\rangle, |1\rangle$ represent the single photons which pass through either slit of the double slit aperture. 
The strategy used for characterizing entanglement consists in performing measurements in the
Schmidt basis, either by directly measuring the Schmidt coefficients
or through a direct connection between concurrence,
given by Eq.~(\ref{csimple}), and the visibility of interference
patterns obtained when one of the detectors is used as trigger. 
The Schmidt basis is selected by letting the photons
propagate through an appropriate lens system and keeping one of the detectors 
fixed at some specific positions, while the other is scanned at the Fourier plane of the lens. 
The Schmidt coefficients are then obtained from these
conditional interference patterns. In the second method the fixed detector used as trigger is completely opened
detecting all photons that crossed one of the double-slits while the second point detector
is scanned at the lens Fourier plane. This corresponds to a measurement of the marginal probability
which is defined as the
probability, $\bar{P}_{1}(x_{i})$, of observing a photon idler
at $x_{i}$ and the signal at any location
($-\infty<x_{s}<+\infty$) \cite{abouraddy01} and is related to the concurrence as
\begin{equation}  \label{singlerate2}
P_{1}(x_{i}) \propto
\left[1+\sqrt{1-[\mathcal{C}(\Psi)]^{2}}
                     \cos(\alpha x_{i})\right].
\end{equation}

In reference \cite{peeters09} different two-qubit states
are generated by using collinear type II SPDC and a double-slit.
Different states are produced by tailoring the imaging
system in between the two-photon source and the double
slit.  More than 30 different two-photon states are characterized 
by measuring their complete two-photon
interference patterns in the far field of the double slit.

\subsection{Hermite-Gaussian modes}
\label{sec:hg}
The mode decomposition of the two-photon state produced in SPDC can also  be performed using the Hermite-Gaussian modes,  which are given by the complex field amplitude \cite{beijersbergen93} 
\begin{align}
U^{HG}_{nm}(x,y,z)=&
C_{nm}\frac{1}{w(z)}
H_{n}\left(\frac{\sqrt{2}x}{w(z)}\right)H_{m}\left(\frac{\sqrt{2}y}{w(z)}\right)\exp\left(-\frac{x^2+y^2}{w(z)^{2}}
\right)\nonumber \\ & \exp\left\{-i\left[\frac{k(x^2+y^2)}{2R(z)}
-(n+m+1)\gamma(z)\right]\right\}.
\label{eq:hg}
\end{align}
Here  $C_{nm}$ are the coefficients, $H_{n}(x)$ is the $n^{\mathrm{th}}$-order Hermite polynomial, which is an even or odd function of $x$ if $n$ is even or odd, respectively.      The \emph{order} $\mathcal{N}$ of the beam is the sum of the indices: $\mathcal{N} = m+n$.  The usual Gaussian beam is the zeroth-order $U^{HG}_{00}$ beam. See plots of HG modes in Fig. \ref{fig:modes}(right).
 \par
  
 \subsection{Down-conversion with Hermite-Gaussian modes}
 \label{sec:hgspdc}
 Consider SPDC in which the pump beam is described by a HG mode $U_{nm}^{HG}$.  The two-photon state can be expanded as \cite{walborn07c,walborn05b} 
\begin{equation}
\ket{\psi_{nm}} = \sum_{j,k,u,t=0}^{\infty}C^{\,nm}_{jkut}\ket{v_{jk}}\ket{v_{ut}},
\label{eq:stateHG}
\end{equation}
where $\ket{v_{ab}}$ are Hilbert space vectors such that $v_{ab}(\vect{q})=\langle\vect{q}|{v_{ab}}\rangle$ corresponds to the HG mode written in $k$-vector space, given by the Fourier transform of Eq. \eqref{eq:hg}.  
The coefficients $C^{\,nm}_{jkut}$ are related to the pump beam and phase matching function through
\begin{equation}
C^{\,nm}_{jkut} =  \iint    d\rmvect{q}_{u}\,d\rmvect{q}_{i} v^{*}_{jk}(\rmvect{q}_{s})v^{*}_{ut}(\rmvect{q}_{i}) \Phi_{nm}(\rmvect{q}_{s},\rmvect{q}_{i}),
\label{eq:hgspdc7}
\end{equation}  
where $\Phi_{nm}(\rmvect{q}_{s},\rmvect{q}_{i})$ is given by the product of the angular spectrum for a HG pump beam and the phase matching function (see section \ref{sec:fundamentals}).  
 An analytical expression for the coefficients $C^{\,nm}_{jkut}$ was calculated in Ref. \cite{walborn05b}, showing that spatial parity is conserved in the $x$ and $y$ directions.  That is, the indices $j+u$ ($k+t$) must have the same parity as the pump beam index $n$ ($m$).  Since the parity of each down-converted photon is undetermined, yet the sum of parities is well-defined, the two-photon state is entangled in parity.  
Thus, assuming that the pump beam has well-defined parity in the $y$ direction, and ignoring the $x$ direction, the two-photon state can be written as
\begin{equation}
\ket{\psi}_{\mathrm{even}} = \frac{1}{\sqrt{2}}\left ( \ket{\psi_{\mathrm{EE}}} + \ket{\psi_{\mathrm{OO}}} \right ),
\label{eq:evenstate}
\end{equation}  
or
\begin{equation}
\ket{\psi}_{\mathrm{odd}} = \frac{1}{\sqrt{2}}\left ( \ket{\psi_{\mathrm{EO}}} + \ket{\psi_{\mathrm{OE}}} \right ),
\label{eq:oddstate}
\end{equation}  
where E and O stand for the $y$ indices of photons 1 or  2 being even (odd) and odd (even) numbers, respectively.  If the pump beam is a HG mode, the states $\ket{\psi_{\mathrm{XY}}}$ ($X,Y=E,O$) are given by
linear combinations of modes with the same $y$ parity.  For example, 
\begin{equation}
\ket{\psi_{\mathrm{EO}}} = \sum_{k\, \mathrm{even}, \,t\, \mathrm{odd}}\sum\limits_{j,u} C^{\,nm}_{jkut}  \ket{v_{jk}}
 \ket{v_{ut}} ,
\label{eq:3}
\end{equation}
and similarly for the other states in \eqref{eq:evenstate} and \eqref{eq:oddstate}.  Yarnall \emph{et al.} \cite{abouraddy07,yarnall07a} have used this type of parity entanglement to violate Bell's inequalities with parity measurements, which we describe in detail below in section \ref{sec:bellinequality}. 
  \par
Using the expansion \eqref{eq:stateHG}, the concurrence of the two-photon state can be calculated as a function of the pump beam profile and the length of the non-linear crystal.  Numerical results reported in \cite{walborn07c} show that in general the entanglement increases with the order $n+m$ of the pump beam.  
       
\subsection{Bell Non-locality}
\label{sec:bellinequality}
The Einstein-Podolsky-Rosen paradox showed an inconsistency between local-realism and the completeness of quantum mechanics.  In 1965, John S. Bell considered the predictions of local realism against those of quantum mechanics \cite{bell65}.  His argument, made in the context of David Bohm's version of the EPR paradox involving two entangled spin-1/2 systems \cite{bohm51}, showed a conflict between the predictions of quantum mechanics and local realistic theories.  This provided an experimentally accessible testing ground for these theories.  The Clauser-Horne-Shimony-Holt inequality \cite{clauser69}, also derived independently by Bell \cite{bell87}, shows that any local realistic theory should obey 
\begin{equation}
S = |E(a,b) + E(a^\prime,b) + E(a,b^\prime) - E(a^\prime,b^\prime)| \leq 2,  
\label{eq:bell}
\end{equation} 
where $-1 \leq E(a,b) \leq 1$ is the correlation function of dichotomic measurements $a$ and $b$ on systems $A$ and $B$.   For a pair of maximally-entangled spin-$\frac{1}{2}$ particles,  quantum mechanics predicts $S_{\mathrm{max}} = 2 \sqrt{2}$. This set the stage for tests of quantum mechanics versus local realism beginning with the work of Freedman and Clauser \cite{freedman72} and Aspect and collaborators \cite{aspect82a,aspect82b}.  With the advent of down-conversion sources, recent experiments have reached a considerable degree of sophistication, with the objective of performing a conclusive loop-hole free test \cite{kwiat95,weihs98,tittel98}, and also to discard more complex alternatives to standard quantum theory \cite{groblacher07,salart08}.  
\par 
An early adaptation of the Bell inequality to continuous variables was provided by John Bell, who considered linear combinations of position and momentum observables of the form $\oper{x}+\alpha \oper{p}$, where $\alpha$ is a continuous parameter  \cite{bell87}.  By binning the results, for instance  into non-negative and negative values, one can translate these continuous measurements into two outcomes and apply the Bell-CHSH inequality (\ref{eq:bell}).      
Bell reached a striking conclusion: quantum states with positive Wigner functions should allow for local realistic descriptions, since in this case the Wigner function itself serves as a classical probability distribution of the $x$ and $p$ variables \cite{bell87}.  A surprising example is the EPR state \eqref{eq:EPR}, with Wigner function given by \cite{bell87}
 \begin{equation}
 W_{\mathrm{EPR}}(x_1,p_1,x_2,p_2) = 2 \pi \delta(x_1-x_2)  \delta(p_1+p_2), 
 \end{equation} 
which is non-negative everywhere. This state displays an infinite amount of entanglement, and as such it is remarkable that it does not violate the Bell-CHSH inequality \eqref{eq:bell}. The EPR state is not physical, however it represents the limit $a_{-}/a{+}\longrightarrow 0$ (and $b_+=b_-=0$) of the Gaussian state \eqref{eq:psirho}.      
\par      
Despite these arguments against the use of continuous variables in this context, it was later shown by Banaszek and W\'odkiewicz \cite{banaszek98,banaszek99} that the EPR state does violate the Bell-CHSH inequality when one considers measurements of the displaced parity operator.  The parity operator does not have a well-behaved Wigner-Weyl representation, and thus expectation values of this operator do not correspond to averages over classical probability distributions.  This point has been explored in more detail by  Revzen {\it et al.} \cite{revzen05}.    
Thus, the state of a given system alone is not enough to determine whether it is suitable or not 
for a Bell-type experiment, but rather one must also take into account the observables measured.   
\par
Extending the idea of the parity operator, it possible to define a set of pseudo-spin operators, forming an analogy with the Pauli operators for a spin-1/2 system. Chen {\it et al.} have shown that the operators \cite{chen02}
\begin{subequations}
\label{eq:pseudospin}
  \begin{align}
   \oper{X} & = \sum\limits_{n=0}^{\infty} \ket{2n+1}\bra{2n} + \ket{2n}\bra{2n+1} \\
    \oper{Y} & = i \sum\limits_{n=0}^{\infty} \ket{2n+1}\bra{2n} - \ket{2n}\bra{2n+1} \\
 \oper{Z} & = \sum\limits_{n=0}^{\infty} \ket{2n}\bra{2n} - \ket{2n+1}\bra{2n+1},
 \end{align} 
 \end{subequations}
 defined in terms of photon number Fock states $\ket{n}$, obey a SU(2) algebra.  One can recognize $\oper{Z}$ as the parity operator.  These operators were originally formulated for single-modes of the electromagnetic field, and their realization presents considerable experimental challenges \cite{chen02}.   
  \begin{figure}
\begin{center}
\includegraphics[width=8cm]{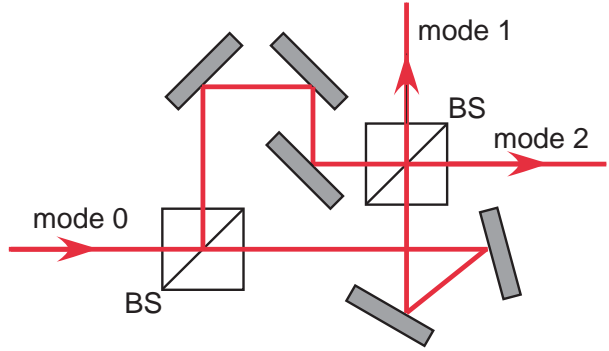}
\caption{Asymmetric interferometer used as a parity projector.  When $\delta_1=\delta_2$ and the beam splitters are 50:50, transverse modes with even parity leave the interferometer in port 1, and odd transverse modes in port 2.}
\label{fig:asymint}
\end{center}
\end{figure}

 \par
 An insightful contribution was made by Abouraddy \emph{et al.}, who recognized that the operators \eqref{eq:pseudospin} could be constructed for the spatial degrees of freedom of photons \cite{abouraddy07}.  This is possible due to the isomorphism between the photon number Fock states $\ket{n}$, and the    
spatial degrees of freedom of a field described by a one-dimensional Hermite Gaussian mode with index $n$ \cite{abouraddy07}.  To employ the full power of the analogy between the spatial version of the pseudo-operators \eqref{eq:pseudospin} and the Pauli operators, one must be able to implement arbitrary rotations, composed of linear combinations of $\oper{X},\oper{Y}$ and $\oper{Z}$.   In Ref. \cite{abouraddy07}, the authors showed that arbitrary rotations of the  ``spatial parity", can be manipulated through a series of linear optical components.  This allows for the use of the spatial parity of single photons as a quantum bit in quantum information tasks \cite{yarnall07b}.           
-For example, $\oper{Z}$ can be implemented by a mirror reflection or a lens system.  
One can discriminate between even and odd spatial modes of a single photon by interfering the field with its mirror image \cite{sasada03}.  
This can be achieved using the asymmetric Mach-Zehnder interferometer shown in figure \ref{fig:asymint}. Photons traveling through arm 1 of the interferometer are subject to $2$ mirror reflections, while photons in arm $2$ are subject to $3$ mirror reflections.  Each reflection transforms the transverse wave vector as $(q_x,q_{y}) \longrightarrow (q_x,-q_{y})$.  Assuming a single-photon detection at an exit port, the photon annihilation operators at the exit ports are related to the input port by 
\begin{equation}
\oper{a}_{1}(\rmvect{q}) = t^{2}e^{i\delta_{1}}\oper{a}_{0}(q_{x},q_{y})-r^{2}e^{i\delta_{2}}\oper{a}_{0}(q_{x},-q_{y})
\end{equation}      
\begin{equation}
\oper{a}_{2}(\rmvect{q}) = itre^{i\delta_{1}}\oper{a}_{0}(q_{x},-q_{y})+itre^{i\delta_{2}}\oper{a}_{0}(q_{x},q_{y})
\end{equation}      
where $t$ and $r$ are the transmissivity and reflectivity of the beam splitters.  $\delta_{1}$ and $\delta_{2}$ are phases due to propagation through arms $1$ and $2$.  Assuming $|\delta_{1} - \delta_{2}| \mod 2\pi = 0$, and $t=r$, an even input mode will leave the interferometer in output  2 while an odd mode  will exit in output 1.    
 The asymmetric interferometer is thus capable of projecting onto even or odd modes, and is the spatial parity analog to the polarizing beam splitter \cite{abouraddy07}.
\par

To perform rotations in parity space, it is necessary to imprint a variable phase on half of the light beam.  In Ref. \cite{abouraddy07} it was shown that this can be done using two phase plates. Each plate is placed in half of the input field, so that plate 1 covers the region $x<0$ and imprints a phase $\exp(-i\theta/2)$, and plate 2 covers the region $x\geq 0$ and imprints a phase $\exp(i\theta/2)$.  The parity rotation depends upon the overall phase difference between the two halves of the field.  
\par
Using a set of variable phase plates and asymmetric interferometers for each of the down-converted fields, Yarnall {\it et al.} violated Bell's Inequality in spatial parity space for pump beams with different spatial parity 
\cite{yarnall07a}.

%% file: section9.tex
\section{Applications to quantum information}
The facility with which one can produce highly-entangled  states
from SPDC and manipulate the spatial characteristics of the pump
and down-converted fields allows for the investigation of a wide
range of quantum information applications and phenomena.   A
principal advantage of the spatial degrees of freedom of photons,
as compared to the polarization degree of freedom, for example, is
the high-dimensionality of the system.  This allows for the
generation of entangled continuous variable states or entangled
$d$-dimensional \emph{qudits}.  Higher dimensional ($D>2$) quantum
systems have been shown to allow for higher security and
transmission rates in quantum cryptography
\cite{bechmann00a,bourennane01,cerf02}, increased security in
two-party protocols such as bit commitment \cite{spekkens02} and
coin tossing \cite{ambainis01,spekkens02b}, and more robust tests
of Bell's inequalities \cite{collins02}.  In sections
\ref{sec:spatialentanglement} and \ref{sec:transversemodes} we
showed that the two-photon quantum state produced by SPDC can be
highly entangled.  Furthermore, there exist schemes for
engineering the two-photon entanglement by properly shaping the
two-photon angular spectrum \cite{monken98a,torres03a,torres03b}.
There is also a wide array of gadgetry devised to manipulate these
spatial degrees of freedom, including lens systems
\cite{howell04,tasca08,tasca09,tasca09b}, holographic masks
\cite{mair01,heckenberg92}, phase plates
\cite{oemrawsingh05,pors08,yarnall07a} and more recently the
development of spatial light modulators have opened up even more
possibilities for the synthesis and control of the spatial degrees
of freedom of photons \cite{yao06b,stutz07,leach09,lima09}. 
\par
Orbital angular momentum is a natural choice for encoding single-
and entangled qudits, and a number of proof of principle
experiments have been performed which employ these degrees of
freedom. These include the demonstration of bit commitment
\cite{langford04} and coin tossing protocols  \cite{molina05} with
entangled qutrits ($D=3$), manipulation of heralded qutrits
\cite{molina04}, violation of Bell's inequality and entanglement
concentration of entangled qutrits \cite{vaziri02,vaziri03a},
violation of Bell's inequality in various two-dimensional
subspaces of the infinite-dimensional OAM space \cite{leach09} and
quantum cryptography \cite{gibson04,groblacher06}.
\par
Higher-dimensional quantum information protocols can also be
implemented using the linear transverse position and momentum.  In
this respect, entangled qudits have been produced using $D$-slit
apertures \cite{neves05,neves04}, and by binning the near and
far-field detection positions \cite{osullivan05} and quantum
cryptography with a 37-dimensional alphabet has been demonstrated
\cite{walborn06a}.
\par
Another notable use of spatial entanglement is the creation of
multiply entangled states, which are entangled in multiple degrees
of freedom \cite{kwiat97,walborn05c}.  In fact, pairs of photons
which are {\it hyperentangled} (entangled in {\it all} degrees of
freedom) have been produced and tomographically analyzed
\cite{barreiro05}.   It has been shown theoretically
\cite{kwiat98a,walborn03c,wei07} and experimentally
\cite{walborn03b,schuck06,barreiro08}, that entanglement in an
auxiliary degree of freedom can improve Bell-state analysis of
photon pairs.  Barreiro, Wei and Kwiat have used auxiliary
OAM entanglement to surpass the channel capacity imposed by linear
optics in the quantum superdense coding protocol
\cite{barreiro08}.  First-order transverse modes and polarization
can be used to define single-photon logical qubits which are
invariant under azimuthal coordinate rotations \cite{aolita07, souza08},
and to implement small-scale quantum algorithms \cite{oliveira05}.
\par
 \subsection{Generation of Entangled Qudits}
\begin{figure}
\begin{center}
\includegraphics[width=5cm]{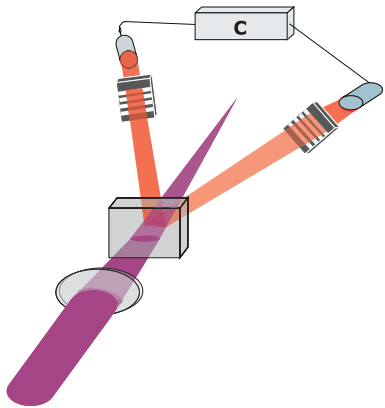}
\caption{Multiple-slit scheme used to produce entangled qudits.}
\label{fig:neves}
\end{center}
\end{figure}

There are a number of ways to generate entangled qudits using the
spatial degrees of freedom of twin photons. A natural choice is using a set of
transverse modes with discrete quantum numbers, such as the Hermite-Gaussian or Laguerre-Gaussian modes.  A number of
experiments in this direction have been performed \cite{vaziri02,vaziri03a,molina04,molina05,langford04}.
\par
Another possibility is through $D$-slit apertures placed in each down-converted field \cite{neves04,neves05}. 
as illustrated in Fig. \ref{fig:neves}.  
By focusing the pump beam in the plane of the apertures, 
one can guarantee that the photons always pass through symmetrically opposite slits of their respective apertures.  
In this fashion, the two-photon state immediately after the apertures can be written as \cite{neves05,neves04}:

\begin{equation}
\ket{\psi} = \frac{1}{\sqrt{D}} \sum\limits_{j=(1-D)/2}^{(D-1)/2}
\exp\left (i k d^2 j^2 / 2 z \right ) \ket{j}_1 \ket{-j}_2,
\end{equation}
where $d$ is the distance between the slits, $z$ is the distance from crystal to slits, $k$ is the wave number and
$\ket{j}$ are states describing the spatial profile of a single photon which has passed through slit $j$, which are given by
 \begin{equation}
\ket{j} =\int d\rho R (\rho - j d) \ket{\rho}.
\end{equation}
\par
Here $R(\rho)$ are the aperture functions describing each individual slit. The position correlation of these photons as well
as the entanglement can be confirmed by measuring in the near-field and far-field of the apertures, respectively.  In
reference \cite{lima09}, it was demonstrated that different
spatial qudit states can be generated by sending one of the
down-converted beams through diffractive apertures with controllable
transmission coefficients, produced with a spatial light modulator.
\par
Another technique for generating spatial qudits is by simply defining a discrete array of spatial regions.  This has been
realized in the single-photon regime\cite{walborn06a} for dimension $D=37$ and entangled-photon regime for up to $D=6$ \cite{osullivan05}.  Using lens
systems, it is possible to create and measure superposition states \cite{walborn08a}.

Possible applications of two entangled qudits in quantum communication rely on the ability of transmitting these
photonic high dimension entangled states between different communicating parties. The propagation of entangled states of
qubits have been subject of recent studies. Optical-fibers links were used to send energy-time entangled qubits for receivers
separated by more than $10$km \cite{Gisin98} and a test of the Bell inequality \cite{bell65} showed that the two-photon state was
still in an entangled state. Free-space distribution of polarization entangled qubits through the atmosphere was
demonstrated for a distance of 144km \cite{ursin07}. The experimental free-space propagation of two entangled four-dimensional
qudits or ``ququarts" produced by four-slit apertures was reported in reference \cite{Lima06}. Free space propagation of the entangled photons was performed on a laboratory scale. The spatial correlation of the propagated state
was observed at the image plane of two lenses and high-fidelity with its original form was measured. Far from the image plane a
conditional interference pattern of the propagated  state was measured. The presence of the conditional interference pattern
shows that the state of the propagated ququarts is entangled \cite{neves05}.  Several studies of free space propagation of single and biphoton fields with OAM have been performed, and are mentioned in more detail in section \ref{sec:qcrypto}.
\par
An important step in future use of entangled spatial qudits is the development of quantum tomography suitable for this degree of
freedom or a method for the determination of the density matrix of the compound system. In references \cite{Lima07,Lima08} a technique
for determination of the density operator was developed for the case of two spatial qubits, generated with two double-slits placed
in the path of the photon pair. Measurements were performed at the image plane and in the focal plane of a lens for different positions of the detector in the
transverse plane.
This method can be used for the tomography of two spatial qubits in a mixed state \cite{Lima08a}. In a second method,
the interference patterns obtained at the intermediate plane between the image and the focal plane are used for reconstruction of the quantum state 
of two spatial qubits \cite{Taguchi08} and qutrits \cite{Taguchi09}.

\subsection{Quantum Cryptography}
\label{sec:qcrypto}
An interesting application for spatial qudits is quantum key distribution, in which the increased dimensionality presents benefits not only in terms of the information transmission rate but also the security of the protocol.  It is straightforward to generalize the well-known BB84 \cite{bb84} or BBM \cite{bbm92} protocols to qudits \cite{bechmann00a,bourennane01,cerf02}.  In this case, it is possible in principle to send on average $\log_{2} D$ bits per sifted qudit, while an eavesdropper employing an intercept-resend strategy would induce a qudit error rate of $E_{D}=\frac{1}{2}\frac{D-1}{D}$, which saturates at $1/2$ for large $D$.  The increased security is due to the increased disturbance, since half the time the eavesdropper measures in the wrong basis, and consequently sends the wrong state with a probability of $(D-1)/D$.  
\par
One can achieve the advantages mentioned above using discrete sets of transverse modes or employing the continuous position and momentum correlations present in the down-converted photons.  For discrete modes, the quantum cryptography protocol and theoretical security analysis should be nearly identical to the general cases of qudits covered in Refs. \cite{bechmann00a,bourennane01}.  Theoretical proposals and experimental investigations in this direction have focused primarily on the orbital angular momentum degree of freedom \cite{spedialieri04,gibson04,groblacher06}.  
\par
\begin{figure}
\begin{center}
\includegraphics[width=9cm]{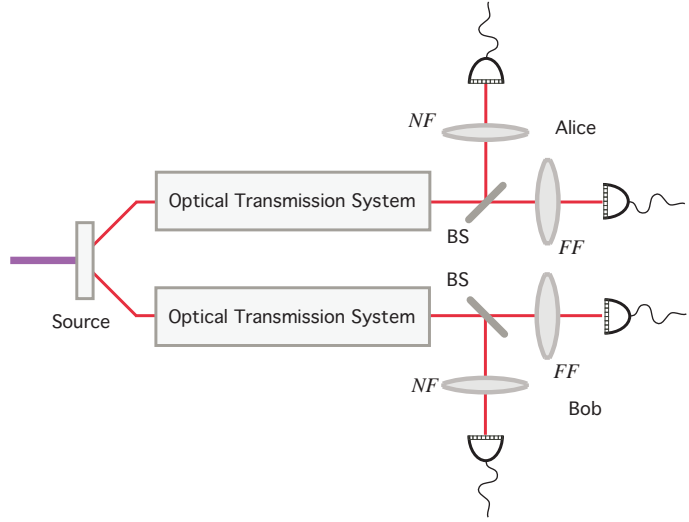}
\caption{Quantum key distribution using transverse position and momentum of entangled photons.  Photons are sent to Alice and Bob's input planes via optical transmission systems.  The random basis selection can be performed using beam splitters (BS).}
\label{fig:qkd}
\end{center}
\end{figure}
The conjugate position and momentum correlations provide an interesting scenario from a practical point of view and in addition to the advantages mentioned above, have been shown to present particular security features.  Figure \ref{fig:qkd} shows an illustration of the main idea.  Alice and Bob receive spatially entangled photon pairs.  Independently, they choose randomly between two conjugate bases: the near field variable $\vect{\rho}$ and far-field variable $\vect{q}$.  The random selection can be performed passively using a 50:50 beam splitter.  Using optical lens systems (represented by lenses $NF$ and $FF$), they map the near and far-field distributions onto their respective detection planes.  Using an array of single-photon detectors, they register the detection position, and thus infer the variable $\vect{\rho}$ or $\vect{q}$.  
Several experimental investigations of this scheme have been performed in the case of entangled photons \cite{almeida05} and single photons \cite{walborn06a}.        
\par
A first motivation for the study of spatial QKD is the increased information transmission rate per photon pair.  In the case that Alice and Bob measure in the $\vect{\rho}$ basis, the secret key rate $\Delta I_{\rho}$ is given by the difference between Alice-Bob and Alice-Eve mutual information:
\cite{grosshans04,zhang08}
\begin{align}
\Delta I_\rho & = I(\vect{\rho}_A|\vect{\rho}_B) - I(\vect{\rho}_A,E), 
\end{align}
where the mutual information is given by
  \begin{equation}
I(\vect{x}_1,\vect{x}_2) = H(\vect{x}_1) - H \left(\vect{x}_1|\vect{x}_2 \right),  
\end{equation} 
and 
\begin{equation}
H(\vect{x}) = - \int d\vect{x} P(\vect{x}) \ln P(\vect{x}), 
\end{equation}
\begin{equation}
H(\vect{x}_1|\vect{x}_2) = - \iint d\vect{x}_1d\vect{x}_2 P(\vect{x}_1,\vect{x}_2) \ln P(\vect{x}_1|\vect{x}_2), 
\end{equation}
are the entropy and conditional entropy respectively, for a continuous distribution \cite{shannon,coverthomas}. $P(\vect{x})$ is the probability distribution for $\vect{x}$, $P(\vect{x}_1,\vect{x}_2)$ is the joint probability distribution for $\vect{x}_1,\vect{x}_2$ and $P(\vect{x}_1|\vect{x}_2)$ is the conditional probability distribution for $\vect{x}_1$ conditioned on measurement of $\vect{x}_2$. A similar expression can be written for the secret key rate $\Delta I_q$ corresponding to $\vect{q}$ measurements.  Here $E$ stands for the Eve's measurement, and we are considering the case of forward reconciliation \cite{grosshans04}.  
Using the above equations, the key rate can be put in the form
\begin{equation}
\Delta I_\rho =  H(\vect{\rho}_A|\vect{\rho}_B) - H(\vect{\rho}_A|E),  
\end{equation}
and similarly for $\Delta I_q$.  
For individual attacks, one can consider that Alice, Bob and Eve share a pure state \cite{grosshans04}.  Thus, upon measurement, Bob and Eve prepare Alice's photon in some state, which must satisfy the entropic uncertainty relation
\cite{bialynicki75}:
\begin{equation}
H(\vect{\rho}_A|E) + H(\vect{q}_A|\vect{q}_B) \geq \ln \pi e.   
\end{equation} 
To establish a secret key requires $\Delta I \geq 0$.  This leads to \cite{grosshans04}
\begin{equation}
\Delta I \geq \ln \pi e - H(\vect{q}_A|\vect{q}_B) - H(\vect{\rho}_A|\vect{\rho}_B)\geq 0. 
\end{equation} 
For any distribution with variance $\sigma^2$, the entropy satisfies $H \leq \ln \sqrt{2 \pi e \sigma^2}$ \cite{shannon,coverthomas}.  Thus, the key rate can be bounded by \cite{zhang08}
\begin{equation}
\Delta I \geq - \ln 2 \Delta(\vect{q}_A|\vect{q}_B)\Delta(\vect{\rho}_A|\vect{\rho}_B),
\label{eq:QKDEPR}
\end{equation}
where $\Delta(\vect{x}|\vect{y})$ is the variance of the distribution of $\vect{x}$ given a measurement of $\vect{y}$.

 This condition is fulfilled when $\Delta(\vect{q}_A|\vect{q}_B)\Delta(\vect{\rho}_A|\vect{\rho}_B) \leq 1/2$, 
which is equivalent to the EPR variance criterion \eqref{eq:reid}. 
For photons produced by SPDC, the EPR criteria depends on the characteristics of the pump beam and the length of the non-linear crystal $L$.   
\par
Implementation of QKD requires preservation of quantum information over long transmission distances, and as such is a technical challenge.  QKD with spatial variables should be most aptly suited to a free-space transmission, as opposed to optical fibers, due to the multi-mode character of the down-converted fields.  There are at least two main factors which limit the distance at which spatial QKD could be a viable endeavor.  The first are fluctuations in the index of refraction along the free-space transmission path. This causes local phase fluctuations which distort the single-photon fields, increasing the error rate.   Atmospheric turbulence causes fluctuations in the index of refraction which transform an initial field $\psi(\vect{\rho}) \longrightarrow \exp[i \delta(\vect{\rho})] \psi(\vect{\rho})$, where the position-dependent phase $\delta(\vect{\rho})$ describes the phase fluctuations in the wave function.  The transmission of spatial quantum information under these conditions has been considered in Refs. \cite{paterson05,smith06} in the case of Laguerre-Gaussian modes and in Ref. \cite{walborn08a} for single-photon spatial QKD.   Assuming that the phase fluctuations are isotropic and a random Gaussian process, the induced error rate is dependent upon the initial field and the length scale of the phase fluctuations known as the Fried parameter \cite{fried66,paterson05}.  Typical values of the Fried parameter are $10-60$cm \cite{wright05}.  For realistic experimental parameters, it was shown in Ref. \cite{walborn08a} that spatial QKD should be feasible up to a few kilometers.  It is important to note that the recent progress in the field of classical optical communication has shown that it should be possible to correct the effects of turbulence in real time with adaptive optics and a reference beam \cite{wright05,tyson02,tyson03}.  A recent experiment by Pors {\it et al.} has shown that OAM entanglement should resist typical atmospheric turbulence for up to a few kilometers \cite{pors09}.     
 \par
The second limiting factor is the presence of losses due to absorption, which reduces the transmission rate as well as the signal to noise ratio, which Eve can exploit to her advantage.  The noise present in spatial QKD comes from spurious ambient photons and detector dark counts.  Following Zhang et al.  \cite{zhang08}, three types of events can contribute to the coincidence count rate, occurring with probabilities $P$(spdc,spdc), $P$(noise, spdc) or $P$(spdc, noise), and $P$(noise, noise).  Focusing only on dark counts, they show that for reasonable experimental parameters,   one can tolerate losses of a few dB and still satisfy Eq. \eqref{eq:QKDEPR}.  This level of loss corresponds to free-space transmission distances on the order of a few km.  However, it is important to note that the left-side of inequality \eqref{eq:QKDEPR} is an upper-bound which is saturated only when Alice and Bob's coincidence distributions are Gaussian.  However, assuming a constant dark count rate for each detector, the coincidence counts in the presence of loss will be composed of a Gaussian distribution of correlated photons,  combined with a \emph{constant} background count, resulting in a non-Gaussian distribution.  The interesting eavesdropping strategy for Eve is to perform measurements so that she can mask her disturbance with the counts due to noise, so that, in this case, her disturbance should appear as a constant background noise.  This can be achieved with a simple intercept resend strategy, so that  when she measures in the same basis as Alice and Bob, she goes undetected.  Zhang et al. \cite{zhang08} have shown that considering the non-Gaussian nature of the total coincidence distribution allows one to increase the allowable losses by at least a factor of ten, giving a secure transmission distance of several tens of kilometers.  
Thus, in a realistic scenario, the noise due to dark counts can increase the allowable losses, while maintaining security of the protocol.                 
\subsection{Quantum Teleportation}
  Quantum teleportation is a method to transfer quantum information between distant locations, without direct transmission of the physical system containing the information. The original quantum teleportation proposal \cite{bennett93} considered discrete $D$-level($D = 2$) systems. Further proposals generalized it to continuous variables \cite{vaidman94,braunstein98,furusawa98}. Quantum teleportation plays an important role in quantum information processing since it is the basis for quantum repeaters \cite{briegel98}, non-local interactions between light and matter \cite{sherson06}, as well as linear optics quantum computing \cite{gottesman99}.
\par
A quantum teleportation protocol which uses transverse spatial properties of intense light fields has also been proposed \cite{sokolov01,magdenko06}, and a teleportation procedure which can transmit the angular spectrum of a single-photon has been proposed in Ref. \cite{walborn07b}.   
\par
\begin{figure}
\begin{center}
{\includegraphics[width=8cm]{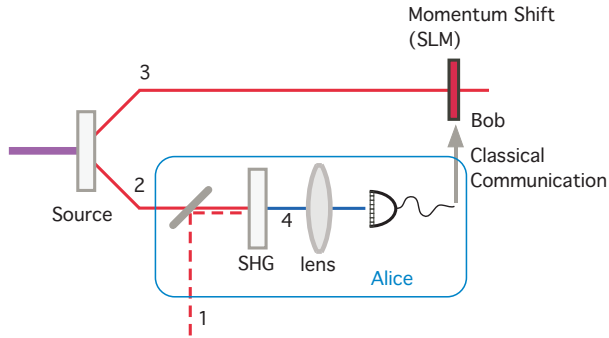}}
\caption{\label{fig:setup} Illustration of teleportation of the spatial degrees of freedom of a single photon. A nonlinear crystal is used to perform second harmonic generation(SHG). Bob's momentum shift can be performed using spatial light modulation (SLM).}
\end{center}
\end{figure}
We will describe the teleportation procedure of Ref. \cite{walborn07b} in more detail. Figure \ref{fig:setup} illustrates the main idea.  
Alice is in possession of a single photon described by the quantum state 
\begin{equation}
\ket{\phi}_1 = \int d\vect{q}\, u(\vect{q}) \ket{\vect{q}}_{1},
\label{eq:uket}
\end{equation}
where $u(\vect{q})$ is the normalized angular spectrum of the field in the $z=0$ plane.  She would like to  
 send the angular spectrum $u(\vect{q})$ of the single-photon field to Bob.  To do this,  Alice and Bob share a pair of spatially entangled photons described by the quantum state

\begin{equation}
\label{eq:standard}
|\psi_{AB}\rangle \propto \int \int d\vect{q}_i d\vect{q}_s v(\vect{q}_i + \vect{q}_s) \gamma(\vect{q}_i - \vect{q}_s),
\end{equation}
where $v$ is the angular spectrum of the laser and $\gamma$ is the phase matching function defined in  Eq. (\ref{eq:quantumstate}).  
Assuming that Alice is in possession of fields 1 and 2 and Bob of field 3,  she follows the usual recipe for quantum teleportation \cite{bennett93}. First, it is necessary for Alice to perform a joint measurement on her fields, and sends the measurement result to Bob using classical communication. Depending on the outcome, Bob performs a unitary operation on his field, and then recovers the initial quantum state of field 1.
\par
Alice can perform the joint measurement on the spatial degrees of freedom of photons 1 and 2 by injecting then into a nonlinear up-conversion crystal cut for second harmonic generation (SHG) or sum frequency generation.  The full multimode Hamiltonian operator for SHG has been described in Ref. \cite{ether06}, and in the thin-crystal limit can be written as
 \begin{equation}
\oper{H} \propto \iiint\hspace{-1mm} d\vect{q}\,d\vect{q}^{\prime}\,d\vect{q}^{\prime\prime}\,\delta^2(\vect{q}+\vect{q}^\prime-\vect{q}^{\prime\prime}) \oper{a}_1(\vect{q})\oper{a}_2(\vect{q}^{\prime})\oper{a}_4^{\dagger}(\vect{q}^{\prime\prime}) + H.c..
 \label{eq:HSHG}
 \end{equation}
Index $4$ denotes the second-harmonic field, which is assumed to be initially in the vacuum state.  After the nonlinear interaction, Alice  measures the transverse wave vector $\vect{q}^{\prime\prime}$ of the up-converted field, and sends the measurement result to Bob.  The detection position is sent to Bob by a classical communication channel, as shown in figure \ref{fig:setup}. The paraxial field operator which describes photodetection at position $\vect{\rho}_D$ in the far-field is given by Eq. \eqref{eq:Eff}.   Applying this operator to the state at the output of the beam splitter, detection of a photon at the position $\vect{\rho}_D$ selects the state
\begin{equation}
\ket{\psi_f} \propto  \iint   d\vect{q}\, d\vect{q}^{\prime}\,   u(\vect{q})  \Phi\left(\frac{\kappa}{f}\vect{\rho}_D-\vect{q},\vect{q}^{\prime}\right)  \ket{\vect{q}^{\prime}}_{3},
\label{eq:jointstate3}
\end{equation}
where all modes in the vacuum state have been excluded, $\kappa$ is the wave number of the SHG field, $f$ is the focal length of the lens, and $\Phi$ is the two-photon angular spectrum defined in Eq. (\ref{amplioo}).
\par
In the limiting case of an EPR-like state, in which the angular spectrum of the SPDC pump field is a plane wave, $v(\vect{q}) \propto \delta(\vect{q})$, and the phase matching function is very broad, so that it is reasonable to approximate $\gamma(\vect{q})\propto 1$. This is the thin crystal approximation described in section \ref{sec:state}.
Under these conditions the state teleported to Bob \eqref{eq:jointstate3} is
\begin{equation}
\ket{\psi_f^{EPR}} = \int d\vect{q}\, u\left(\vect{q}+\frac{\kappa}{f}\vect{\rho}_D\right) \ket{\vect{q}}_{3},
\label{eq:1ket}
\end{equation}
which is equivalent to the original state \eqref{eq:uket} except for a translation in $\vect{q}$. Applying a momentum translation:
\begin{equation}
\oper{P}(\vect{\alpha})\ket{\vect{q}} = \ket{\vect{q}+\vect{\alpha}},
\label{eq:P}
\end{equation}
Bob will then have $\oper{P}(\kappa \vect{\rho}_D/f)\ket{\psi_f^{EPR}}=\ket{\phi}_3$, which corresponds to perfect teleportation of the initial state $\ket{\phi}_{1}$. The momentum translation can be implemented using a spatial light modulator.
\par
In a more general case where $\gamma$ cannot be approximated by unity, the final state after momentum translation is 
\begin{align}
\ket{\psi_f} =  C \iint & d\vect{q}\,  d\vect{q}^\prime  u(\vect{q})v(\vect{q}^\prime-\vect{q})\gamma\left(2\frac{\kappa}{f}\vect{\rho}_D-\vect{q}^\prime-\vect{q}\right)\ket{\vect{q}^\prime}_{3},
\label{eq:jointstate4}
\end{align}
where $C$ is a normalization constant. Assuming that Alice detects a photon in beam 4, the fidelity measures the success of the teleportation protocol \cite{chuang00}, and is defined as $\mathcal{F}=|\langle\phi| \psi_f\rangle|$.  The fidelity depends on the format of the angular spectrum $v(\vect{q})$ and phase matching function $\gamma(\vect{q})$, respectively.  
\par
This teleportation procedure has not yet been realized experimentally, and prevents several serious technical challenges.  The principal obstacle is the low up-conversion efficiency for single-photon fields.  New materials, such as periodically-poled crystals \cite{tanzilli05} or photonic crystal structures may improve these conversion rates.  In spite of the low conversion efficiency, a similar experiment using pairs of non-linear crystals to implement a complete Bell-state measurement in the polarization degree of freedom has been performed \cite{kim01}.   
\par
The same scheme can be used to swap entanglement between spatially entangled photons pairs, which would be necessary for long-distance spatial quantum communication using a large portion of the available Hilbert space.   Another application of this teleportation procedure is the frequency tuning of one photon of a spatially entangled pair. This can be done by replacing photon 1 with an intense field, increasing the up-conversion rate.  
Frequency tuning of entangled photons has been realized using a periodically-poled crystal \cite{tanzilli05}.      

%% file: section10.tex
\section{Conclusion}
\label{sec:conc}

Parametric down-conversion has definitely played a very important role in the development of Quantum Optics
and more recently Quantum Information. This review focused on the spatial degrees of freedom of the down-converted fields, which 
have been an important ingredient.  The quantum correlations between down-converted photons have allowed for
the production of quantum states of light, leading to the observation of non-classical effects and the study of entangled states and their application in Quantum Information tasks. Initially, the concept of spatial coherence in Optics was generalized to include
two-photon coherence. Counter-intuitive aspects of the quantum world have been experimentally observed
with the transverse spatial degrees of freedom, such as the measurement of the De Broglie wavelength of a 
two-photon wave packet, the anti-bunching of photon pairs, conditional two-photon interference patterns, quantum erasure, ghost images and ghost interference.  
These investigations have led to a considerable conceptual improvement and a real advance in the knowledge of the quantum world
through the study of the spatial correlations in parametric down-conversion. Applications immediately
arose from these new concepts, such as quantum lithography and metrology, quantum cryptography schemes,
quantum teleportation and spatial qudits for quantum computation.
\par
This field is still very active and contributions in the fundamental and applied domains appear regularly in the literature.
Among the most promising research in the field, we would like to mention the study of the transverse correlations as a resource for quantum
information in the domain of continuous variables and high dimension discrete states. Transverse spatial variables are natural entangled continuous
variables that can be prepared in a vast range of entangled states whose properties and applications are just beginning to be explored.  